\documentclass[iop,superscriptaddress]{emulateapj}
\usepackage{amsmath,epsfig,url,natbib}
\usepackage{graphicx,epstopdf,float,color,array}
\usepackage{rotating,multirow}
\journalinfo{\@Submitted to the Astrophysical Journal}
\submitted{}
\shortauthors{Hemmati et al.}
\begin{document}
\title{kpc-scale properties of emission-line galaxies}
\author{
Shoubaneh Hemmati\altaffilmark{1}, 
Sarah H. Miller\altaffilmark{1, 2, 3},
Bahram Mobasher\altaffilmark{1},
Hooshang Nayyeri\altaffilmark{1},
Henry C. Ferguson\altaffilmark{4, 5},
Yicheng Guo\altaffilmark{6},
Anton M. Koekemoer\altaffilmark{4},
David C. Koo\altaffilmark{6},
Casey Papovich\altaffilmark{7}
}
\email{shoubaneh.hemmati@ucr.edu}
\altaffiltext{1}{University of California, Riverside, CA 92512}
\altaffiltext{2}{University of California, Irvine, CA 92697}
\altaffiltext{3}{California Institute of Technology, Pasadena, CA 91125}
\altaffiltext{4}{Space Telescope Science Institute,  Baltimore, MD 21218}
\altaffiltext{5}{Johns Hopkins University,  Baltimore, MD 21218}
\altaffiltext{6}{UCO/Lick Observatory and Department of Astronomy and Astrophysics,
  University of California, Santa Cruz, CA 95064, USA}
\altaffiltext{7}{Texas A\&M University, College Station, TX 77843}

\begin{abstract}

We perform a detailed study of the resolved properties of
emission-line galaxies at kpc-scale to investigate how small-scale
and global properties of galaxies are related. We use a sample of 119
galaxies in the GOODS fields. The galaxies are selected to cover a
wide range in morphologies over the redshift range $0.2 < z <
1.3$. High resolution spectroscopic data from Keck/DEIMOS observations
are used to fix the redshift of all the galaxies in our sample. Using
the HST/ACS and HST/WFC3 imaging data taken as a part of the CANDELS
project, for each galaxy we perform SED fitting per resolution
element, producing resolved rest-frame U-V color, stellar mass, star
formation rate, age and extinction maps. We develop a technique to
identify ``regions'' of statistical significance within individual
galaxies, using their rest-frame color maps to select red and blue
regions \footnotemark[a]. As expected, for any given galaxy, the red
regions are found to have higher stellar mass surface densities and older ages
compared to the blue regions. Furthermore, we quantify the spatial
distribution of red and blue regions with respect to both redshift and
stellar mass, finding that the stronger concentration of red regions
toward the centers of galaxies is not a significant function of either
redshift or stellar mass. We find that the ``main sequence'' of star forming galaxies
exists among both red and blue regions inside galaxies, with the
median of blue regions forming a tighter relation with a slope of $1.1\pm0.1$ and a scatter of $\sim
0.2 $ dex compared to red regions with a slope of $1.3\pm0.1$ and a
scatter of $\sim 0.6$ dex. The blue regions show higher specific Star
Formation Rates (sSFR) than their red counterparts with  the sSFR
decreasing since $z\sim1$, driver primarily by the stellar mass
surface densities rather than the SFRs at a giver resolution element.

\footnotetext[a]{a broader defenition for what are called ``clumps''
  in other works}

\end{abstract}

\section{Introduction}

Recent photometric and spectroscopic studies of galaxies show various
trends in the evolution of their structural properties with redshift,
leading to the populations of galaxies we see today.  In particular,
at intermediate redshifts, there is evidence for the growth of central
bulges (e.g.  \citealt{Elmegreen2008}; \citealt{Daddi2007};
\citealt{Hopkins2010}; \citealt{Lang2014}), formation and development
of the Hubble sequence ( e.g. \citealt{wuyts2011};
\citealt{Bell2012}), disk growth \citep{Miller2011}, potential ``disk
settling'' \citep{Kassin2012}, and quenching of star formation
(e.g. \citealt{Bell2004}; \citealt{Faber2007}), all affecting observed
global properties of galaxies.

However, despite extensive observational studies and theoretical
simulations over the last decade, the specific details behind many of
these processes are not well understood. For example, we do not yet
know the underlying reason for the correlation between star formation
rate and stellar masses of star forming galaxies and why the bulk of
the star formation activity occurred early in the most massive
galaxies (e.g., \citealt{Noeske2007}; \citealt{Elbaz2007};
\citealt{Daddi2007}). This relates to outstanding questions of how
mass is assembled in galaxies and the efficiency with which gas is
being converted into stars within different galaxy environments and
feedback processes from star formation and active galactic nuclei
\citep{Voort2011}. It has also been shown that while passive galaxies
play a significant part in measuring the global stellar mass density
at higher redshifts, they have a minimal effect in the star formation
density measurements, especially at later times (e.g.,
\citealt{Dickinson2003}; \citealt{Rudnick2003};
\citealt{Nayyeri2014}). Also, despite significant progress, it is not
yet clear which parameters govern the star formation activity in galaxies and their dependence
on look-back time (e.g., \citealt{Lilly1996}; \citealt{Madau1998};
\citealt{Bell2005}, \citealt{Mobasher2009}).  Processes
potentially responsible for this are a drop in the galaxy major merger rate
(e.g., \citealt{patton2002}; \citealt{Lin2004}; \citealt{Bell2006}; \citealt{Lotz2011}), exhaustion of the cold molecular
gas (e.g., \citealt{Tacconi2013}; \citealt{Daddi2010}) or the
occurance of disk instabilities that cause the migration of gas into
galaxy centers, creating bulges that stabilize the disk against
further clump formation (e.g., \citealt{Elmegreen2008};
\citealt{Dekel2009}; \citealt{Genzel2011};  \citealt{Wuyts2012}, \citealt{Guo2012}).

While a number of studies have been performed to address these questions, they
are largely based on the integrated properties of galaxies,
without considering the details of the individual components and processes internal to galaxies. A physical
understanding of the relations between various global properties of galaxies  
would not be complete without knowledge of the processes involved at kpc
scales or smaller. Furthermore, one could achieve fundamental insights to the
formation and evolution of galaxies by studying the relations between
the integrated and internal (i.e. resolved) properties of galaxies over time. This requires deep
multi-waveband images with high spatial resolution for representative
samples of galaxies with available measurements of their global
properties (i.e. stellar mass, star formation rate). With the advent of high resolution imaging detectors on 
the Hubble Space Telescope (HST), one can resolve kpc-scale structures
in galaxies along with their photometrically derived properties. 

Over the last decade, extensive multi-waveband photometric surveys of galaxies
have become available. The Cosmic Assembly Near-infrared Deep Evolution Legacy Survey 
(CANDELS; PI. S. Faber and H. Ferguson; see \citealt{Grogin2011} and
\citealt{Koekemoer2011} have provided high-resolution
and deep images of galaxies in different passbands spanning the optical and near-IR
wavelengths. Using the imaging data in the GOODS-S field, \citet{Wuyts2012} studied resolved
colors and stellar populations of a complete sample of star-forming galaxies 
at $0.5 < z < 2.5$. The galaxies were selected to be massive ($>
10^{10}$ M$_\odot$) with high specific star formation rates. The
Spectral Energy Distributions (SEDs) were constructed per bins of pixels of constant 
signal-to-noise and analyzed to study variations in rest-frame colors,
stellar surface mass density, age and extinction as a function of
global properties of galaxies. They identified off-center clumps in
galaxies and studied their contribution to the integrated star
formation rate and integrated stellar mass. However, to fully understand the nature of
these structures, one needs resolved spectroscopy.

\citealt{Forster2009} studied the properties of these stellar 
``clumps'' for a sample of six star-forming galaxies at $z\sim 2$ 
using near-infrared integral field spectroscopy from SINFONI at the
Very Large Telescope (VLT). This allowed measurement of the dynamical
mass of the clumps and their spectral line emission diagnostic of star
formation activity. Furthermore, they identified the ``clumps'' based
on their emission lines (i.e. H$\alpha$ emitting clouds) and used this
information to study the nature of these structures. While this
provides a superior technique for studying kpc-scale structure of
galaxies, it is technically challenging. One could acquire such data with adequate S/N ratios
for only a handful of galaxies, making it difficult to generate a statistically large sample.

In the present paper we take a combined approach. Using the latest 
optical and infrared imaging data in the CANDELS fields, we perform
resolution element-by-element (i.e. pixel-by-pixel) SED fits to a sample of star-forming disk galaxies at $
0.2 < z < 1.3$ with deep, spatially-resolved high-resolution
spectroscopy from Keck/DEIMOS (i.e. rotation curves) along their major
axis. The spatially-resolved spectra will make it
possible to unerstand metallicity gradients and dynamics across
galaxies, as well as the relative time scales of various processes
predicted to drive the formation of disks. We develop techniques to
generate kpc-scale resolved, self-consistent photometric maps allowing
for differences in the image PSF and resolution. We investigate how
star formation rate and stellar mass surface density correlate at
kpc-scale structure and whether there is a difference in evolution
with time between different regions inside galaxies. A full
utilization of spectroscopic data is coming in subsequent papers in
this series.

The structure of this paper is as follows. \S 2 presents the sample
selection. In \S 3 we develop the method to perform resolved SED fitting, producing high resolution photometric maps of
different observables in galaxies. In \S 4, we compare the integrated
and resolved properties of galaxies. We present a technique for identifying
physical regions in galaxies in \S 5. Results are
presented in \S 6, discussed in \S 7, and summarized in \S
8. Throughout this paper all magnitudes are in AB system
\citep{Oke1983} and we use standard cosmology with $H_{0}=70\:kms^{-1}
Mpc^{-1}$, $\Omega_{M}=0.3$ and $\Omega_{\Lambda}=0.7$.

\section{Sample Selection}

The sample for this study is selected in GOODS-S and GOODS-N fields
\citep{Giavalisco2004}, with available ground-based spectroscopy from DEIMOS (DEep Imaging Multi-Object Spectrograph; \citealt{Faber2003}) on Keck II. 

\begin{figure}[htbp]
\centering
\includegraphics[trim=0.1cm 0cm -0.2cm 0cm,width=3.1in]{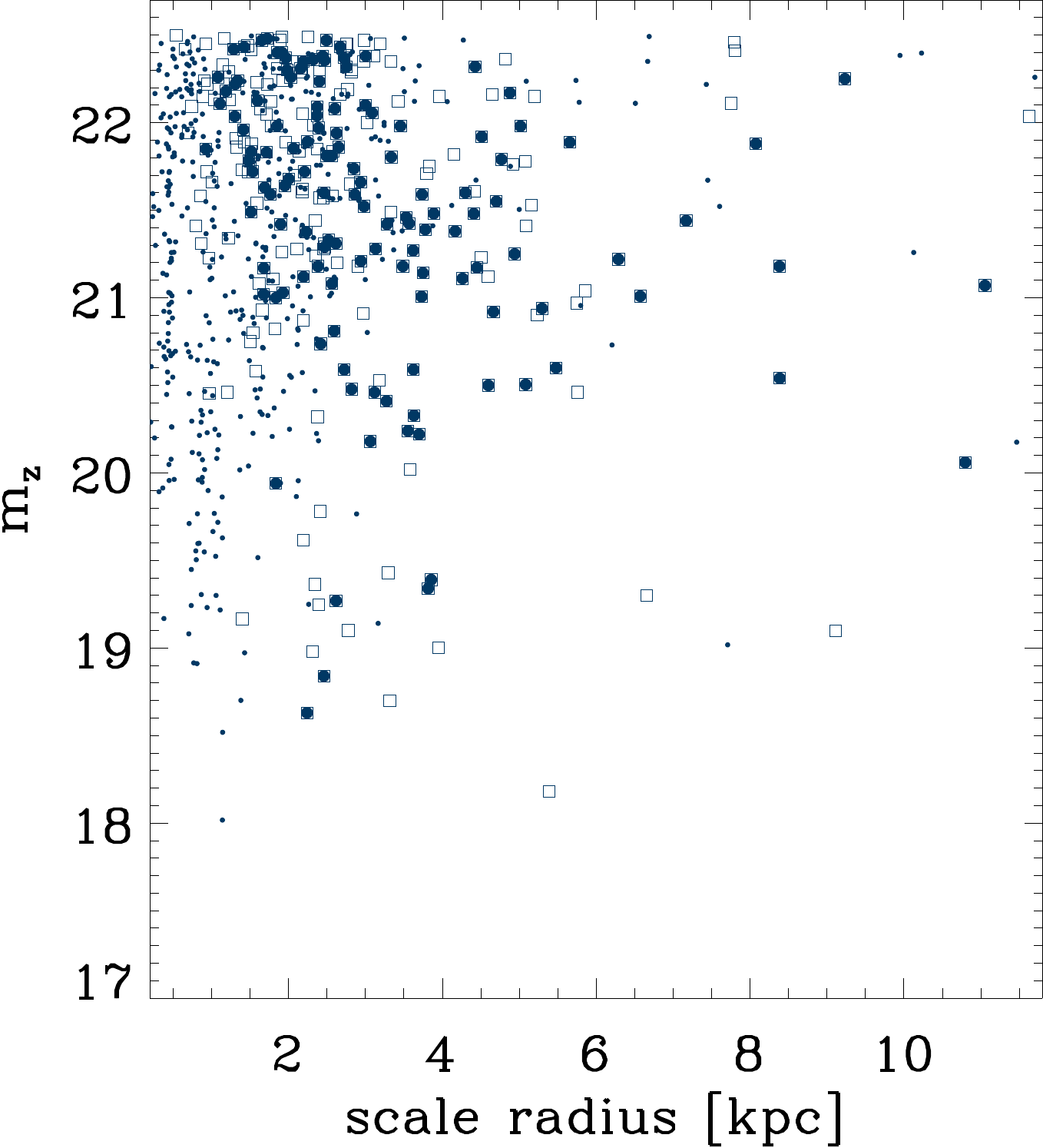} \\
\vspace{0.1in}
\includegraphics[trim=0.1cm 0cm -0.2cm 0cm,width=3.3in]{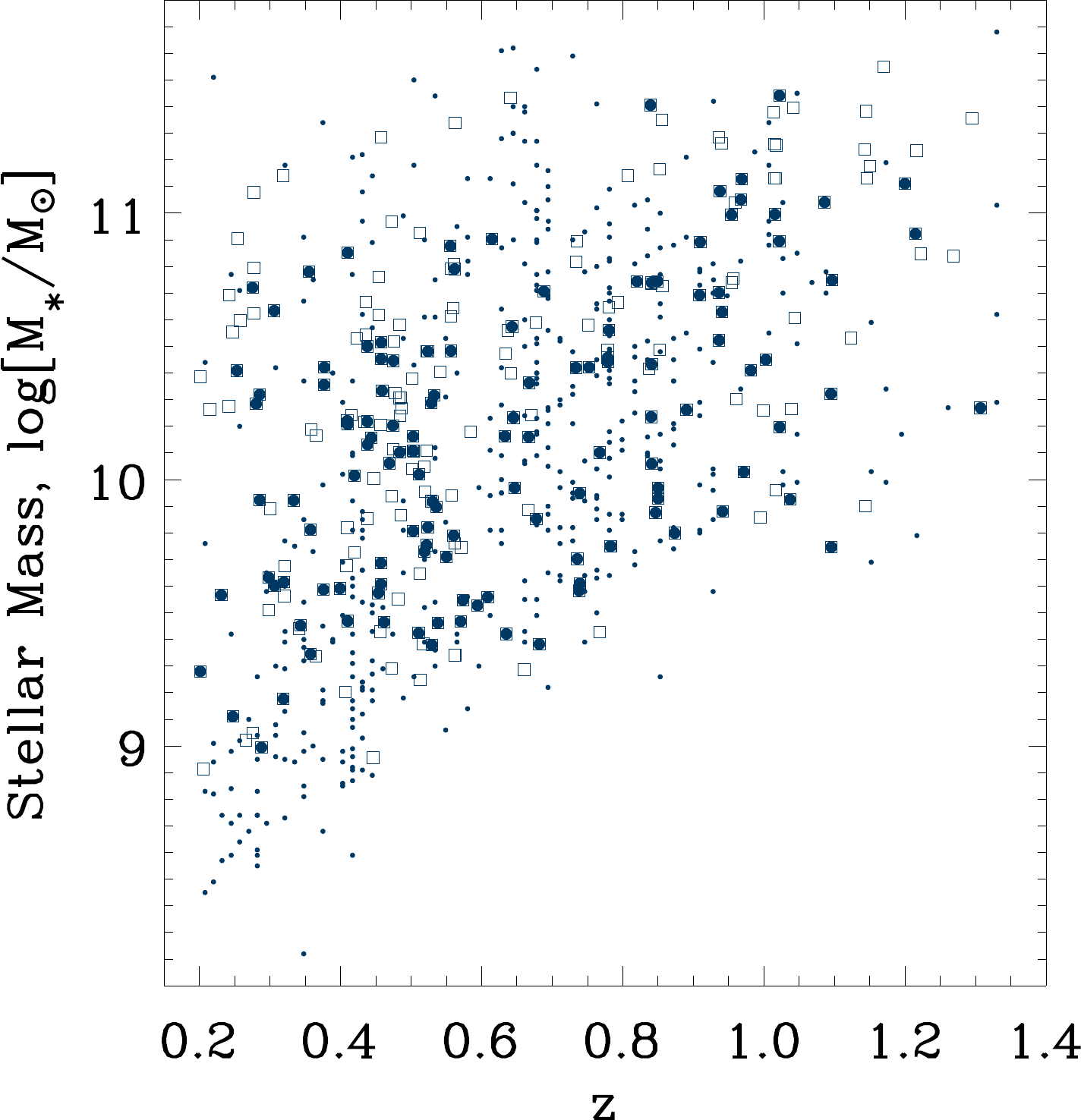}
\caption{Apparent magnitude ($z_{F850lp}$)  vs. disk scale radii (in
  kpc) and stellar mass vs. redshift for galaxies selected for the
  present study, consisting of 129 resolved emissions (large filled
  squares). There are 107 galaxies which also satisfy the selection
  criteria here, with available spectroscopic data
  but unresolved or undetected emission (empty squares). These are not
  included in the analysis in this study. A control sample is also
  shown with similar photometric redshift and magnitude cuts (small
  dots).}
\end{figure}

The galaxies in this sample are selected to be brighter than 22.5
magnitude in the ACS F850lp filter and lie within the spectroscopic redshift range $0.2 < z < 1.3$. They have resolved, disk-like
structures, including not only well-formed spirals but also disturbed and morphologically irregular and abnormal
systems. Early-type spheroids and unresolved objects
were excluded from the sample. Finally, a magnitude limit of $K_{s}
\leq 22.2$ is imposed to ensure a high fraction of reliable stellar masses. 

Keck DEIMOS spectra were obtained for 236 galaxies that were selected
using the above criteria \citep{Miller2011}. A total integration time of 6-8 hours were
acquired  with the $1200$ $ l mm^{-1} $ grating and $1\arcsec$ slits and a
central wavelength of $7500$ \AA, achieving a spectral resolution of
$\sim1.7$ \AA. Of the 236 targets, 129 of the galaxies (about $55\%$ of the
original sample) that were observed spectroscopically, revealed
resolved emission lines for which accurate rotation curves are
measured (the emission extends beyond the seeing-dispersion PSF in the
spectrum). Figure 1 shows that the distribution of sizes or masses of our selected
  targets is not biased with redshift or magnitude compared to galaxies with the same redshift and
  magnitude cut.

This sample is most desirable for studying resolved properties of
galaxies for the following reasons: 

\begin{enumerate}
\item 
Availability of seven high resolution Hubble Space Telescope (HST)
optical and infrared images taken by ACS and WFC3 from the
CANDELS. It has been shown that using optical+NIR filters
reduces the uncertainties in studying resolved properties in galaxies
compared to using optical filters only \citep{Welikala2011}. In this
study we use HST/ACS observations in the F435W, F606W, F775W and
F850LP (hereafter $B_{F435W}$, $V_{F606W}$, $I_{F775W}$ and
$Z_{F850lp}$) and HST/WFC3 observations in the F105W, F125W and F160W
(hereafter $Y_{F105W}$, $J_{F125W}$ and $H_{F160W}$) filters. The ACS
images have been multi-drizzled to the WFC3 pixel scale of 0\farcs06
\citep{Koekemoer2011}.

\item 
The availability of spectroscopic redshifts for all the galaxies. One of
the major sources of uncertainty in measuring galaxy properties through fitting their spectral
energy distribution is the uncertainty in their photometric
redshift. By knowing the precise spectroscopic redshift of the
galaxy we overcome this issue.

\item 
Kinematic information are available with high spectral and spatial
resolution for rotation curves and dynamical models of all the galaxies in this sample. This
allows direct comparison between high resolution photometric maps
(of stellar mass, SFR, etc.) and dynamical information at
kpc-scales. 

\end{enumerate}

\section{Resolved Maps of galaxies}

In this section we explain the step-by-step method we develop to make resolved
rest-frame optical color, stellar mass, star formation rate, age and
extinction  maps by performing spectral synthesis fitting per
resolution element for each galaxy.

\subsection{Cutouts}

The very first step in making resolved maps of galaxies is to make sure that we are 
combining the same resolution and pixel scale in different images. The pixel
scales of CANDELS HST images are all set to $0.06''$. However 
we need to PSF-match these images across the range of the filters used. We
take all lower resolution images to the resolution of F160W which has
the longest wavelength coverage.

\begin{figure*}[htbp]
\centering
\includegraphics[trim=3.0cm 11.0cm 4.0cm 11.0cm,scale=0.8]{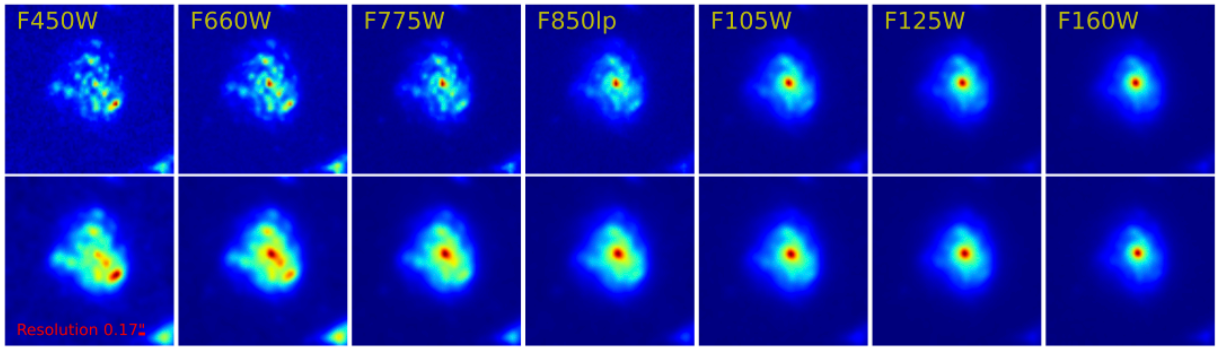}
\caption{7-band HST $80\times 80$ pixel cutouts of a typical galaxy from
    our sample at z=1.1 (80 pixel = 39.7 kpc) before and
  after PSF-matching in top and bottom rows respectively. }
\end{figure*}

We make 14 cutouts (7 band-HST images + their corresponding weight
maps) each of $200\times200$ pixels. We construct the PSF for each band using the IDL routine
\textsc{psfgen.pro}, and perform PSF-matching using the IRAF \textsc{PSFMATCH}
task. This gives a spatial resolution of $\sim 0.17\arcsec$. After
matching the resolutions we make the smaller cutouts of $80\times80$
pixels which at the redshift range of our sample corresponds to a box size of $\approx 15-40$  kpc (see Figure 2).

We removed $10$ galaxies located at the edge of the fields because
the constructed PSF is not a good representation of the true PSF at these locations, due to
distortion, lower exposures, and signal-to-noise level. The final
results are sensitive to this as a wrong PSF kernel would lead to
ripples in the final maps. Fitting Gaussians to the PSFs before
performing the psf-matching would clear this issue, however for the
sake of consistency, we discard these 10 galaxies. This leaves us with
119 galaxies.

To define the boundary of the galaxy, we run the Source Extractor code \citep{Bertin1996}
 on the z band cutouts (seen in Fig. 2, of all the ACS bands, the z
 band reveals the deepest, smoothest, most extended light. This
 feature is seen in most of the galaxies in our sample). The
 segmentation maps of each galaxy, produced by SExtractor, are then
 multiplied by the cutouts to mask the low surface brightness
 outskirts of galaxies.

\subsection{SED Fitting per Resolution Element}

One of the most-widely used methods to study physical properties of
galaxies (e.g. stellar masses, star formation histories and rates,
metallicities, ages) at different redshifts is to fit their observed
spectral energy distributions with stellar population synthesis
models. While there still exists many sources of uncertainties in the
  models and fitting methods, there has been huge progress in
  finding statistical properties of galaxies at all redshifts.
(e.g. \citealt{Reddy2012}, \citealt{Conroy2013} ) 

Here, we apply this technique to the SEDs per resolution element in each galaxy instead of its integrated light
(e.g. \citealt{Zibetti2009}, \citealt{Welikala2011}, \citealt{Wuyts2012},
\citealt{Lanyon2007}). At the most basic level, resolved SED fitting
allows the analysis of substructures in any given galaxy and reduces
the uncertainties caused by using an ``average''  dust attenuation law
and star formation history for the whole galaxy. For each galaxy we build a multi-wavelength catalog. Each row in
the catalog corresponds to a resolution element inside the galaxy with four band ACS, three
band WFC3, their corresponding RMS errors and spectroscopic redshift of the galaxy. 

\begin{table}
\centering
\caption{Parameters for BC03 Models}
\begin{tabular}{*{3}{c}}
\hline
\hline
Parameters & number & Range \\
\hline
\\
Age (Gyr) & 57 & $0.01-13.5$\footnotemark[1] \\
Extinction E(B-V) & 15 & $0.0-1.0$\footnotemark[1] \\
$\tau$ (Gyr) & 21 & $0.001-10.0$ \footnotemark[1]\\
metallicity &\multicolumn{2}{c}{ $40\%\  Z_{\odot}$} \\
\hline
\end{tabular}
\footnotetext[1]{Not equally spaced}
\end{table}

The first phase of the SED-fitting process is to generate an inclusive library of model SEDs which span a wide
range in parameter space. We use the PICKLES library for
stars \citep{PICKLES1998}, different synthetic and composite quasar
libraries available in the LePhare package (\citealt{Arnouts1999};
\citealt{Ilbert2006}) and BC03 models \citep{BC03} for the galaxy
library. While the relative contribution of thermally pulsating
asymptotic giant branch (TP-AGB) stars is still uncertain, it has been shown in previous
studies that the Maraston \citep{Maraston2006} models over predict the
near-infrared luminosity \citep{Kriek2010} compared to BC03
. We assume a Chabrier \citep{Chabrier2003} stellar Initial Mass
Function (IMF). Extinction values range from zero to one using the
Calzetti Starburst and 57 different ages ranging from zero to the age
of the universe at the spectroscopic redshift of the galaxy. We
adopted single burst, constant and exponentially declining star
formation histories. We fix the metallicity to  $40\%$ solar to
mitigate the fitting degeneracies by shrinking the library size. We explore the effects
  of choosing this particular grid in the appendix by comparing it to
  models with different metallicities, star formation histories and
  different resolutions in extinction and age.

\begin{figure*} [htbp]
\centering
\begin{tabular}{|c|c|}
\hline
 \bf{$z = 0.56$}&\bf{$z = 1.1$}\\
\hline
\includegraphics[trim=0cm -0.1cm 0cm -0.2cm,clip,width=3.cm]{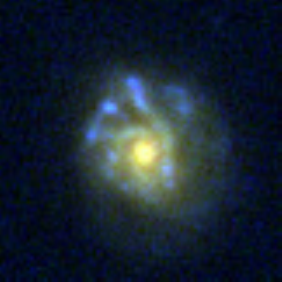}&
\includegraphics[trim=0cm -0.1cm 0cm -0.2cm,clip,width=3.cm]{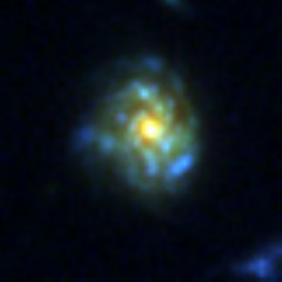}\\
\hline
\includegraphics[trim=0.5cm 16.4cm 1.8cm 2.2cm,clip,width=7.5cm]{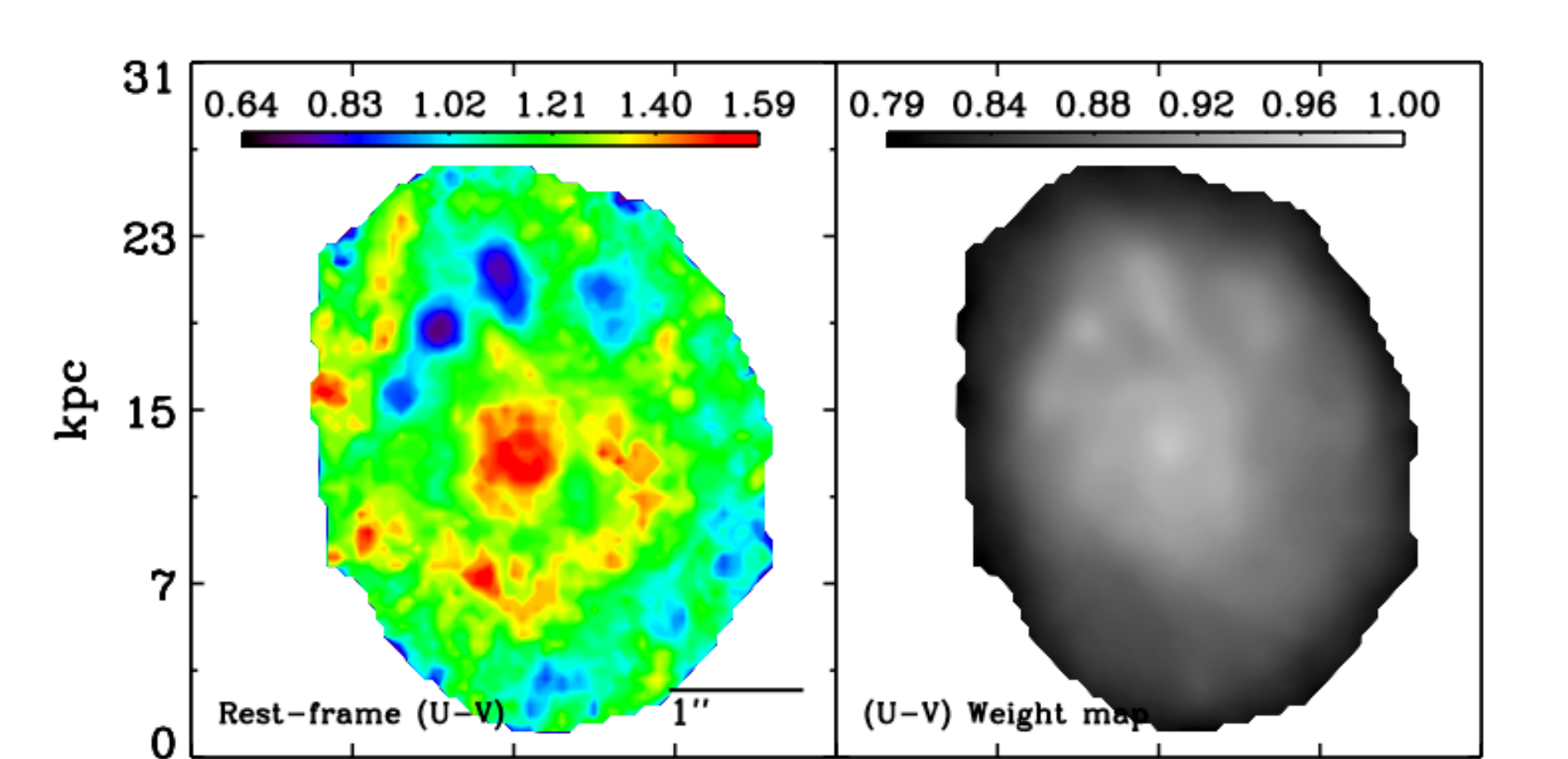}&
\includegraphics[trim=0.5cm 16.4cm 1.8cm 2.2cm,clip,width=7.5cm]{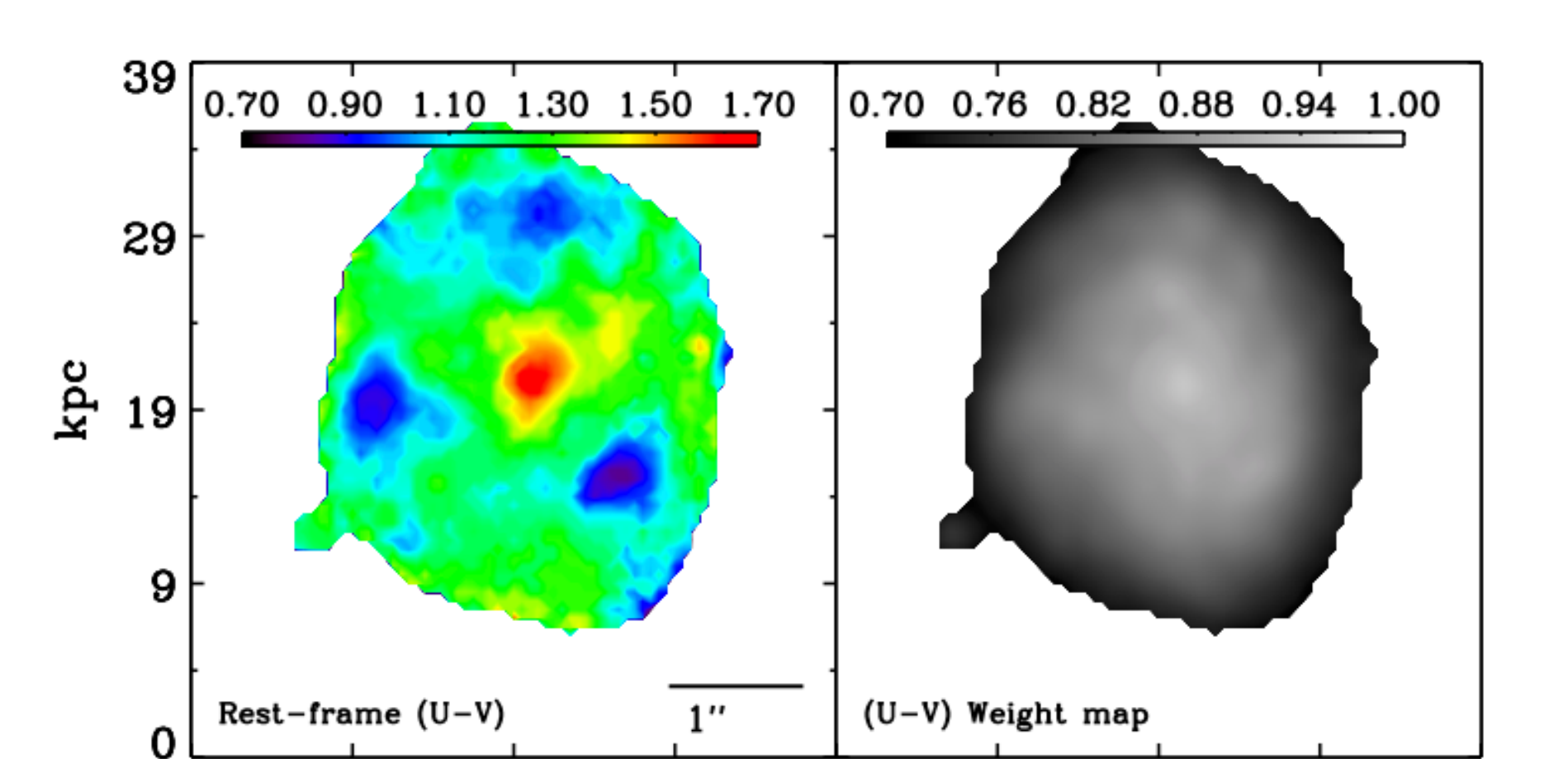}\\
\includegraphics[trim=0.5cm 16.4cm 1.8cm 2.2cm,clip,width=7.5cm]{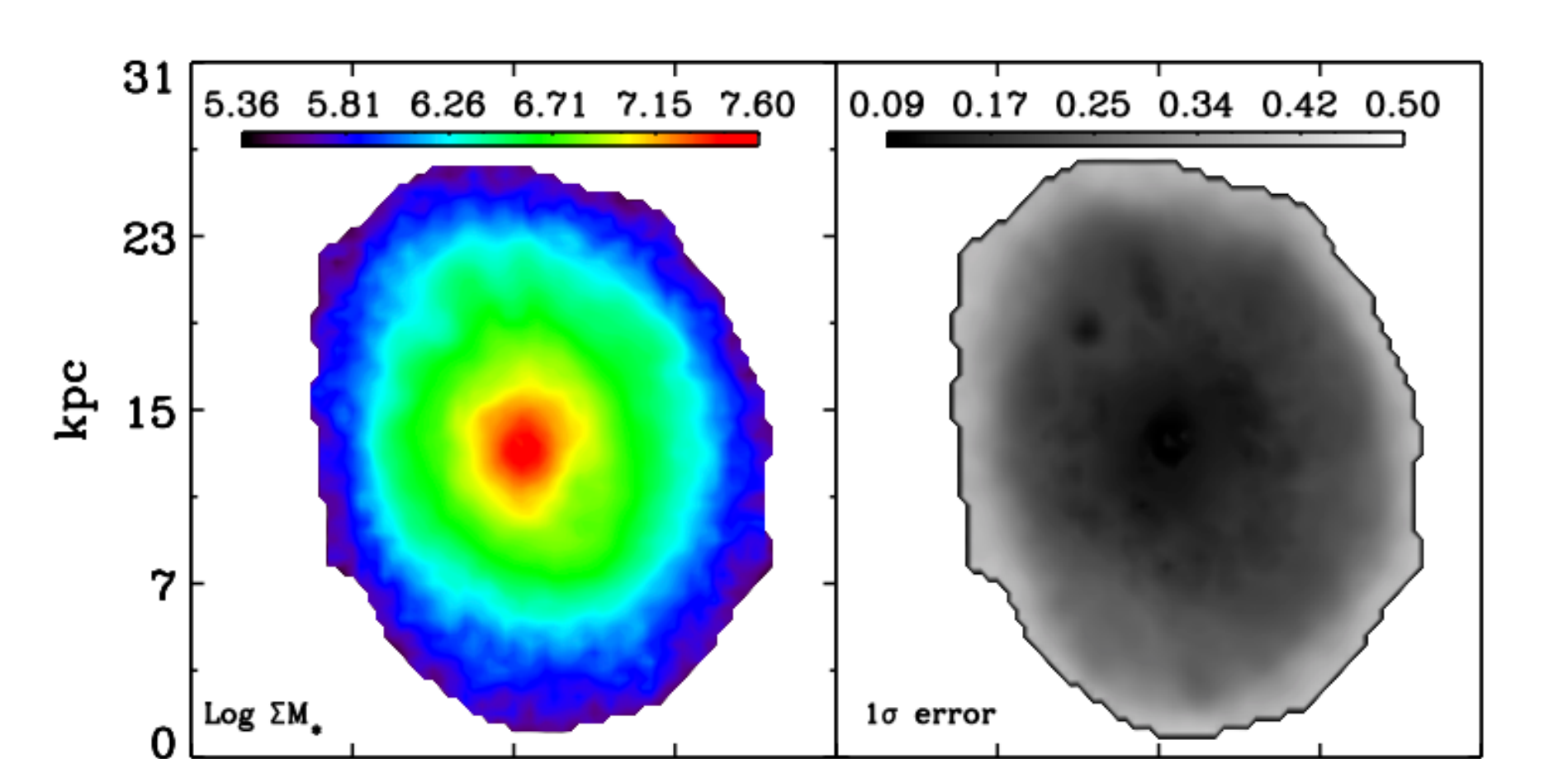}&
\includegraphics[trim=0.5cm 16.4cm 1.8cm 2.2cm,clip,width=7.5cm]{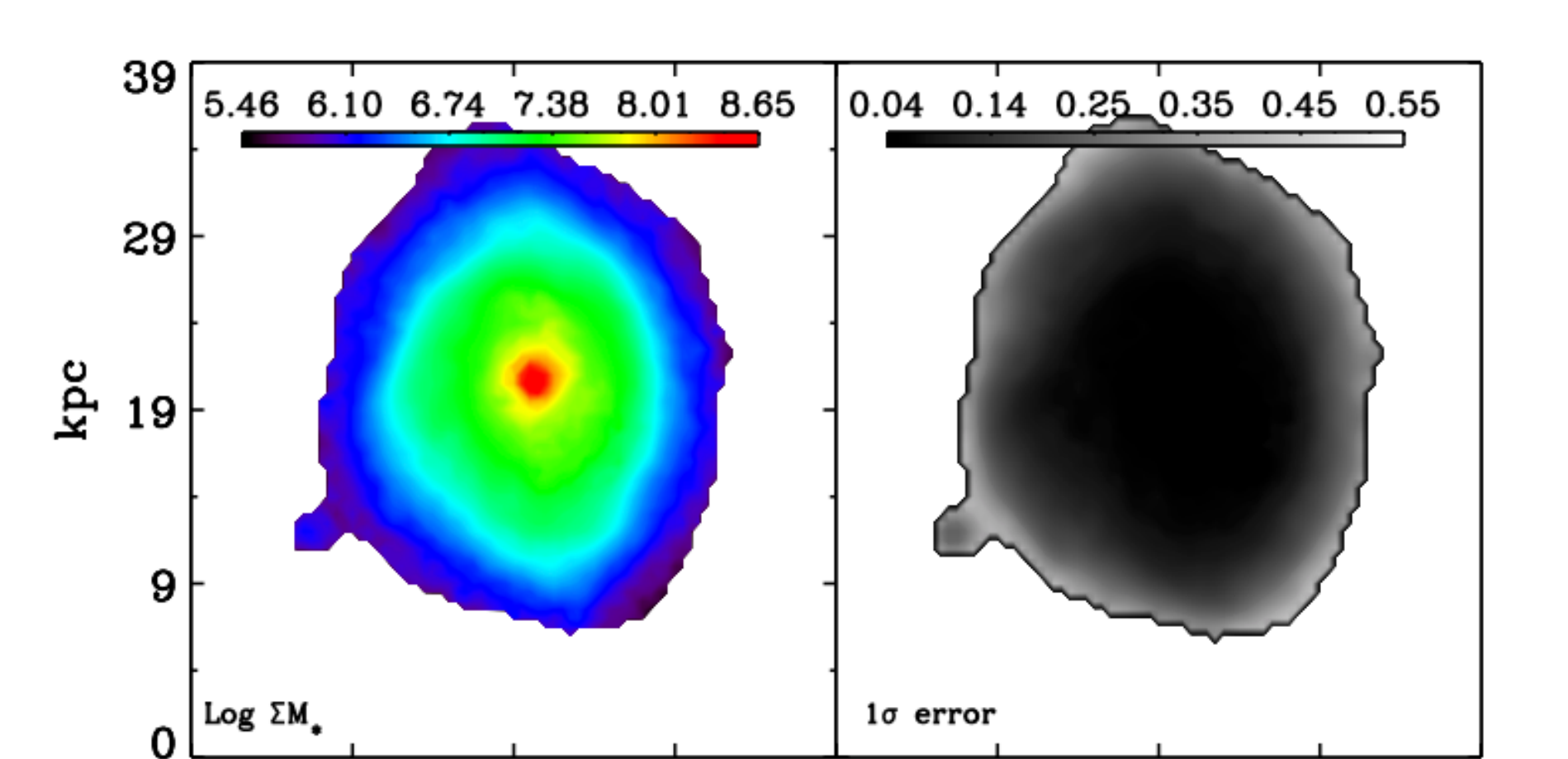}\\
\includegraphics[trim=0.5cm 16.4cm 1.8cm 2.2cm,clip,width=7.5cm]{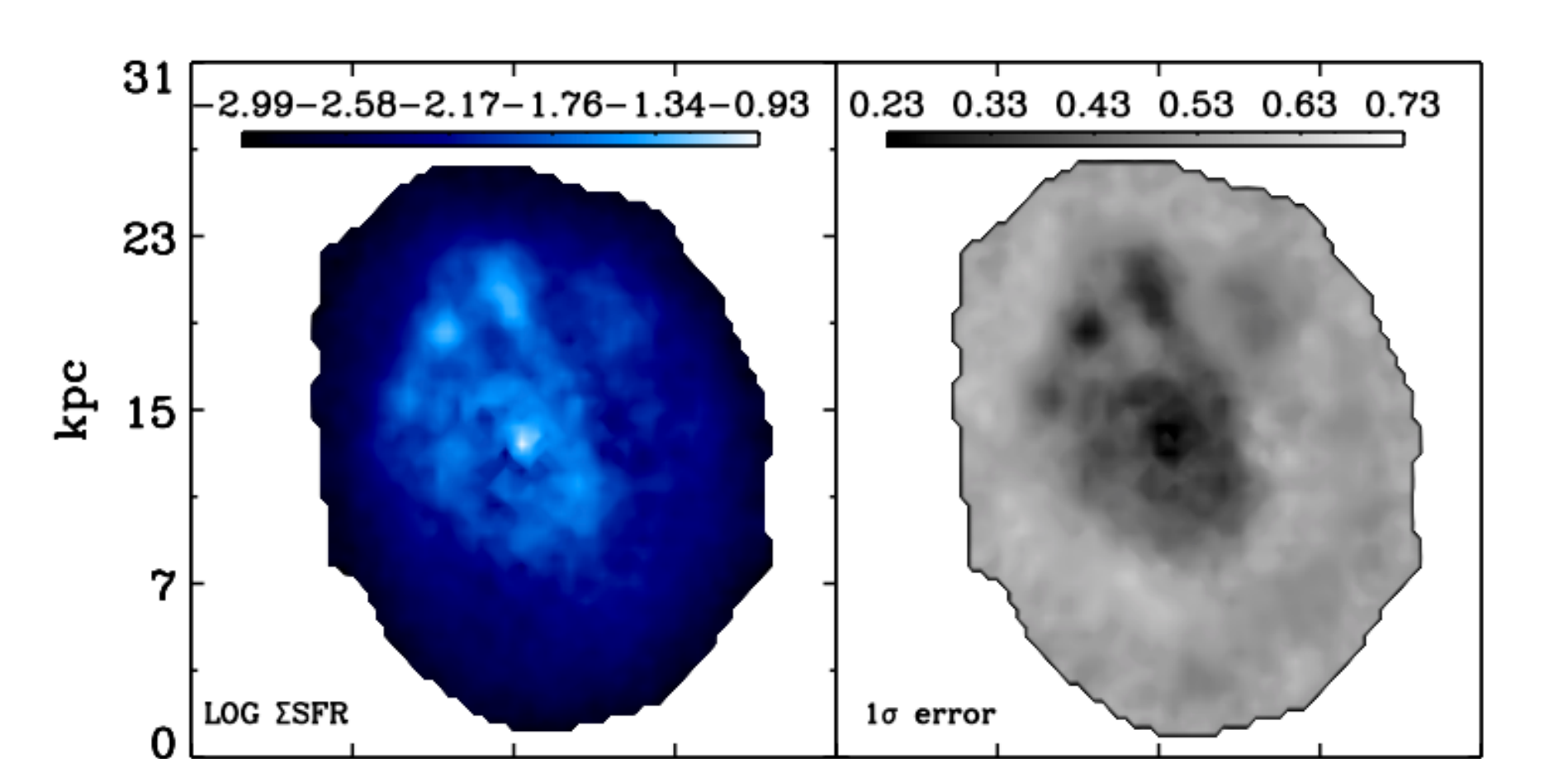}&
\includegraphics[trim=0.5cm 16.4cm 1.8cm 2.2cm,clip,width=7.5cm]{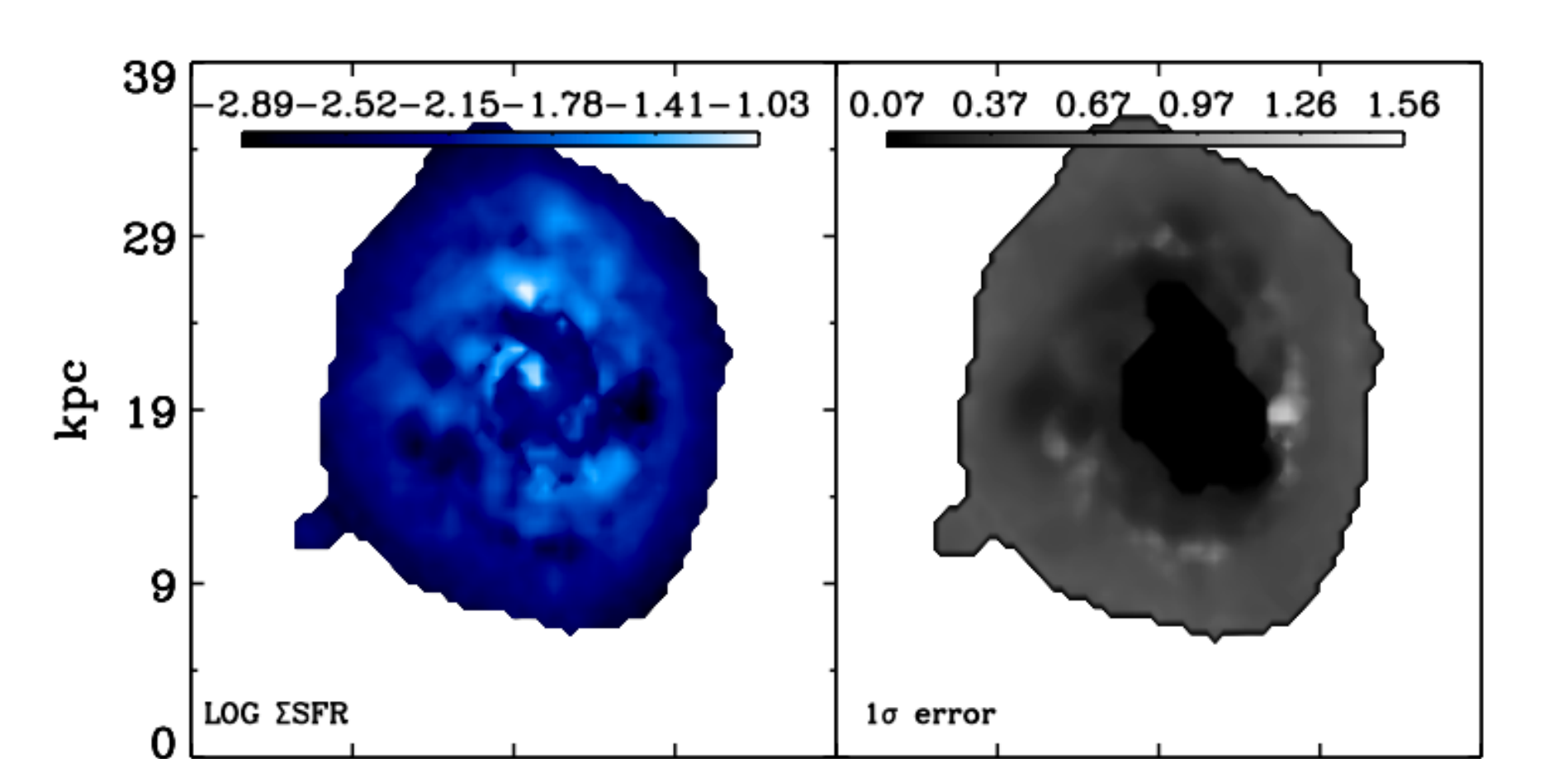}\\
\includegraphics[trim=0.5cm 16.4cm 1.8cm 2.2cm,clip,width=7.5cm]{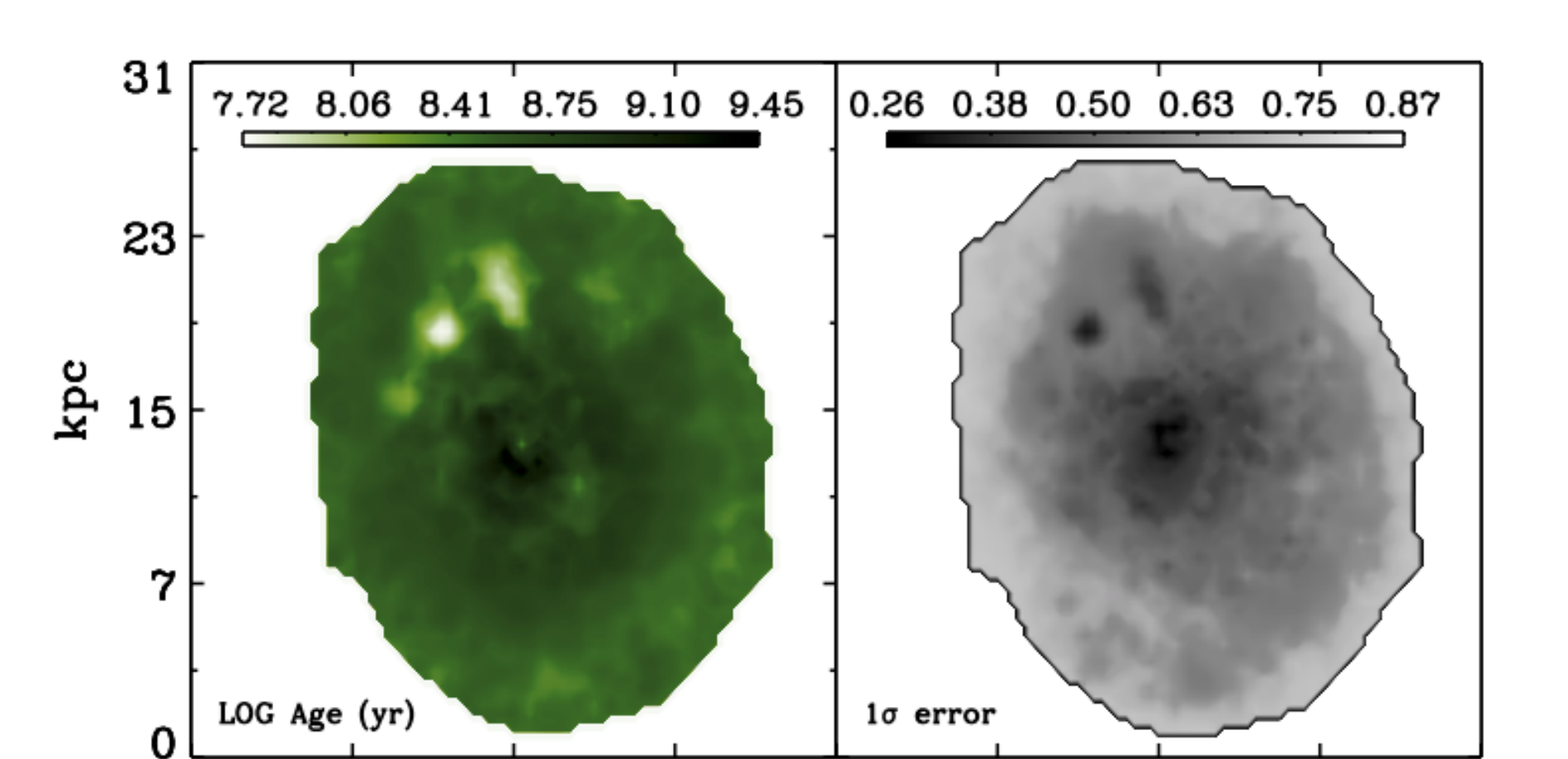}&
\includegraphics[trim=0.5cm 16.4cm 1.8cm 2.2cm,clip,width=7.5cm]{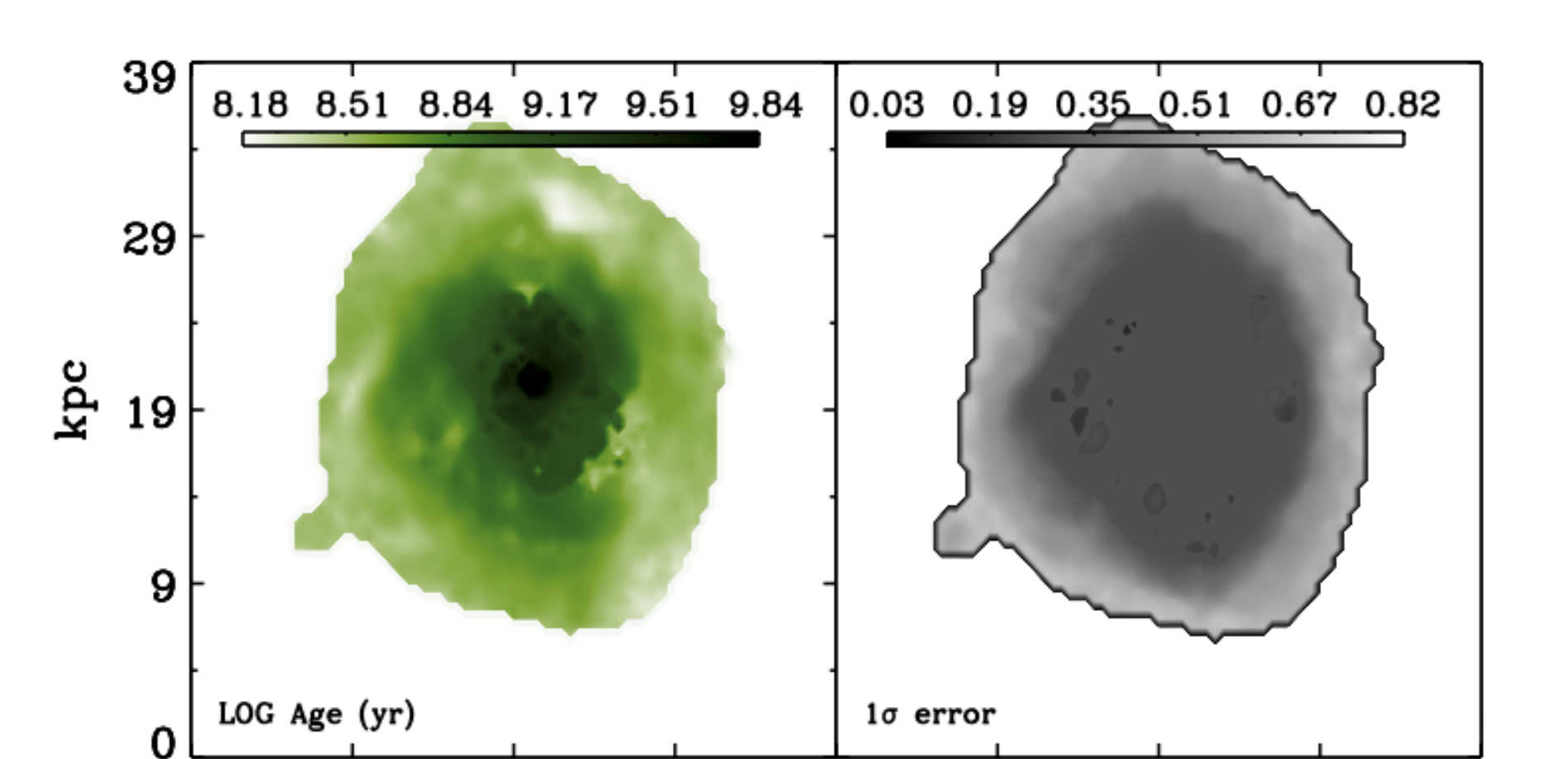}\\
\includegraphics[trim=0.5cm 14.9cm 1.8cm 2.2cm,clip,width=7.5cm]{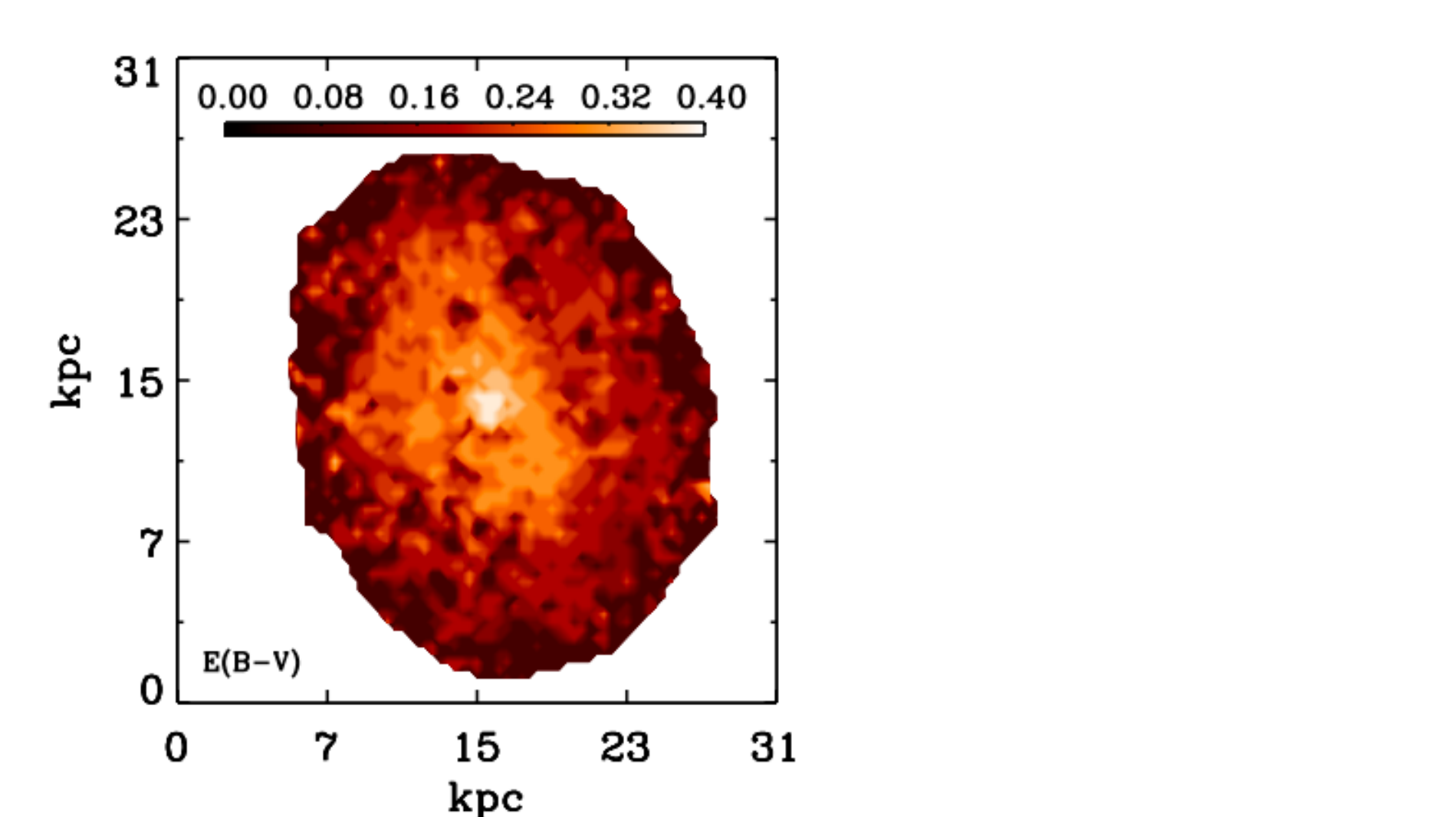}&
\includegraphics[trim=0.5cm 14.9cm 1.8cm 2.2cm,clip,width=7.5cm]{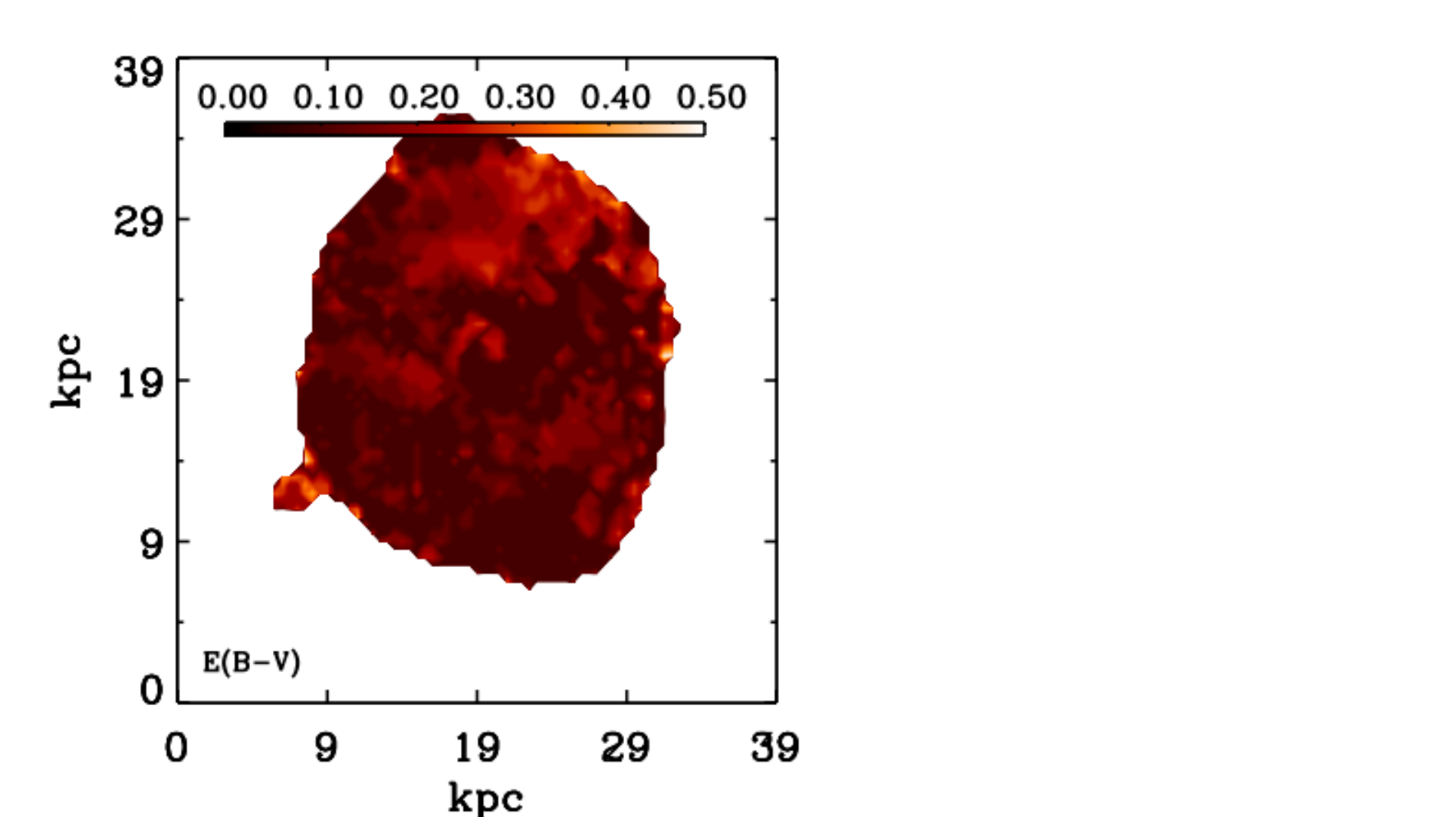}\\
\hline
\end{tabular}
\caption{Top to bottom: Stacked BZH color image, Rest-frame (U-V) color,
  Rest-frame color weight, stellar mass surface density and $1\sigma$ error, star
  formation surface density and $1\sigma$ error, stellar age and
  $1\sigma$ error and extinction maps (LePhare does not produce
  uncertainties for extinction) for 2 galaxies
      at z= 0.56 (left) and 1.1(right).}
\end{figure*}

\begin{figure*} [htbp]
\centering
\begin{tabular}{|c|c|}
\hline
 \bf{$z = 0.31$}&\bf{$z = 0.94$}\\
\hline
\includegraphics[trim=0cm -0.1cm 0cm -0.2cm,clip,width=3.cm]{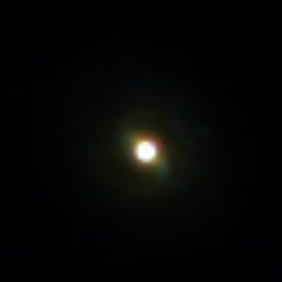}&
\includegraphics[trim=0cm -0.1cm 0cm -0.2cm,clip,width=3.cm]{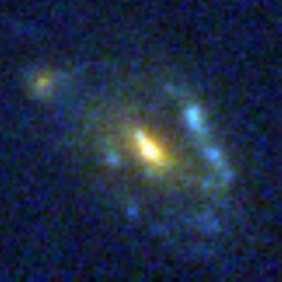}\\
\hline
\includegraphics[trim=0.5cm 16.4cm 1.8cm 2.2cm,clip,width=7.5cm]{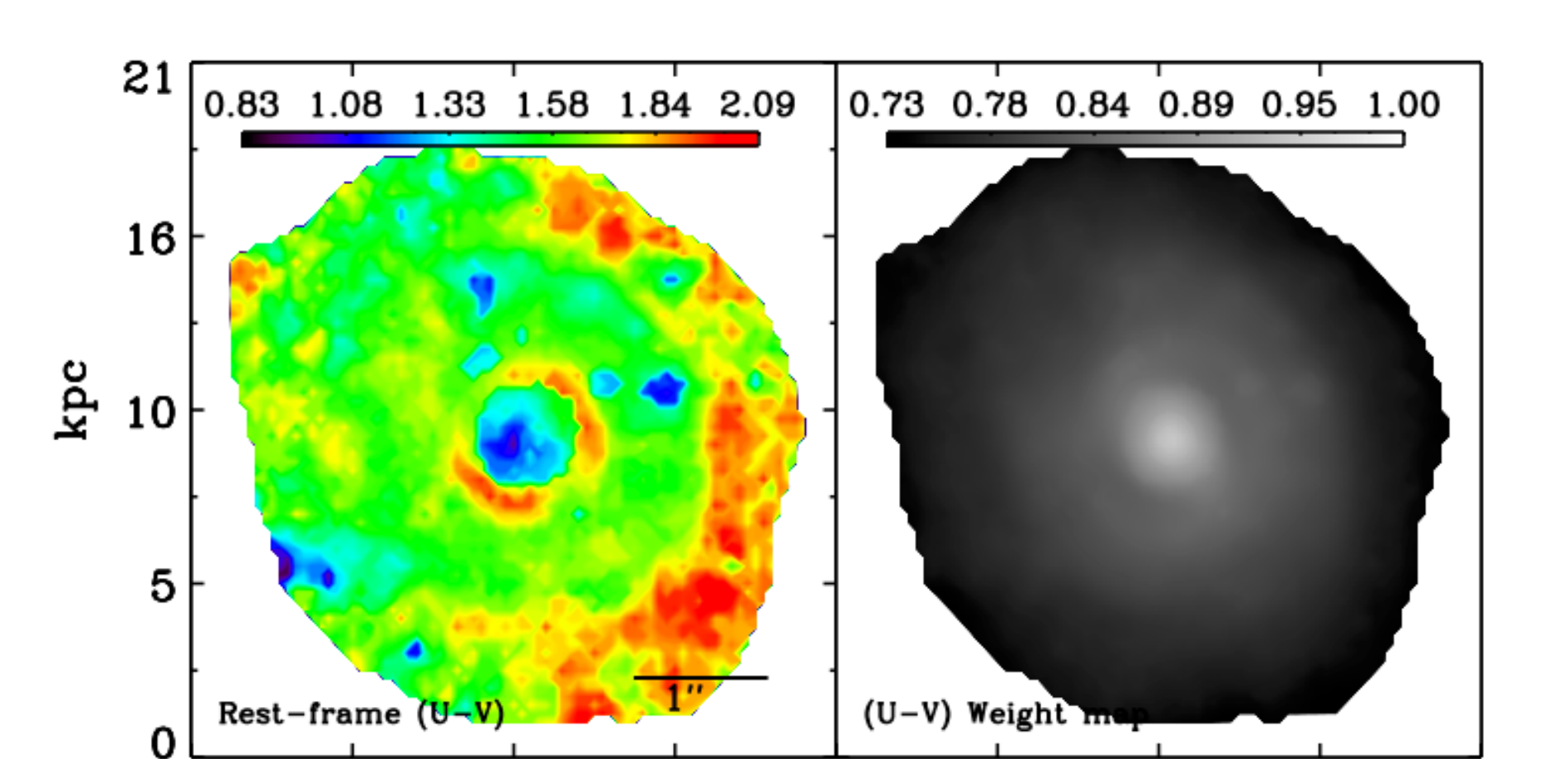}&
\includegraphics[trim=0.5cm 16.4cm 1.8cm 2.2cm,clip,width=7.5cm]{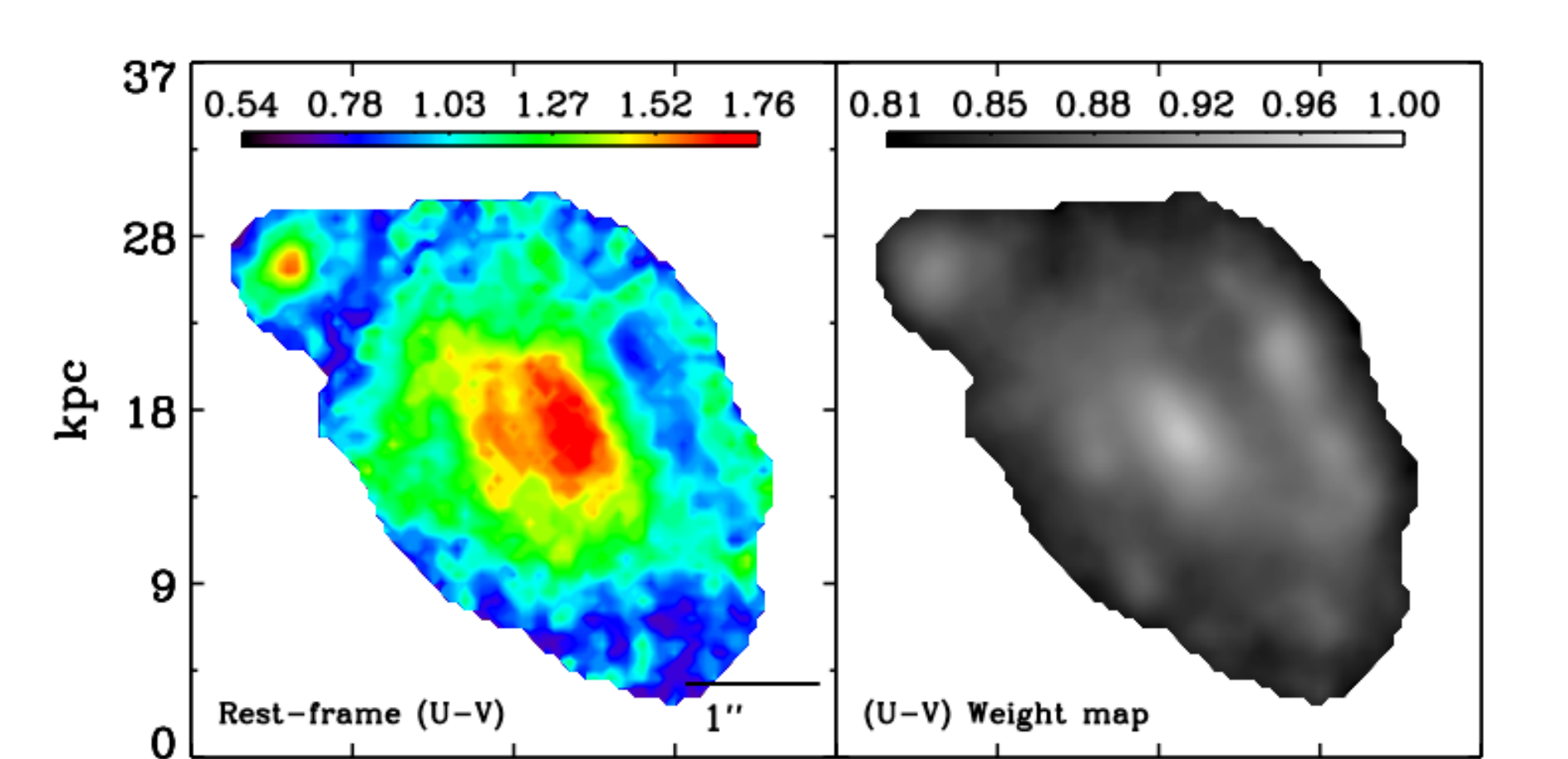}\\
\includegraphics[trim=0.5cm 16.4cm 1.8cm 2.2cm,clip,width=7.5cm]{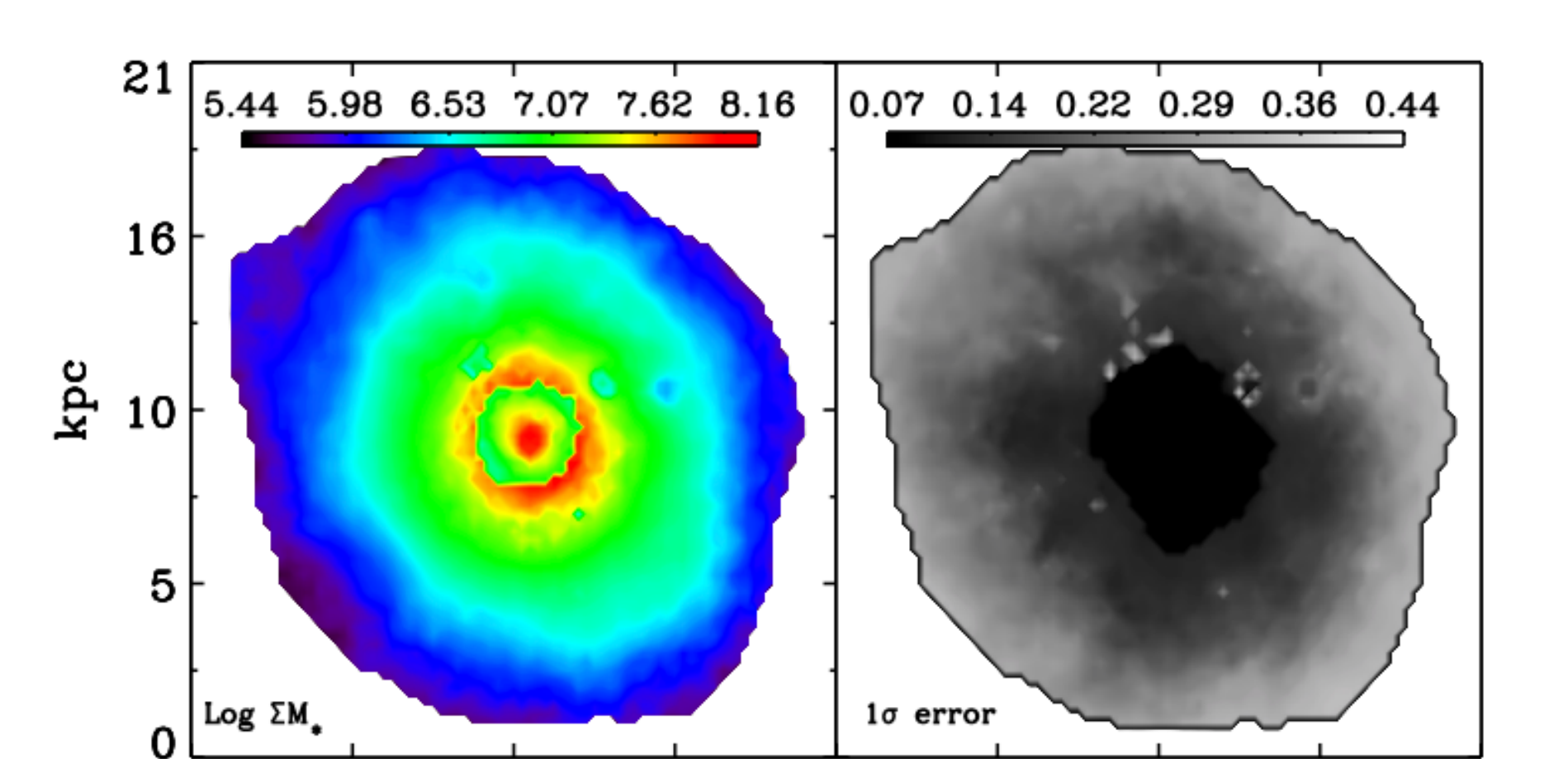}&
\includegraphics[trim=0.5cm 16.4cm 1.8cm 2.2cm,clip,width=7.5cm]{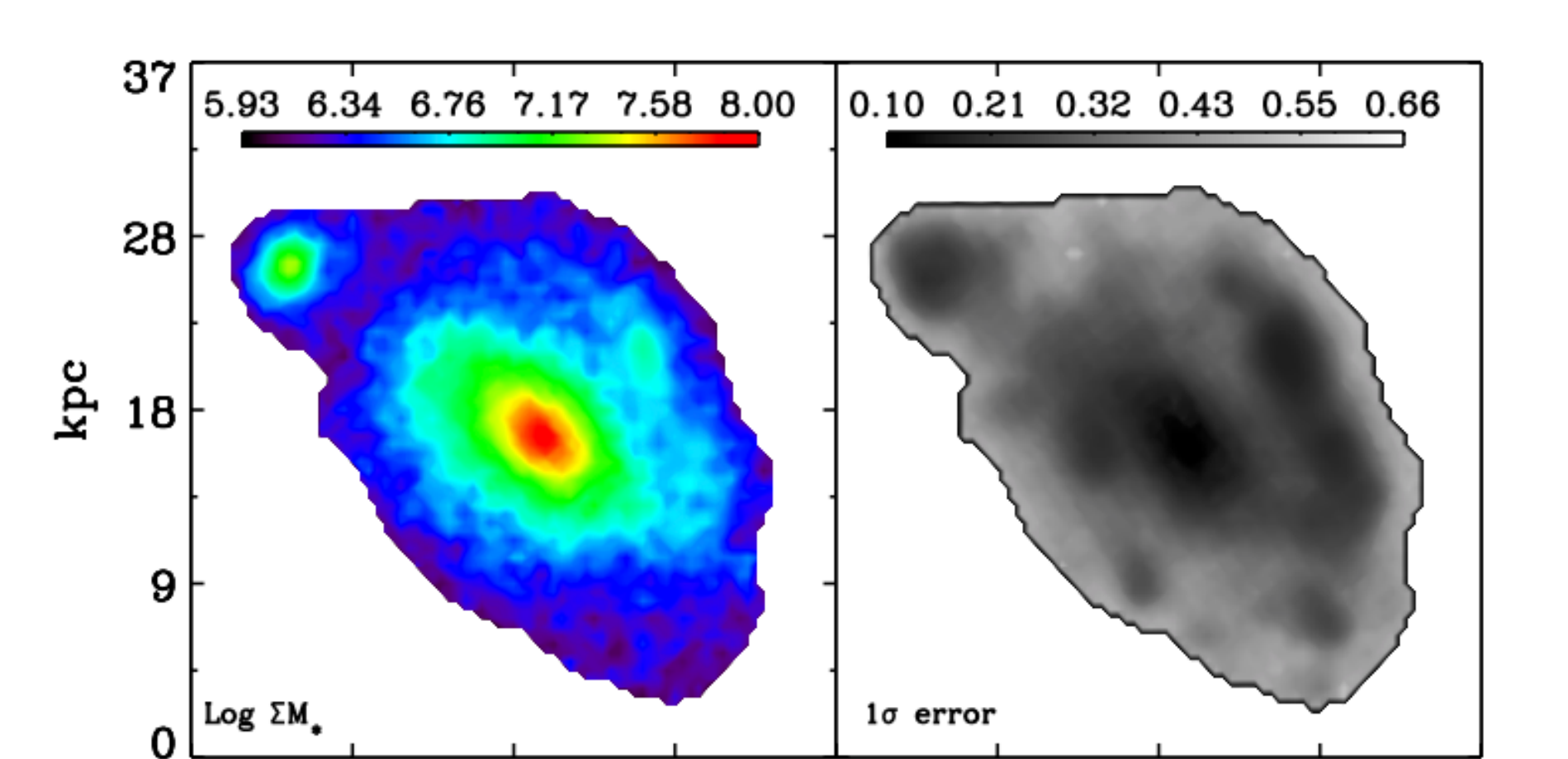}\\
\includegraphics[trim=0.5cm 16.4cm 1.8cm 2.2cm,clip,width=7.5cm]{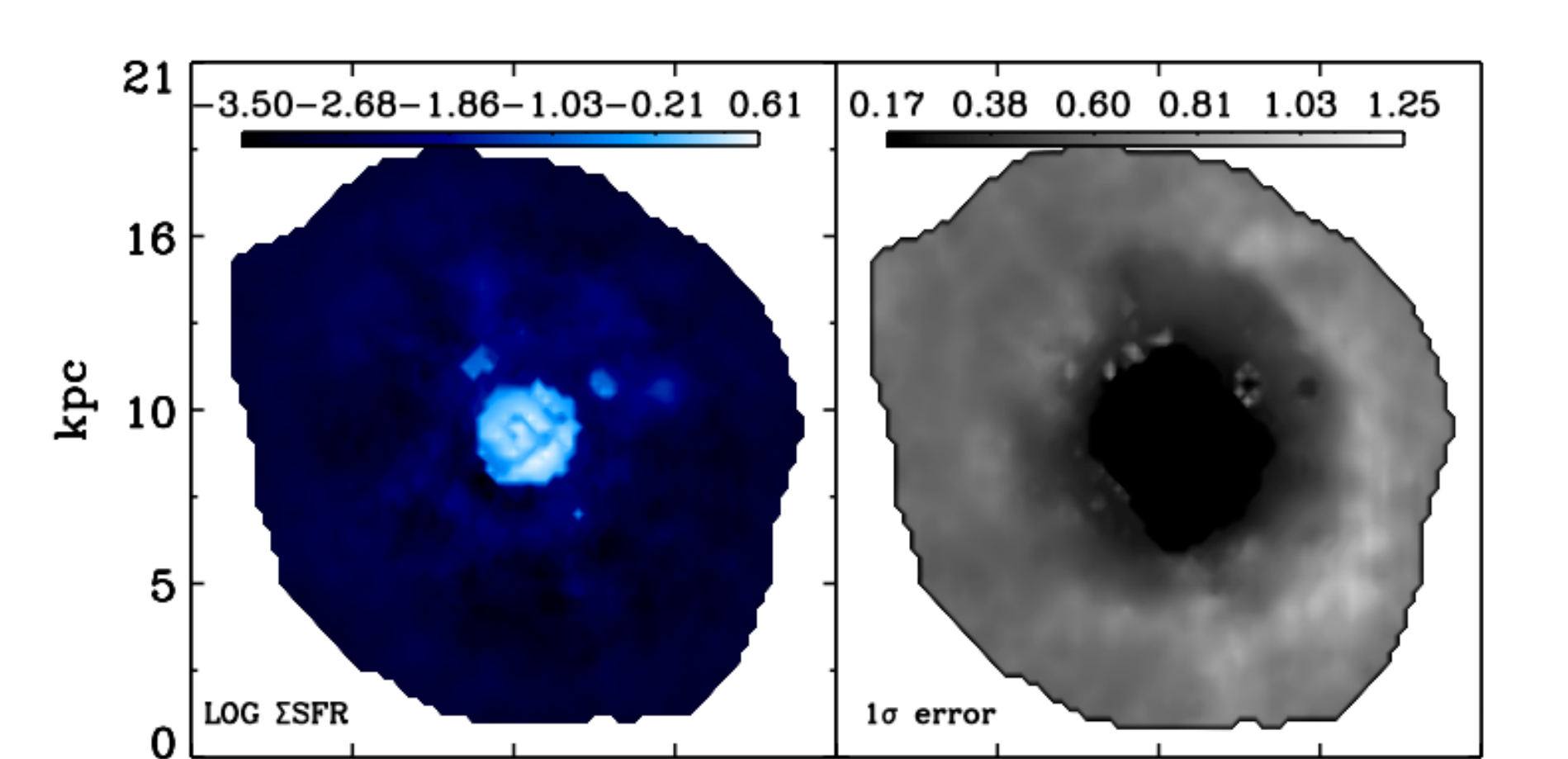}&
\includegraphics[trim=0.5cm 16.4cm 1.8cm 2.2cm,clip,width=7.5cm]{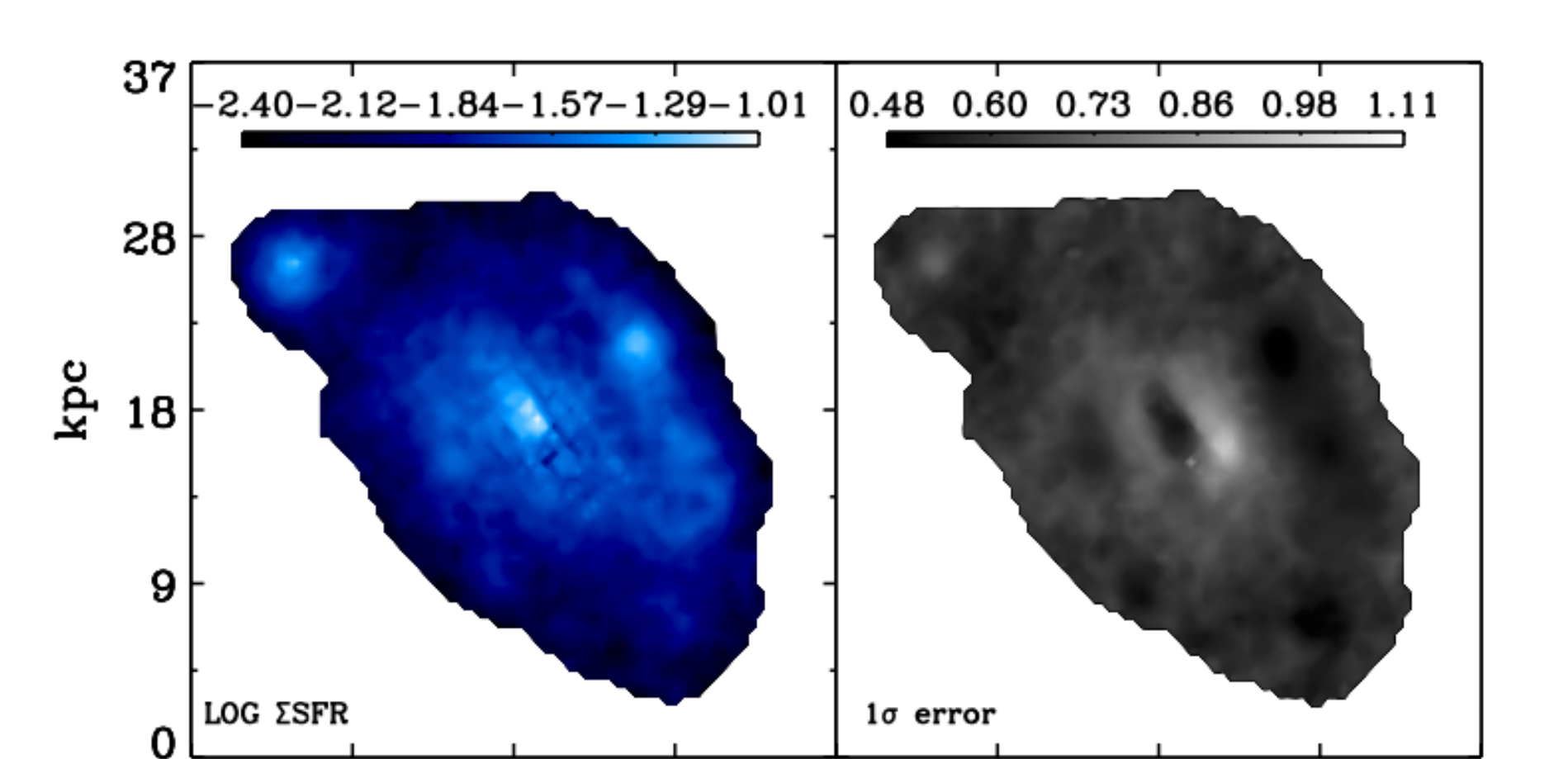}\\
\includegraphics[trim=0.5cm 16.4cm 1.8cm 2.2cm,clip,width=7.5cm]{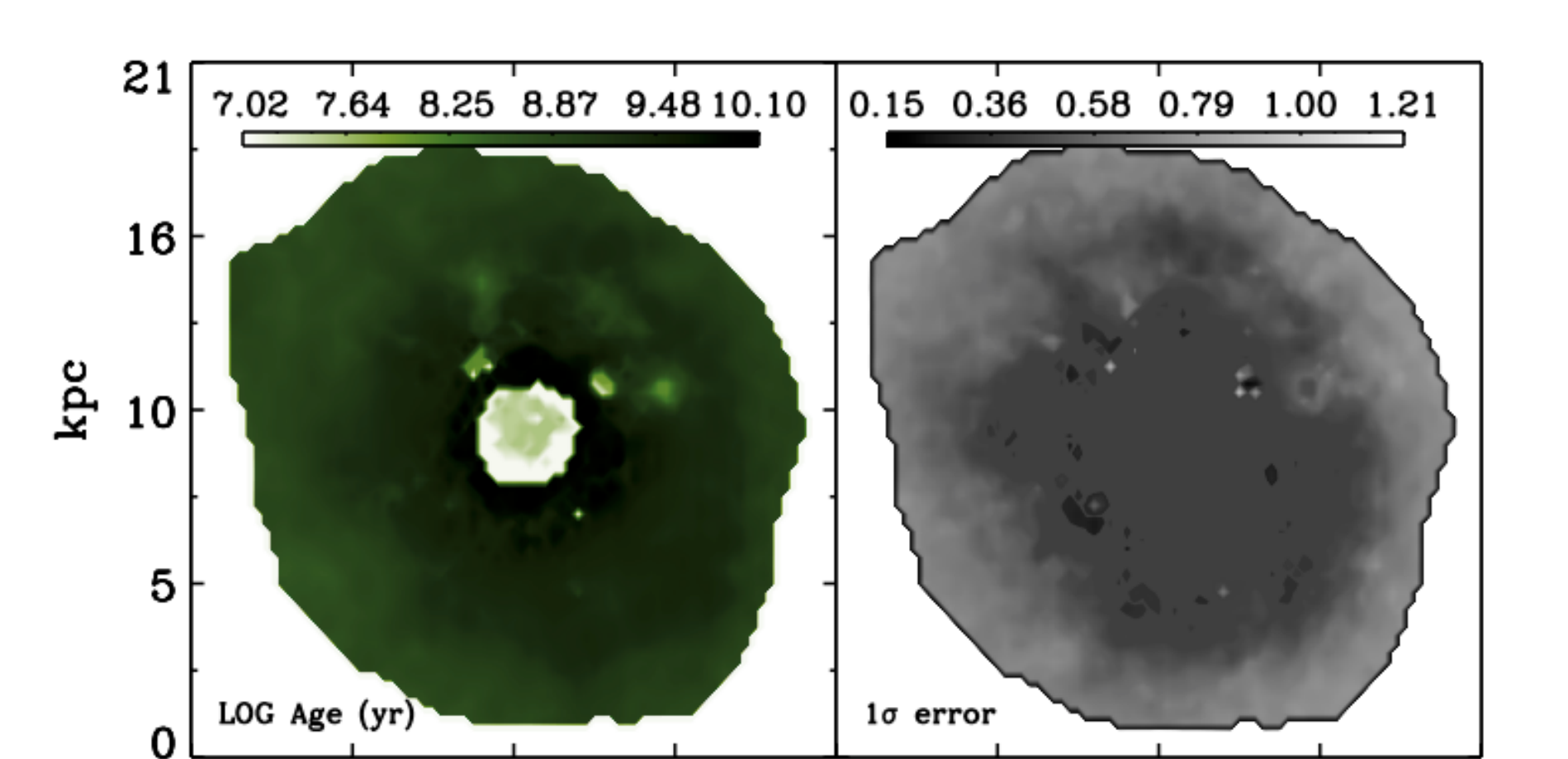}&
\includegraphics[trim=0.5cm 16.4cm 1.8cm 2.2cm,clip,width=7.5cm]{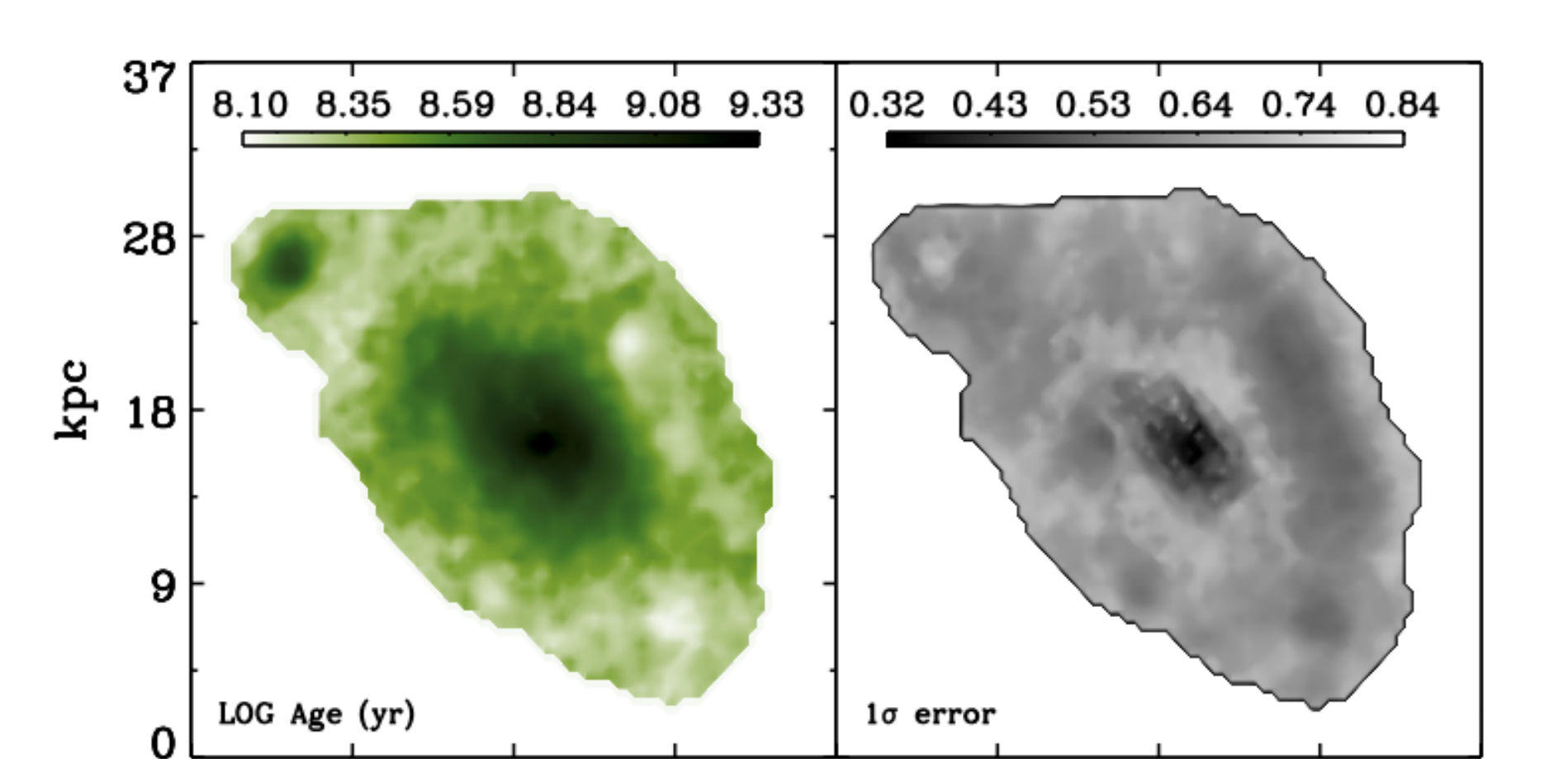}\\
\includegraphics[trim=0.5cm 14.9cm 1.8cm 2.2cm,clip,width=7.5cm]{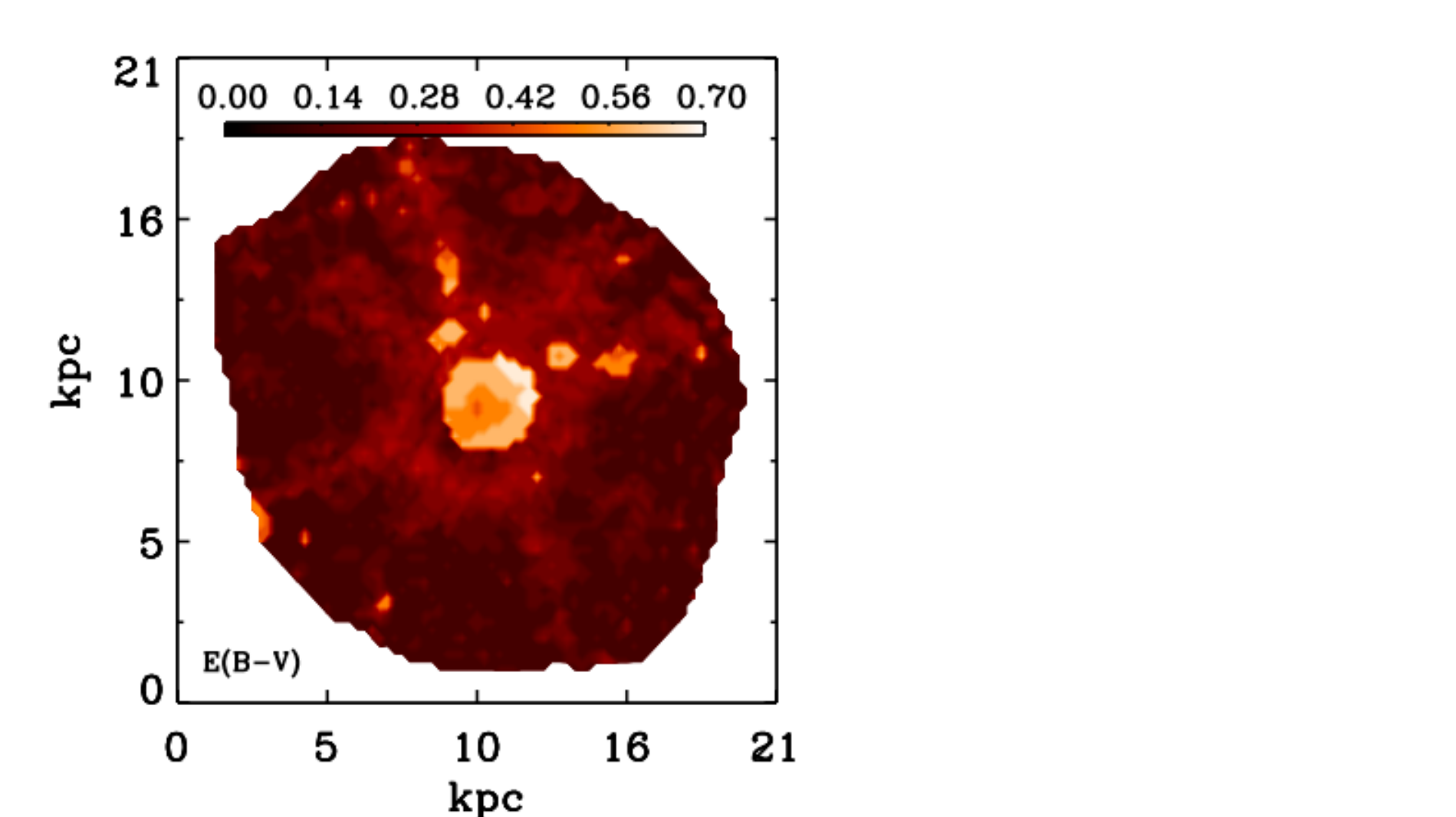}&
\includegraphics[trim=0.5cm 14.9cm 1.8cm 2.2cm,clip,width=7.5cm]{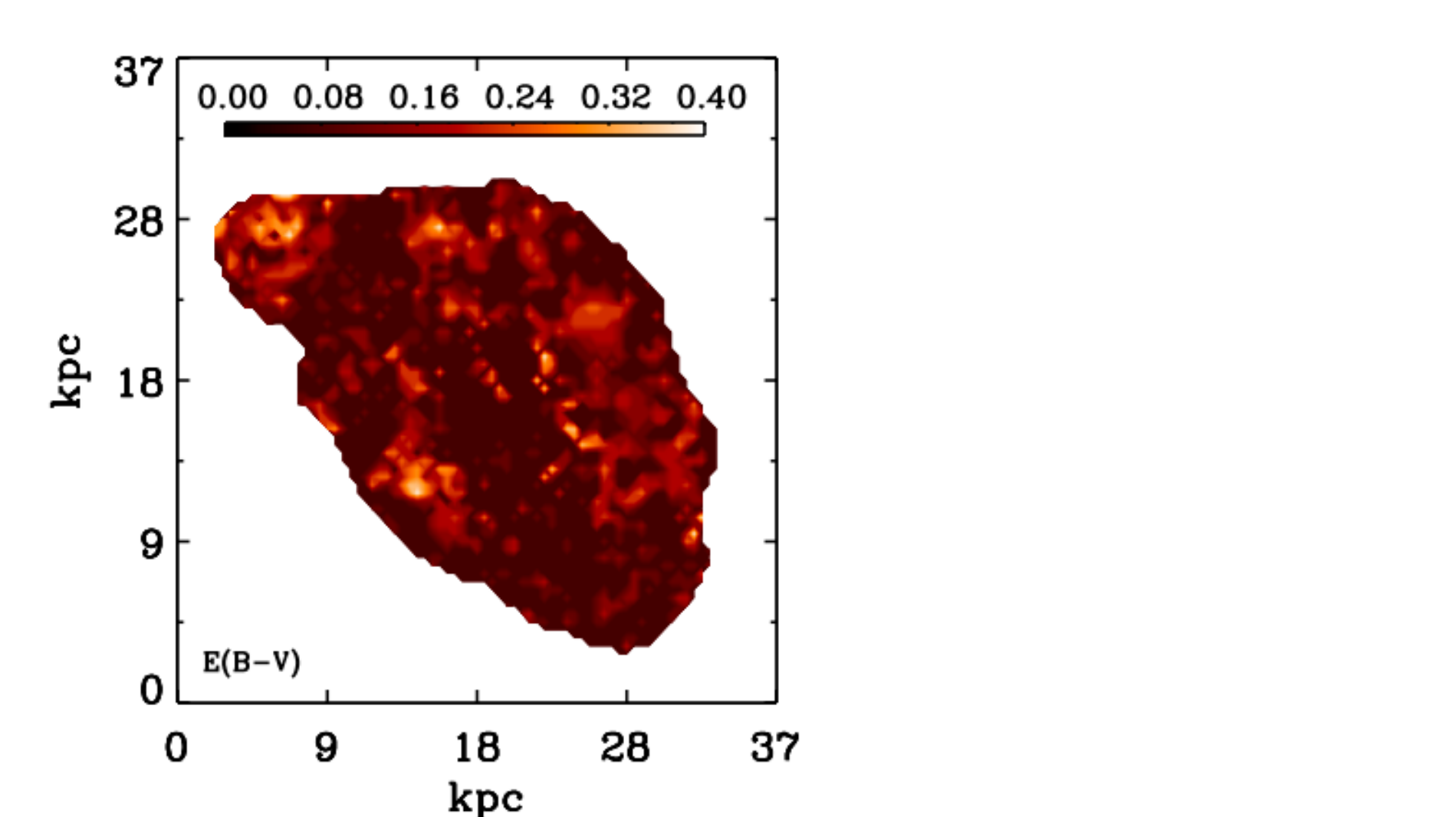}\\
\hline
\end{tabular}
  \caption{Same as Figure 3, but for 2 of the galaxies with abnormal or disturbed
    features or signs of interaction. Respectively from left to right,
    galaxy at z-0.31 with blue (U-V) nucleus hosts an AGN at its
    center (NB artificial ripples in the maps induced by the bright AGN source). 
    Right pannel shows galaxy at z = 0.94, with signs of interaction or  minor merging.}
\end{figure*}

After finalizing the library, we fit the observed flux per resolution
element to our model SEDs. For this we use the Lephare code, an SED
fitting code based on $\chi ^2$ minimization.
After fixing the redshifts to their spectroscopic
values, we fit the SEDs to find the closest match from the model library created for
that specific redshift and measure the physical information associated
to the galaxy (absolute magnitudes, stellar mass, star formation rate, age and
extinction). The Physical parameters from the SED fitting output
  in this work correspond to the median of the probability
  distribution function marginalized over all other parameters. We use
  $16\%$ lower and $16\%$ higher values from the Maximum Likelihood analysis to measure the $1\sigma$ error for each parameter.

We use the best-fit template to estimate the $(U-V)_{rest-frame}$
colors per resolution element. The U-band response function here corresponds to
Johnson filter. We choose this rest-frame optical color to span the Balmer/4000\AA\ break. It has been
shown that the Balmer/4000\AA\ break feature is
not only very sensitive to stellar ages (e.g., \cite{Kauffmann2003}, \cite{Nayyeri2014}), but that the strength of
this break strongly correlates with the width of H$\alpha$ and the
best-fit specific star formation rate \citep{Kriek2011}. Therefore, using passbands straddling the Balmer break is optimized for our
purpose, and provides great insight on the star formation histories
of substructures in galaxies. 

The high resolution maps measured through the SED fits per resolution element, are
presented in Figures 3 and 4 for a subsample of galaxies \setcounter{footnote} {0}\footnote{Maps of all galaxies in the
  sample are at: http://www.shouby.com/research/disk-galaxies/}. These
show the $(U-V)_{rest-frame}$ color, weight, stellar mass surface density, star
formation surface density, extinction and age and
their $1\sigma$ error maps for example galaxies from the sample. Unlike
the more normal Hubble-type galaxy morphologies in Figure 3, Figure 4
shows two examples of galaxies with abnormal features such as blue nuclei (AGN) or disturbed,
interacting morphologies.

In $(U-V)_{rest-frame}$ maps, there is a general tendendency towards a central red bulge and some blue
regions in the overall green disk (Figure 3). There are three clear
exceptions in our sample with the nuclear region in the galaxy
significantly bluer than its surroundings (e.g., see the first column in Figure
4). These are examples of ``blue-nucleated galaxies''
\citep{Schade1995, Abraham1999}. These three galaxies have Chandra
X-ray detections (matched with a $1\arcsec$ radius), which suggests
the presence of an AGN at their centers. Interestingly, even though we include quasar
libraries while performing the SED fitting, none of the central elements in these
galaxies were ``best'' fitted to these templates. In future works we
will investigate whether adding dust to the quasar models alter the
fitting/color of the central resolution elements in these galaxies. The $(U-V)
weight$ maps which will be used for identifying different regions inside galaxies in
the following sections are calculated from the square root of sum of
squares of fluxes in U and V band and normalized to the maximum value.

Contrary to the UV-optical color maps which often show ``clumpy'' structures, the stellar mass surface
density maps are mostly smooth with most of the mass concentrated in the
bulge. The same trend was also reported in \citet{Wuyts2012} for
galaxies at somewhat higher redshift from 1 $< z <$
2. \cite{Lanyon2012} however claims that, the smoothness of the
stellar mass maps neither holds true for all galaxies nor for all features
seen. 
 
To further examine the absence of structure in the stellar mass maps, we smooth them with a Gaussian of
width $\sigma$ (where $\sigma$ is the dispersion of stellar mass
surface density  over all resolution elements) and subtract them from their respective unsmoothed stellar mass
maps. Weighting the residual by the stellar mass error, the structure residuals in all galaxies
account for less than $1.2\%$ of their total stellar mass surface
density. \citet{Wuyts2012} find a similar 2-3\% of the total stellar
surface density in the clumps as defined by either rest-frame U- or
V-band maps of Voronoi-binned pixels for galaxies from $0.5 < z <
1.5$.
 
 The star formation rate surface densities are directly derived
from the SED fitting. Estimating the
SFRs from the SED is very challenging due to  both the age-dust-metallicity
degeneracy  and the choice of SFH prior, especially at
high-redshifts \citep{Conroy2013}. However, it has been shown that at $z<1$ this is less of an issue and SFRs
derived by fitting the SED agree very well with the other indicators
of SFR with less than 0.23 dex scatter \citep{Salim2009}.

\section{Integrated vs. Resolved properties}

 In this section we examine the consistency between resolved
 measurements and those derived for the entire galaxy. We measure the
 photometry of galaxies in the same seven HST bands by adding up the
 fluxes within the boundary defined by the segmentation map of the
 galaxy. We estimate the ``global'' stellar mass, star formation rate
 and mean stellar age for each galaxy by fitting the integrated fluxes
 with the same model library used in the previous sections. 

\begin{figure*}[htbp]
\centering
\begin{tabular}{c}
\includegraphics[trim=0cm 0cm 0cm 0cm,height=2.5in]{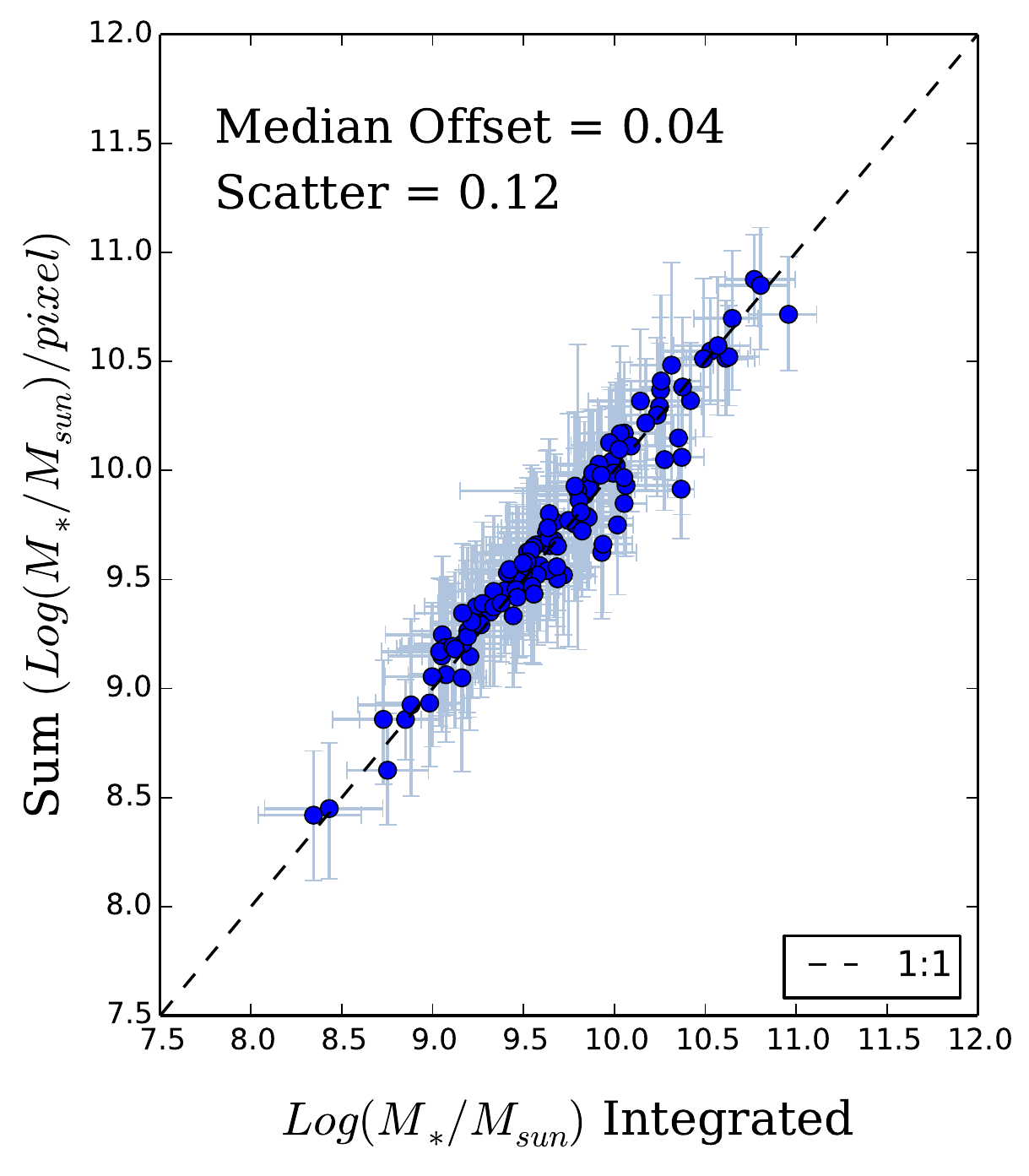}
\includegraphics[trim=0cm 0cm 0cm 0cm,height=2.5in]{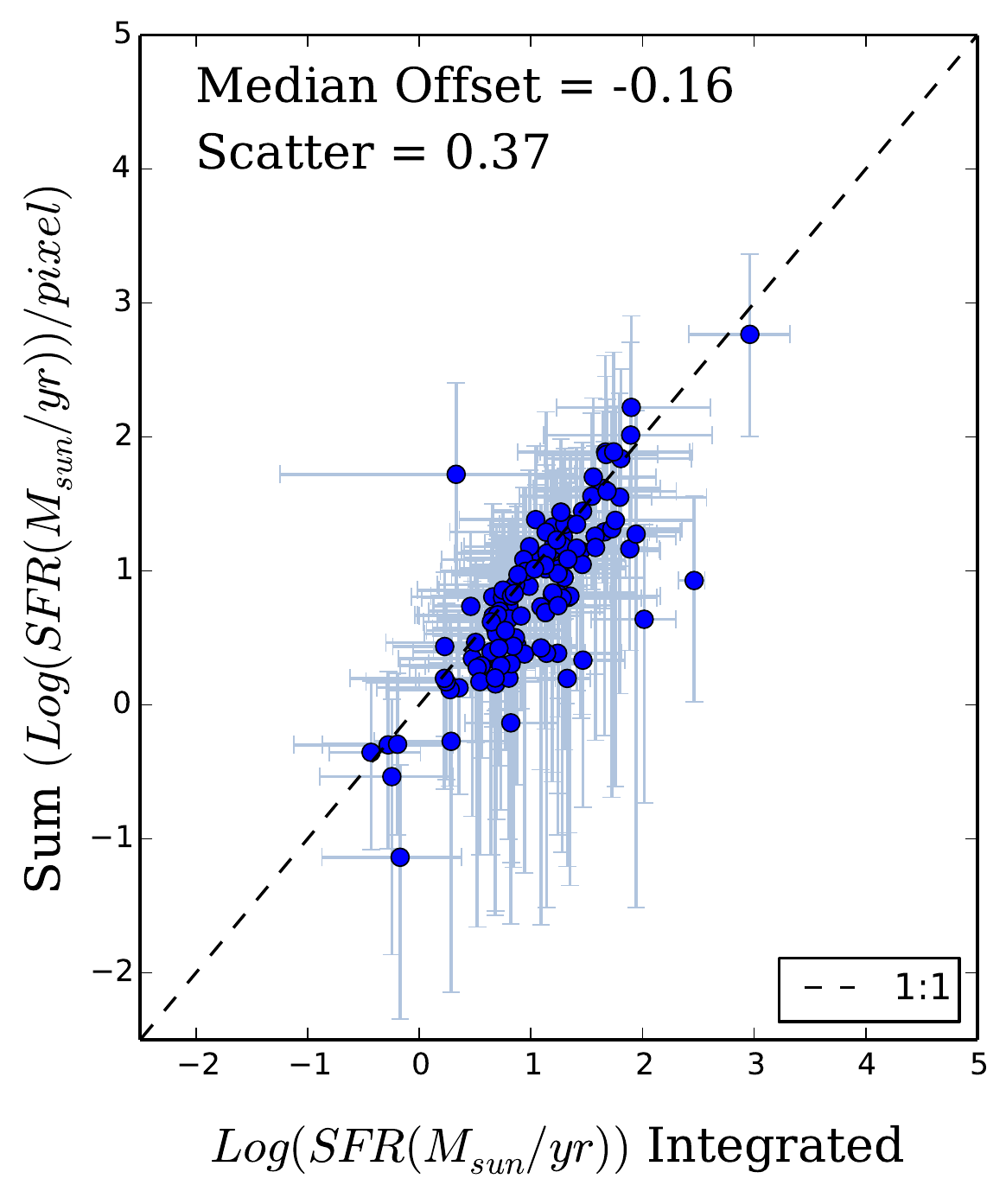}
\includegraphics[trim=0cm 0cm 0cm 0cm,height=2.5in]{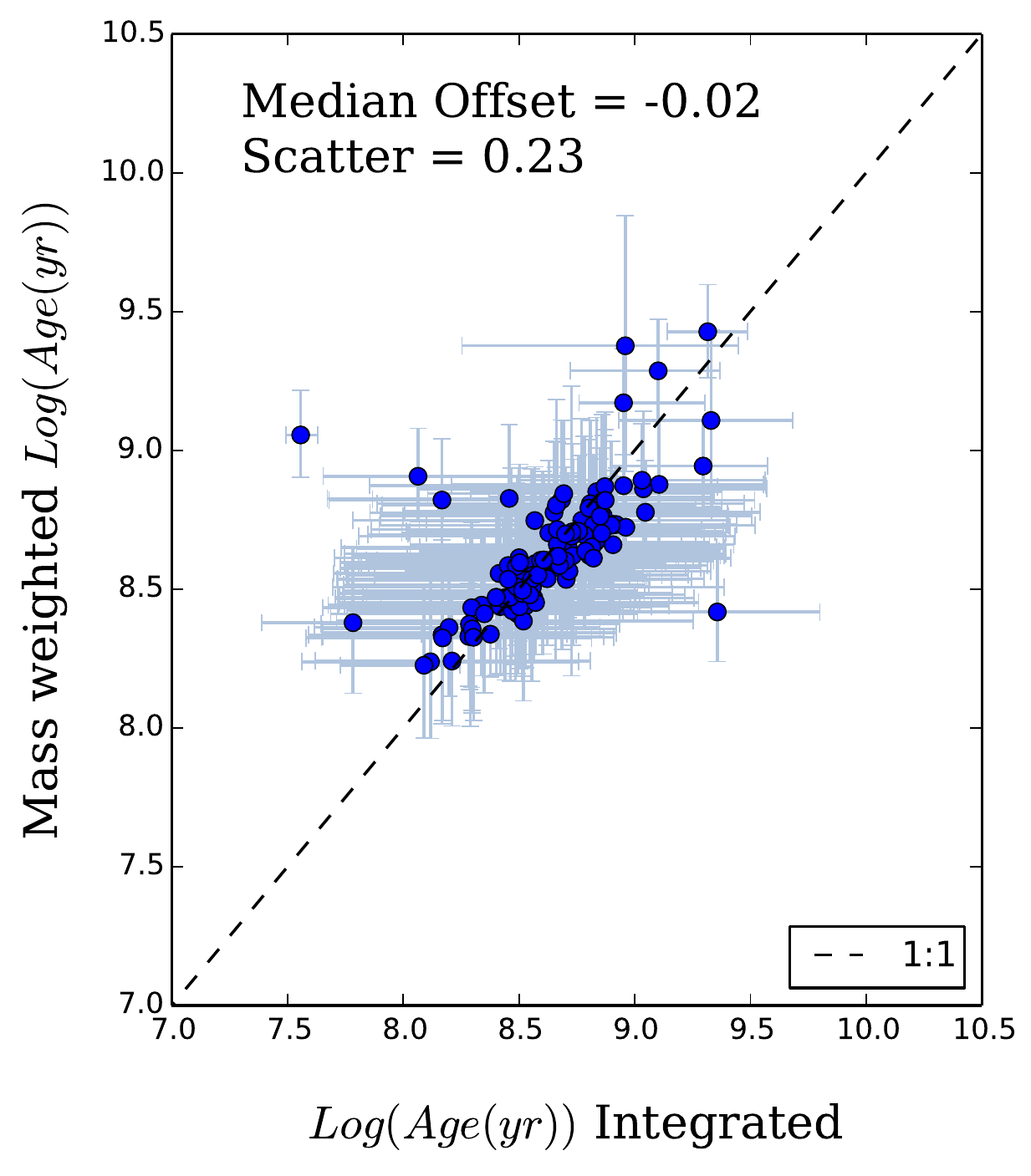}
\end{tabular}
\caption{Comparison between resolved and the integrated properties:
  (left) integral of stellar mass surface density per resolution element
  vs. integrated stellar mass, (middle) integral of SFR surface
  density per resolution element vs. integrated SFR, (right) mass weighted
  stellar age vs. integrated age of the galaxies.}
\end{figure*}

 In the first panel of Figure 5 we compare the integrated stellar
masses with the sum of stellar masses measured per resolution element. We find a median offset of 0.04
dex and a scatter of 0.12 dex for stellar masses. \cite{Wuyts2012} reported
no median offset using Voronoi-binned pixels rather than individual
pixels. This is contrary to some of the previous works which found
larger disagreement, with the global mass being lower than the integral of
masses over resolution elements up to $40\%$ for a sample of nearby
galaxies \citep{Zibetti2009}. The
small offset ($\sim$1 $ M_{sun}$)  can be explained by lower estimates of
M/L from the integrated fluxes due to more luminous young stars in the
disk (e.g., \citealt{Papovich2001}; \citealt{Shapley2001}). 

Comparing the resolved and the integrated SFRs, we find a
median offset of 0.16 dex with a scatter of 0.37 dex (Figure 5- middle
plot). In a similar study \citep{Wuyts2012}, a
larger offset of 0.27 dex in the opposite direction was found. The SFR estimates from the SED-fitting have
different sources of uncertainty. This makes explaining the
offset very challenging. One possible source of the offset is due to the overcorrection for dust extinction and hence
slightly larger SFRs (in $90\%$ of the galaxies in the sample the
average E(B-V) over resolution elements is less or equal to the E(B-V) estimate for
the galaxy).

For stellar age comparison, we calculate the stellar mass
weighted age from the resolved SED fits. The right panel in Figure 5
shows the comparison of resolved and integrated stellar ages. A very small offset of 0.02
dex with a scatter of 0.23 dex is observed. This median offset is 10
times smaller than that of \cite{Wuyts2012}. The plausible explanation
for this small offset is that using the stellar mass weighted age
rather than light weighted age would make outshining of older more
massive stars by younger less massive stars(e.g.,
\citealt{Papovich2001}; \citealt{Maraston2010}) less of an issue.

\section{Identification of Red and Blue Regions in Galaxies}

To allow the study of substructures in galaxies at kpc scales, we
identify regions based on the resolved photometric maps. This is based on the
$(U-V)_{rest-frame}$ color maps. Most of these regions are kpc size with comparable characteristics to what
are called ``clumps'' in many previous studies (e.g.,
\citealt{Guo2012}). However, we remain agnostic regarding the nature
of these regions, preferring this descriptor over the term ``clump,''
given that the latter connotates a distinct entity within the galaxy,
potentially even a distinct self-gravitational mass, which may not be
accurate. This becomes clear when comparing the physical differences
between the regions identified, based on rest-frame optical color,
rest-frame UV image, and stellar mass maps (Hemmati et al. in prep).
Using the available deep spectra and kinematic information, we aim in
future work to more precisely evaluate which regions are
gravitationally-bound ``clumps'' and which regions are simply
differentiated by their relative star-formation or dust properties
with respect to the surrounding parts of the galaxy. 

 In this section we first describe our method of identifying regions
 in our galaxies, red and blue in this case, and proceed by looking at
 their properties.

\subsection{Region Identification}

We select red and blue regions inside galaxies by fitting weighted
rest-frame color distributions in each galaxy to Gaussian
functions. We used the (U-V) weight maps described in previous section
to avoid selecting insignificant regions at the very edges of galaxies with very low
surface brightness. We plot the distribution of weighted rest-frame color for each galaxy. We
choose the bin size following  Scott's rule, which is optimal for
normally distributed data \citep{Scott1979}. We then fit the color distribution with three skewed
Gaussians, one for fitting the median color of the galaxy and the
other two allowing for the bluer or redder regions with centers constrained at two sides of the central peak.  This is done to
build high-pass (red) and low-pass (blue) filters based on the
relative color distribution for each galaxy. Specifically, we do this by taking the $middle peak +
1.5\sigma$ for the red regions and $middle peak - 1.5\sigma$ for the
blue regions (here after $1.5\sigma$ regions). 
\begin{figure} [htbp]
\centering
\begin{tabular}{cc}
\includegraphics[trim=3.1cm 15.3cm 9.8cm 3.2cm,clip ,width=4.1cm,height=3.7cm]{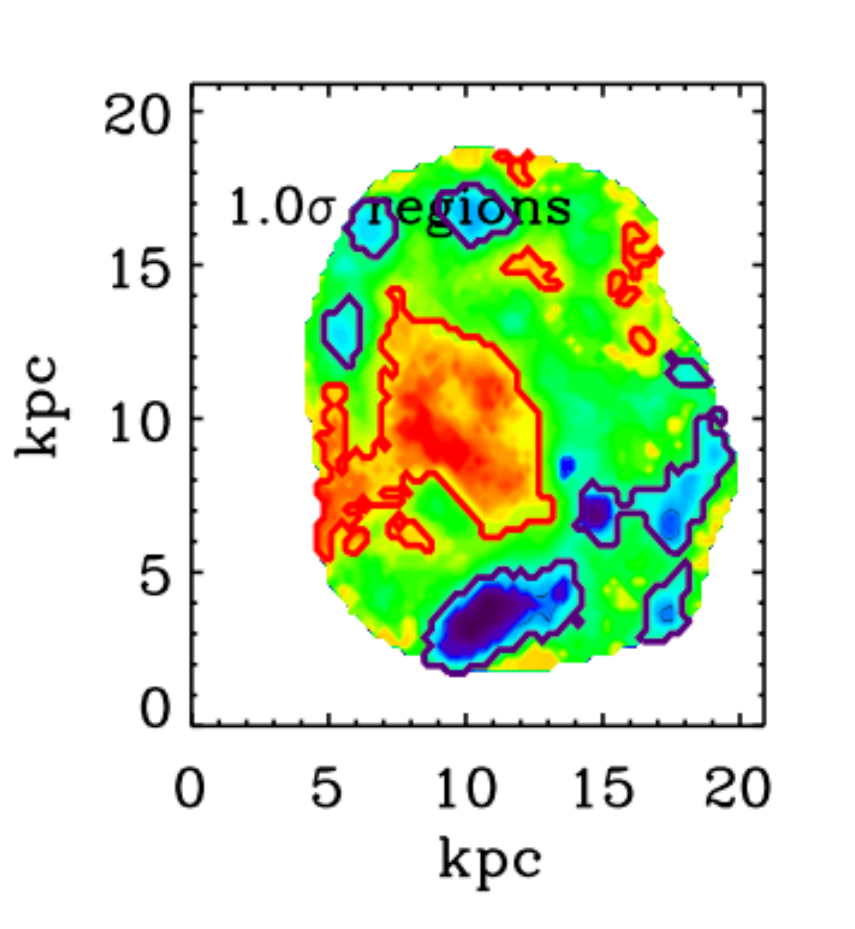}&
\includegraphics[trim=3.1cm 15.3cm 9.8cm 3.2cm,clip ,width=4.1cm,height=3.7cm]{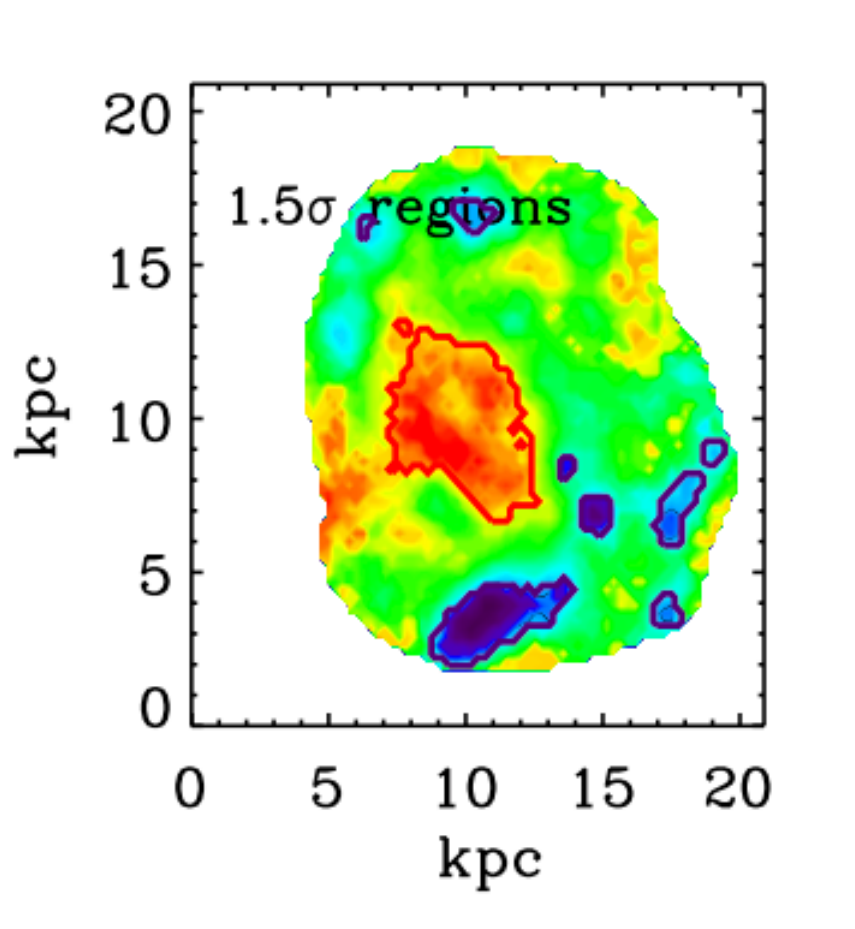}\\
\includegraphics[trim=3.1cm 15.3cm 9.8cm 3.2cm,clip ,width=4.1cm,height=3.7cm]{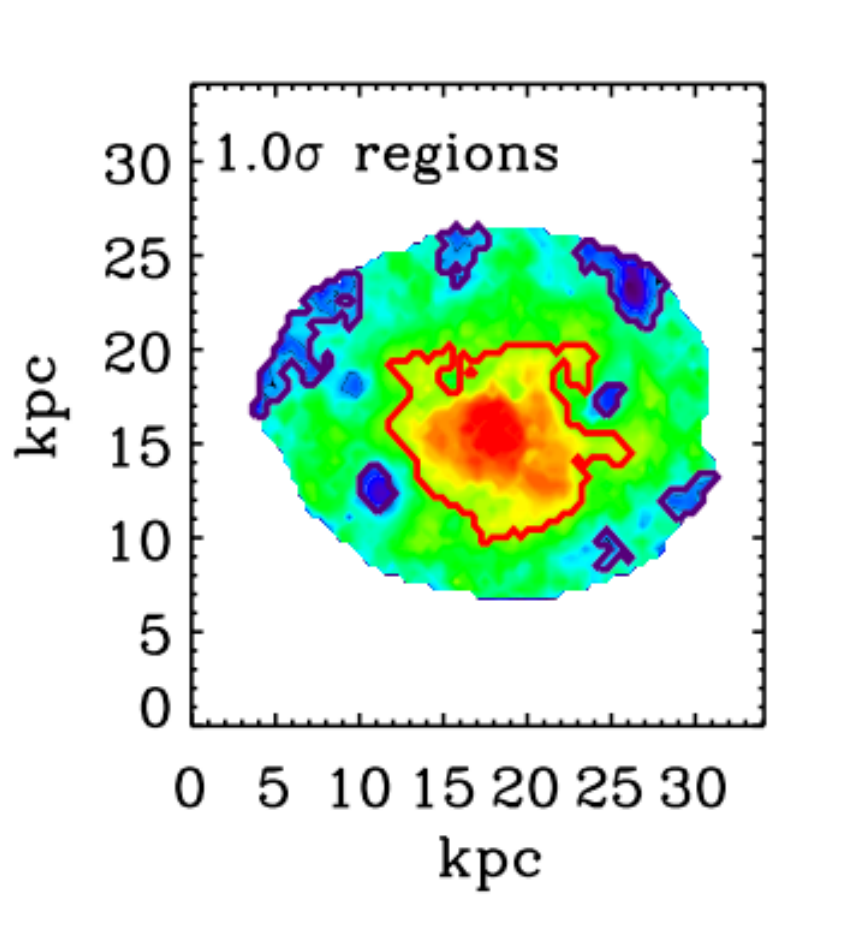}&
\includegraphics[trim=3.1cm 15.3cm 9.8cm 3.2cm,clip ,width=4.1cm,height=3.7cm]{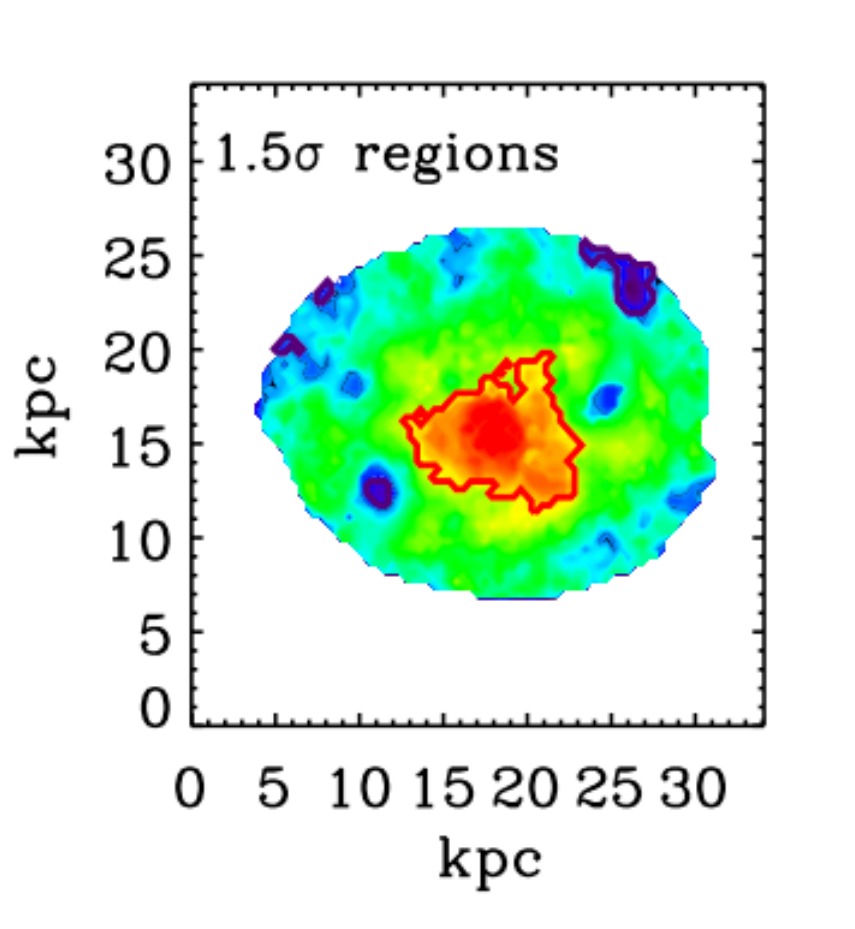}\\
\includegraphics[trim=3.1cm 15.3cm 9.8cm 3.2cm,clip ,width=4.1cm,height=3.7cm]{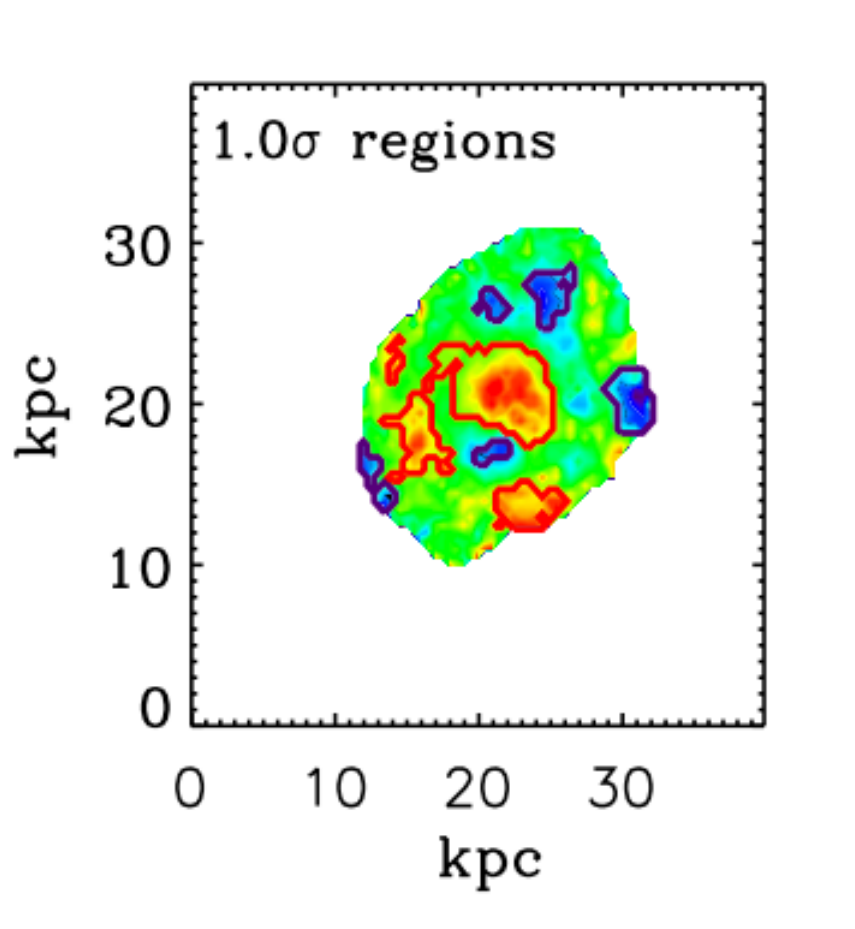}&
\includegraphics[trim=3.1cm 15.3cm 9.8cm 3.2cm,clip ,width=4.1cm,height=3.7cm]{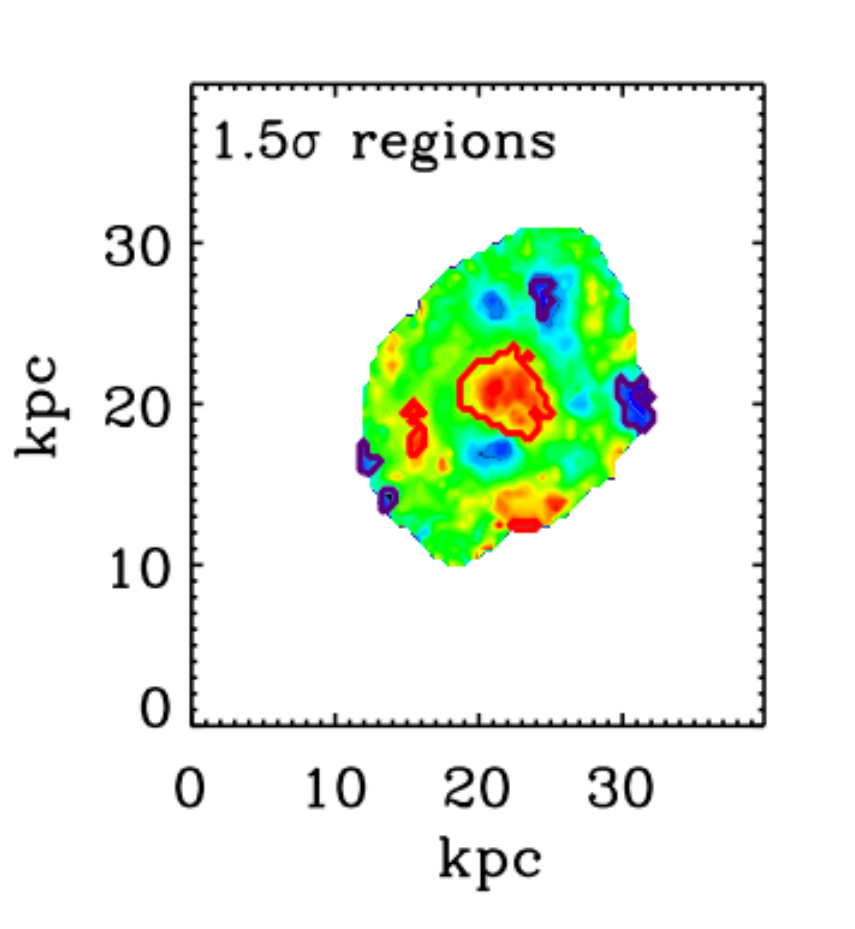}\\
\end{tabular}
  \caption{Red and blue regions identified by contours on (U-V)
  rest-frame color maps for three example galaxies at z=0.29, 0.69 \&
  1.21 respectively from top to bottom. Left panels correspond to regions above and below
$middlepeak\pm1\sigma$ and right panels correspond to regions above
and below $middlepeak\pm1.5\sigma$}
\end{figure}

  We examine different filtering sigma thresholds identifying the regions. Setting higher sigma values would only select
  regions with the most contrast while choosing a smaller
  sigma factor would be sensitive to lower levels of ``unsmoothness''
  closer to the noise level. Another advantage of this method is that
  the middle Gaussian identifies the ``green'' or intermediate color properties
  of the distribution. For our purposes in this paper, we find that $1.5\sigma$
  regions most closely match our visual intuition for red and blue
  regions of color (See Figure 6). 

We then run SExtractor on the high and low pass filtered color images with
the SExtractor configuration file carefully optimized by visual
inspection of all selected regions. The same configuration file for blue and red regions was used. We did not use any filter
(e.g. Gaussian, Top-Hat) to avoid introducing any resolution elements
from outside of the region area identified by the filters. SExtractor
will provide us with two segmentation maps per galaxy, for the red and blue filtered images,
which we use to build the region map for each galaxy. 

Here we take advantage of the dual mode features of SExtractor to
estimate properties of the regions. In dual mode the sources are detected in the first image and the
photometry is then performed on the second image based on those
detections. We use the color maps as our detection image and use the
mass maps as our second image. This way we not only know the area and
position of each identified region based on the segmentation maps, we also know
their corresponding stellar masses. We measure this by converting our stellar mass
surface density maps, from logarithmic to linear scale,
measuring the isophotal flux output from SExtractor for each galaxy
(this is the best choice for our purpose with no assumption depending
on the shape). In this work we do not a priori decompose the bulge from the disk, to minimize 
assumptions on region definitions however, there are GALFIT bulge-to-disk decompositions 
for all galaxies in the sample from \cite{Miller2011}, and we confirm that bulges are
consistently identified as red regions between the two methods with
the exception of the three galaxies with blue nuclei mentioned above.

\subsection{Redshift-dependent Bias}

Since the pixel size is fixed for all the images, the apparent change
due to redshift in the intrinsic angular size of galaxies is likely to introduce a bias in the way the regions are identified and in their
estimated size. Here we examine the presence of such biases by
simulating the images of galaxies. We artificially redshift galaxies
from $0.3<z<0.5$ to $z=1.0$ and re-identify their regions using the
same technique. 

\begin{figure}[htbp]
\centering
\includegraphics[trim=0.5cm 0.0cm 0.0cm 0.2cm,width=3.5in]{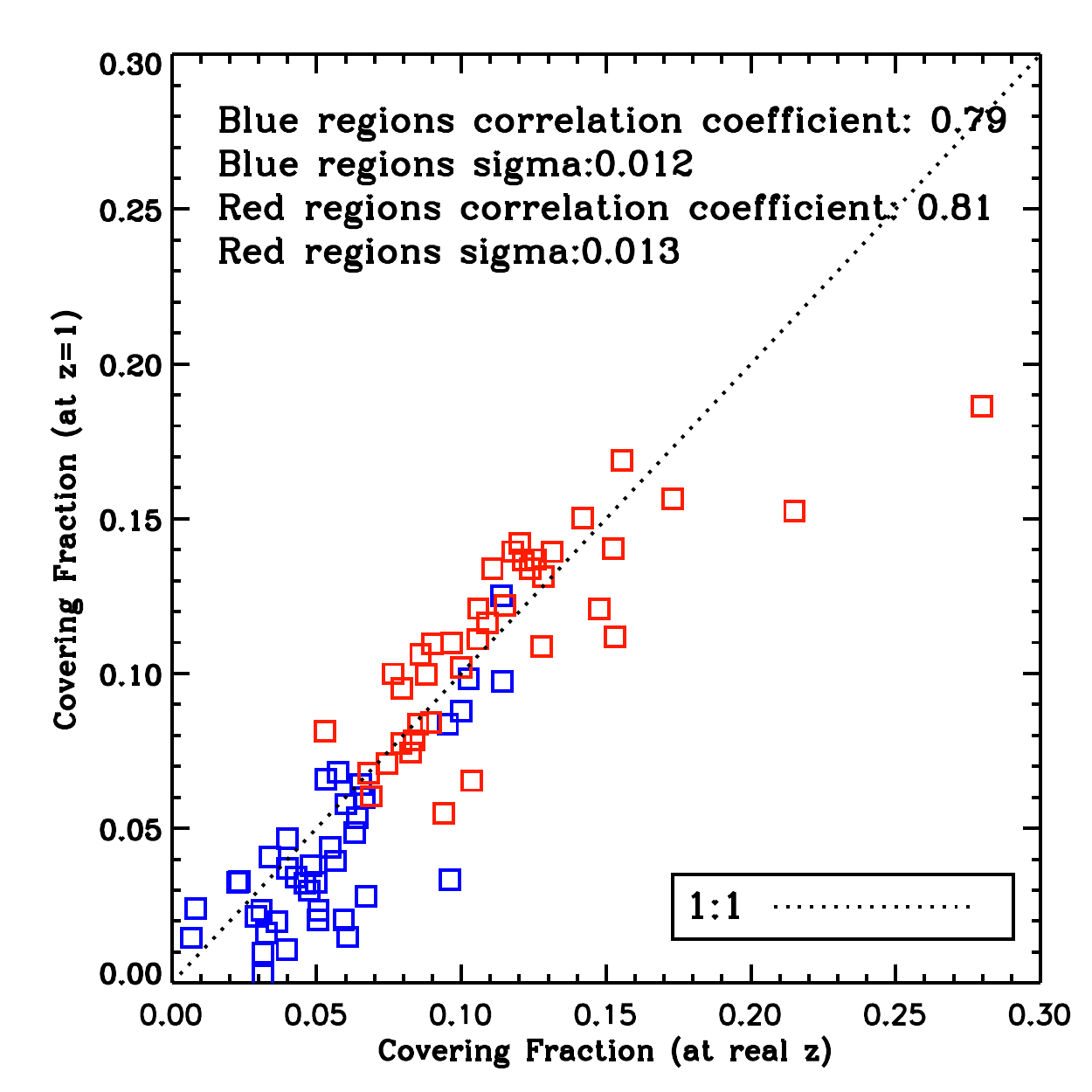}
\caption{The covering fraction of red and blue regions inside galaxies
  redshifted to z=1 versus the covering fraction at their real,
  observed redshift. Blue/Red squares represent covering fraction of
  Blue/Red regions respectively.}
\end{figure} 

By identifying regions based on color maps rather than a parameter
sensitive to surface brightness dimming, any redshift dependent bias
could be better revealed. The dimming factor is redshift
dependent (not wavelength dependent) and thus, the primary redshifting
effect we need to account for is the relative change in observed angular size, {\it a}, with
redshift. Assuming {\it $a_0$} and {\it $a_i$} to be the angular sizes of
the same galaxy at redshifts {\it $z_0$} and {\it $z_i$} respectively, and
{\it $d_0$} and {\it $d_i$} being the luminosity distances
corresponding to these redshifts \citep{Ferengi2008}, we have
$\frac{a_o}{a_i} = \frac{d_i/(1+z_i)^2}{d_o/(1+z_o)^2} $ .

We shift galaxies within the observed redshift range $0.3 < z < 0.5$ to
$z_{rest-frame}=1$, and for each galaxy identify the blue and red regions at the two
redshifts (the observed redshift and at z=1). We then estimate the
covering fractions of these regions and compare them in Figure 7. We define covering 
fractions by normalizing the area within blue and red regions to the total area of a galaxy,
defined based on the number of resolution elements in the segmentation
maps of both the regions and galaxies. The linearity of the relation and small
scatter ($\sigma \simeq 0.01 $) around the 1:1 line in Figure 7 confirms that
there is no significant bias due to cosmological angular size
evolution for our region selection method at higher redshifts
($z\simeq 1$) as compared to lower redshifts in our sample.

\section{Results}
\subsection{Physical properties of the regions and
  small-scale properties of galaxies}

\begin{figure*}[htbp]
\centering
\includegraphics[trim=1cm 9.5cm 1cm 9.7cm,clip,width=17cm]{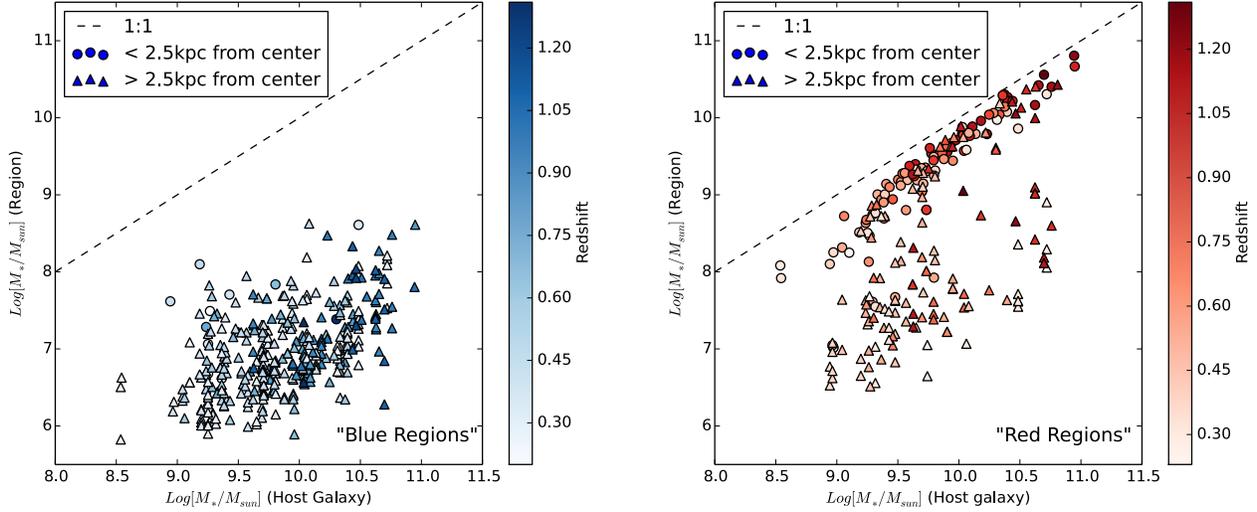}
\caption{Stellar mass of red and blue regions versus stellar mass of
  the host galaxy in right and left panels respectively. Filled
  circles (triangles) represent regions with distances less (more)
  than 2.5. In both panels lighter color corresponds to lower redshift. }
\end{figure*}
\begin{figure}[htbp]
\centering
\begin{tabular}{c}
\includegraphics[trim=0.8cm 0cm 0cm 0.2cm, width=3.in]{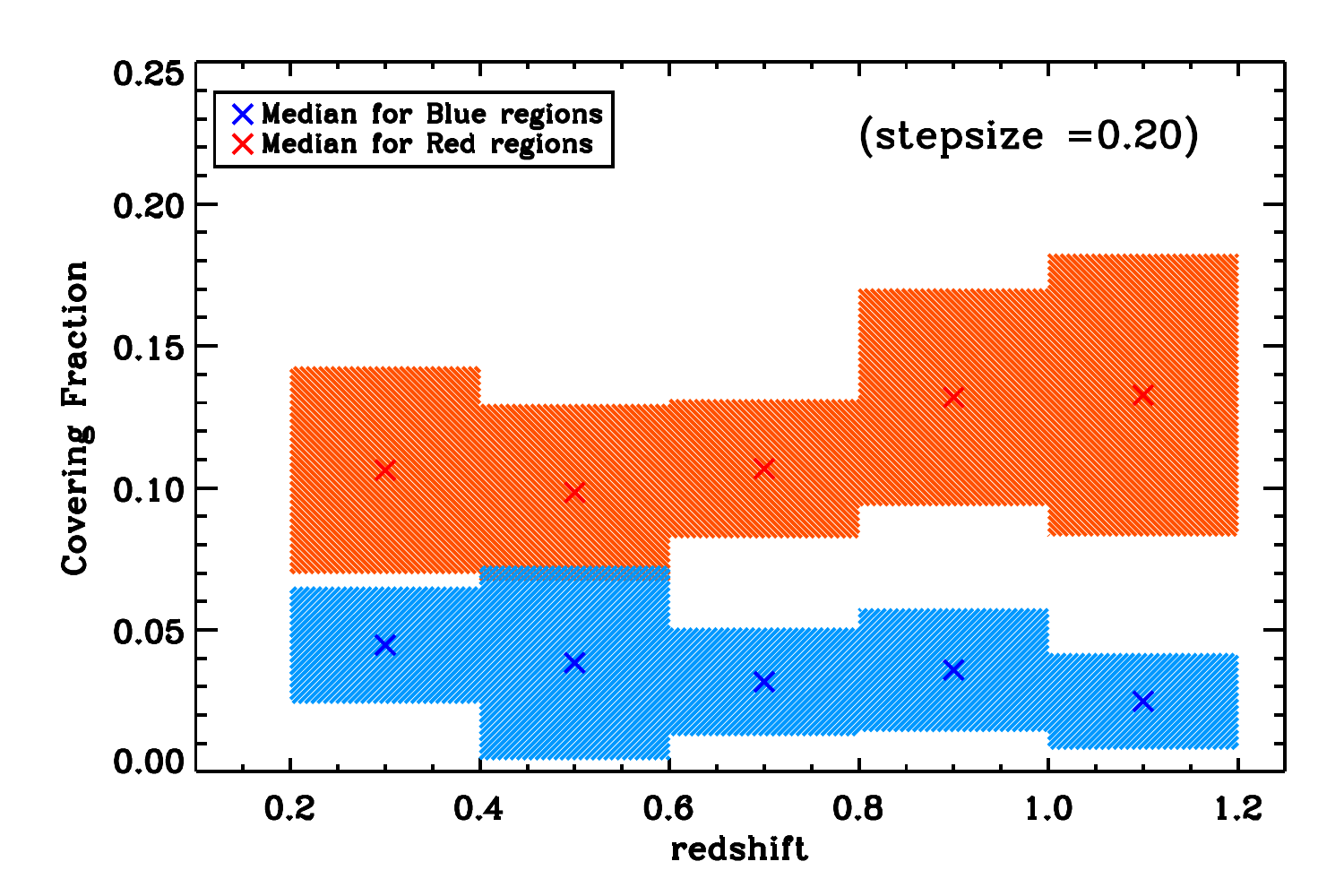}\\
\includegraphics[trim=0.8cm 0cm 0cm 0.2cm,width=3.in]{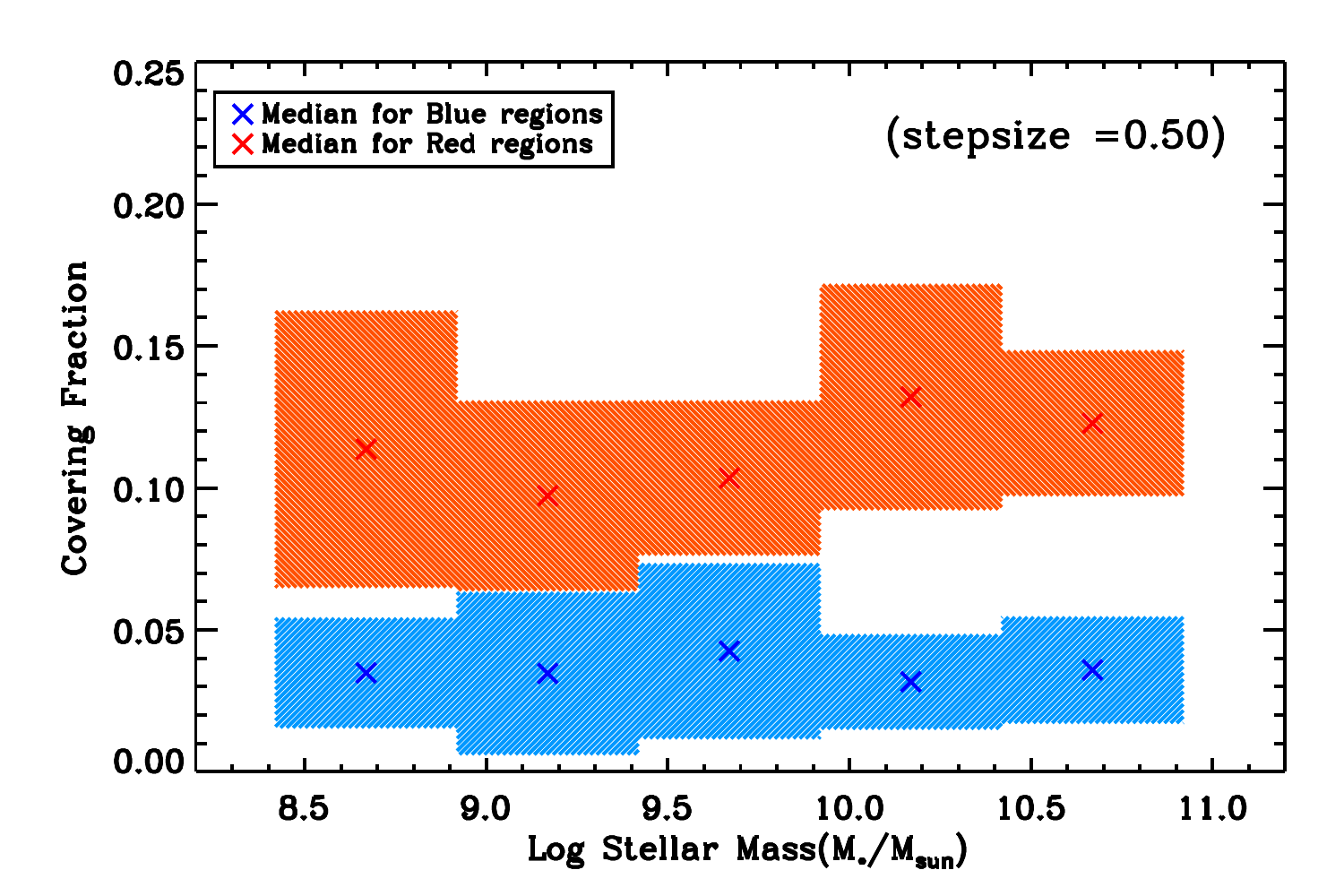}
\end{tabular}
\caption{Covering fraction of red and blue regions inside galaxies
  (area of identified red or blue regions divided by the area of the galaxy) as a
function of redshift and stellar mass in top and bottom panels,
respectively. Crosses shows the median in each bin and the shaded area shows
the 1$\sigma$ dispersion.}
\end{figure}
\begin{figure}[htbp]
\centering
\includegraphics[trim=0.8cm 0cm 0cm 0.2cm, width=3.in]{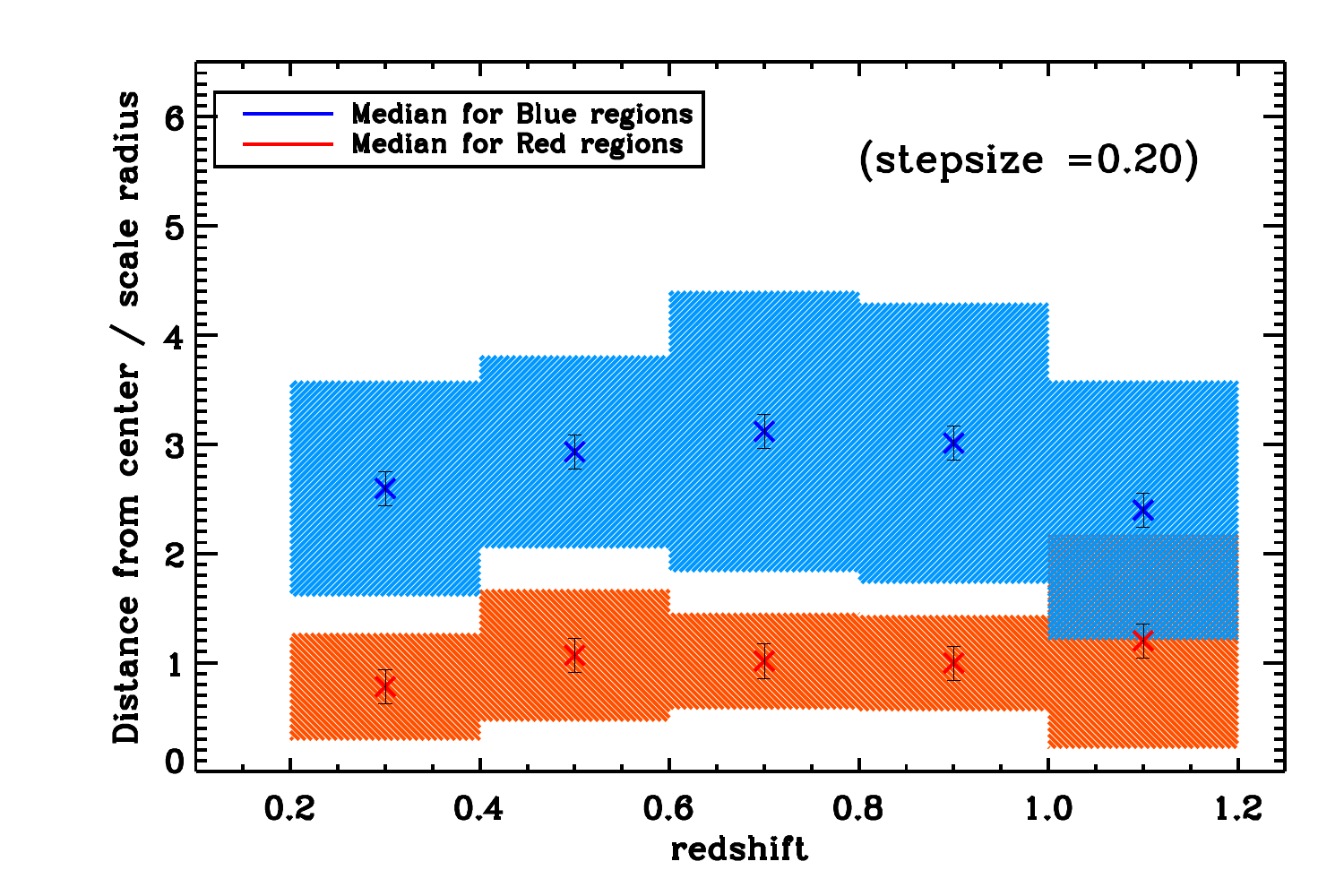}\\
\includegraphics[trim=0.8cm 0cm 0cm 0.2cm, width=3.in]{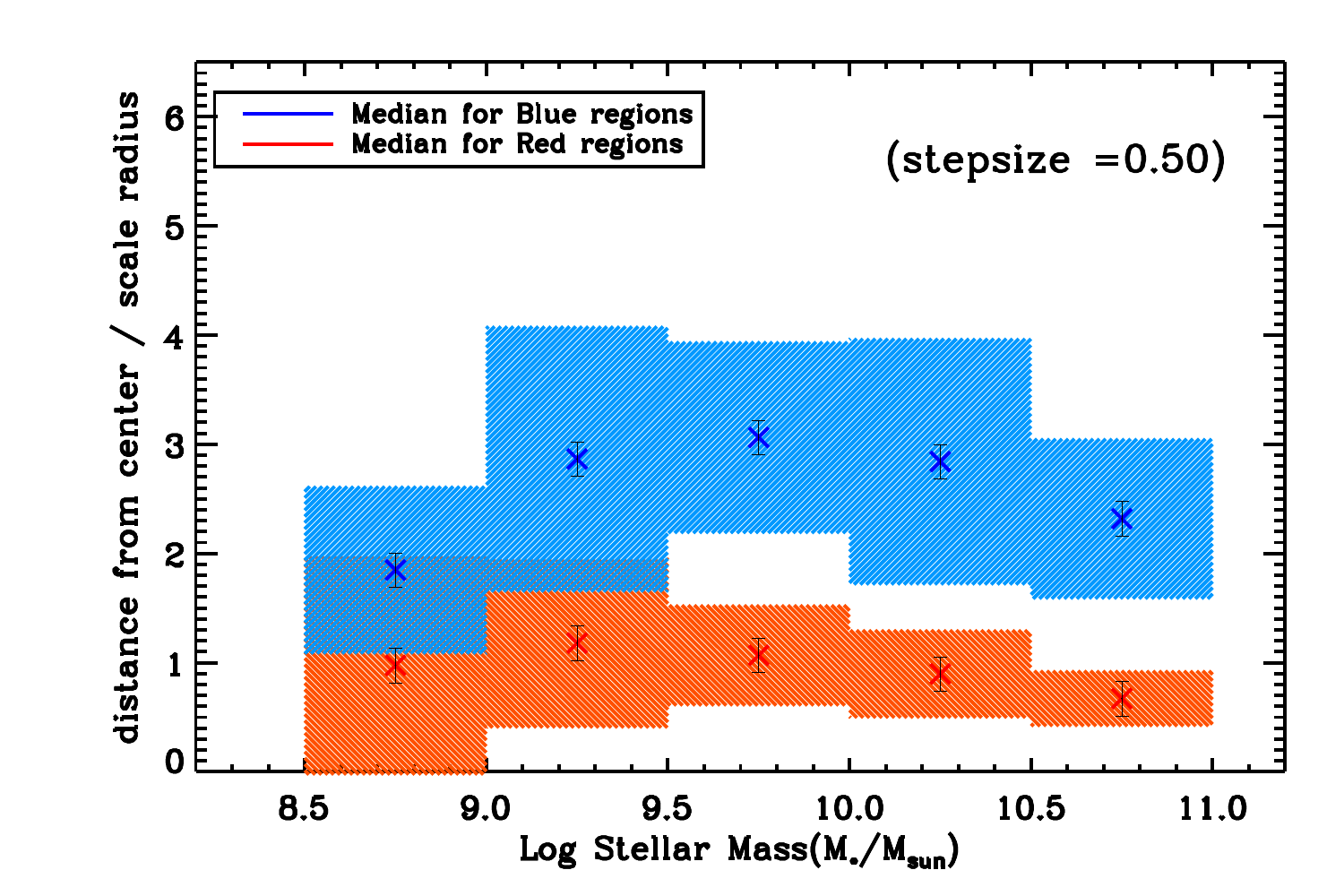}
\caption{Spatial distribution of red and blue regions in their host
  galaxies. The figure shows the distance of blue and red
  regions from the dynamical center of galaxy as a function of
  redshift and stellar mass of the host galaxy. Cross symbols
  represent the median in each bin, error bars are measurement errors and shaded region shows the
dispersion of distances of resolution elements in the identified regions for
 galaxies at each bin. Blue regions are always at larger distances from the
center of galaxy compared to red regions.}
\end{figure}

In Figure 8, we plot the stellar mass of each identified region as a function of
total stellar mass, color-coded by redshift. There is an overall trend of more massive regions
being in more massive hosts and an interesting bimodal behavior seen in the
distribution of red regions. To explore the extent to which the tight
sequence of high-fraction red regions are bulges or central spheroidal
components, we divide the regions into two groups based on their
distance from the center of galaxy (Figure 8). Circles show regions
closer than 2.5 kpc to the center and triangles represent regions
further than 2.5 kpc from the center (where region centers are
calculated using SExtractor). More than 98\% of blue regions
are further than 2.5 kpc from the center. 45\% of the red regions
are located within the 2.5 kpc of the center, 90\% of which contribute
around 30--90\% to the total stellar mass of the host galaxy. It is worth
noting that 2.5 kpc is an arbitrary cut chosen for simplicity rather
than binning in distance. The red regions that are further than
2.5 kpc from the center (red triangles) are likely to be dustier regions in
the disk. This will be explored further in future work by cross
correlating the mass and extinction maps. It is possible many central
components could be dusty star forming regions as well, rather than
relatively passive spheroidal components like classical bulges.

We now examine the evolution of the relative area covered by blue and
red regions inside each  galaxy, with respect  to both redshift and
the total stellar mass of each galaxy. Figure 9 plots the covering
fraction of blue and red regions as a function of redshift and stellar
mass. Clearly, the covering fractions of red regions are always larger
than blue regions. Also, the dispersion in the distribution of the
covering fractions for red regions is consistently larger than that for
the blue. Furthermore, we find no significant evolution with either
redshift or stellar mass in the covering fraction of blue or red
regions. This supports the paradigm that small-scale properties of
galaxies on average are not evolving much over intermediate redshifts.

For the blue and red regions, we measure distances from
each of their associated resolution elements to the center of their respective galaxy
as defined by the rotation curve. We then normalize to the scale radius of the galaxy to avoid
biases due to the evolution of galaxy size across redshift. Figure
10 shows the median and dispersion of distances from the center of the
galaxy for both blue and red regions vs. redshift and total stellar
mass respectively. Blue regions are almost always further away from the
center of the galaxy than red regions with a larger dispersion in the distribution
of distances. Since the red regions are often found to be synonymous with
the bulge, the increase of the weight and strength of this red, bulge
region (manifested by the decrease in scatter) with the stellar mass can be seen in Figure 10. There is no significant evolution in
the median distance (scaled by the scale radii of the galaxy) of red
or blue regions from the center with time in the redshift range
covered in this study. However, there is a a slight increase in the
median distance between blue and red regions in more massive galaxies
compared to the least massive galaxies.

\begin{figure*}[htbp]
\centering
\begin{tabular}{cc}
\includegraphics[trim=0.3cm 0cm 0.3cm 0cm,width=2.8in]{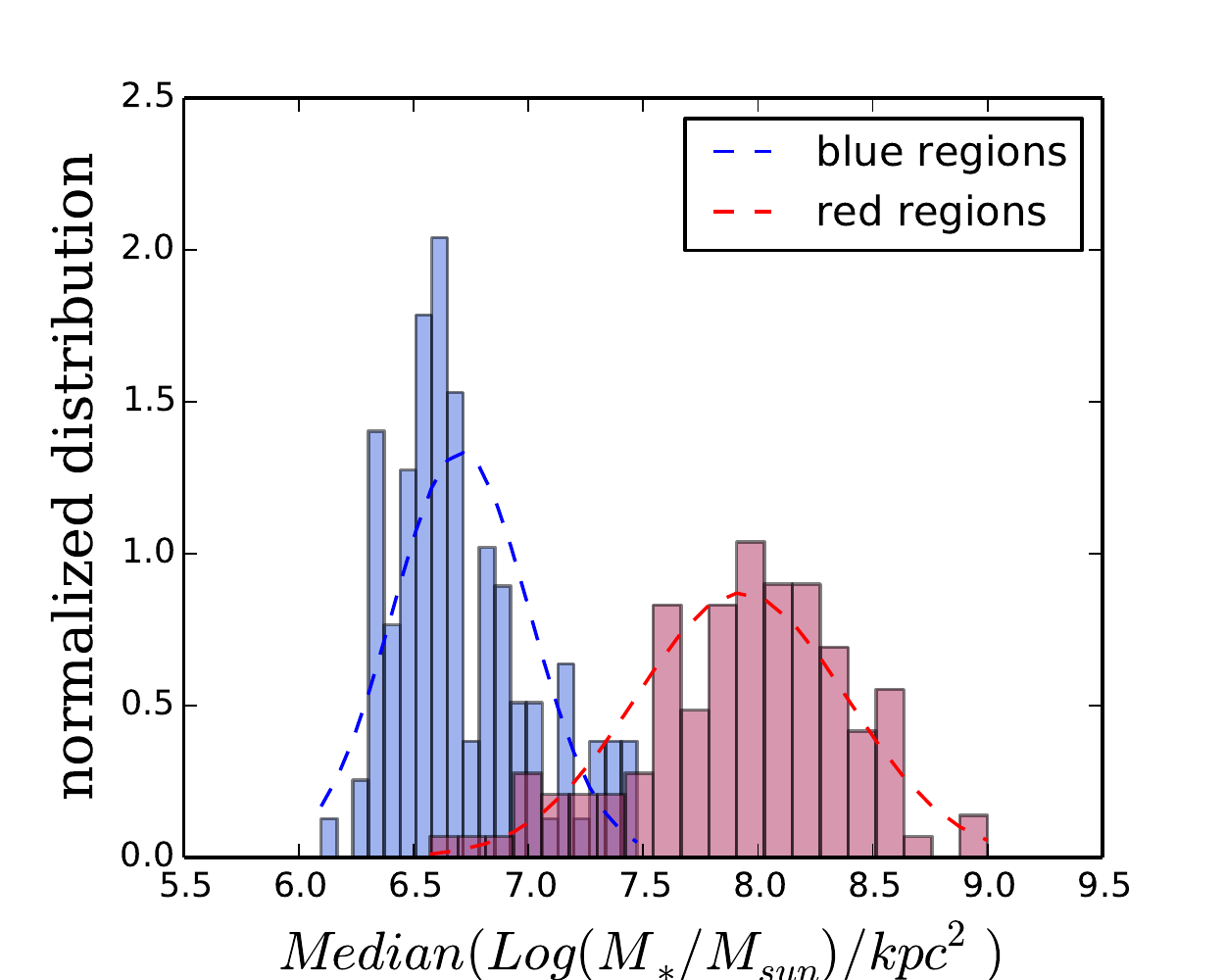}
\includegraphics[trim=0.3cm 0cm 0.3cm 0cm,width=2.8in]{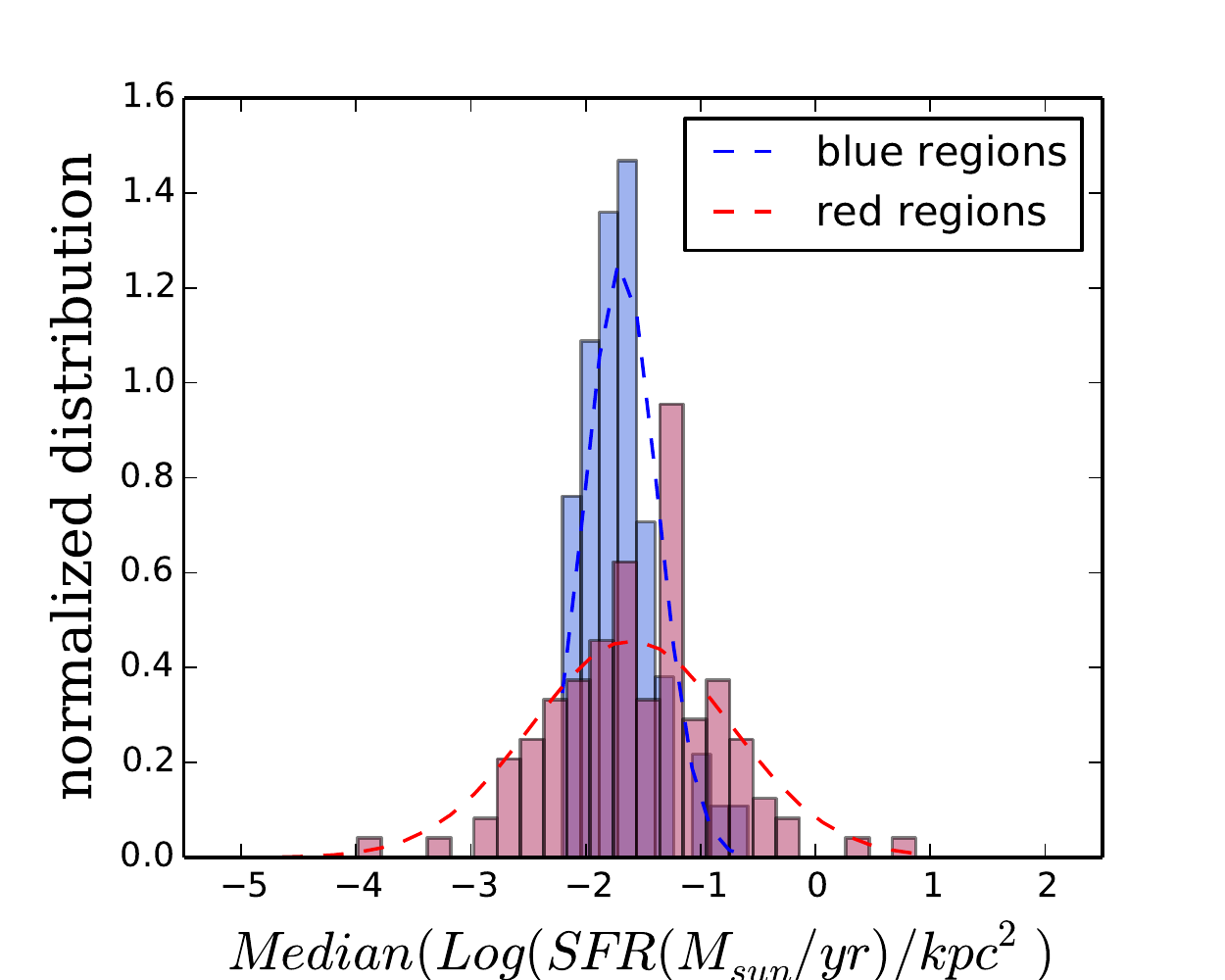}\\
\includegraphics[trim=0.3cm 0cm 0.3cm 0cm,width=2.8in]{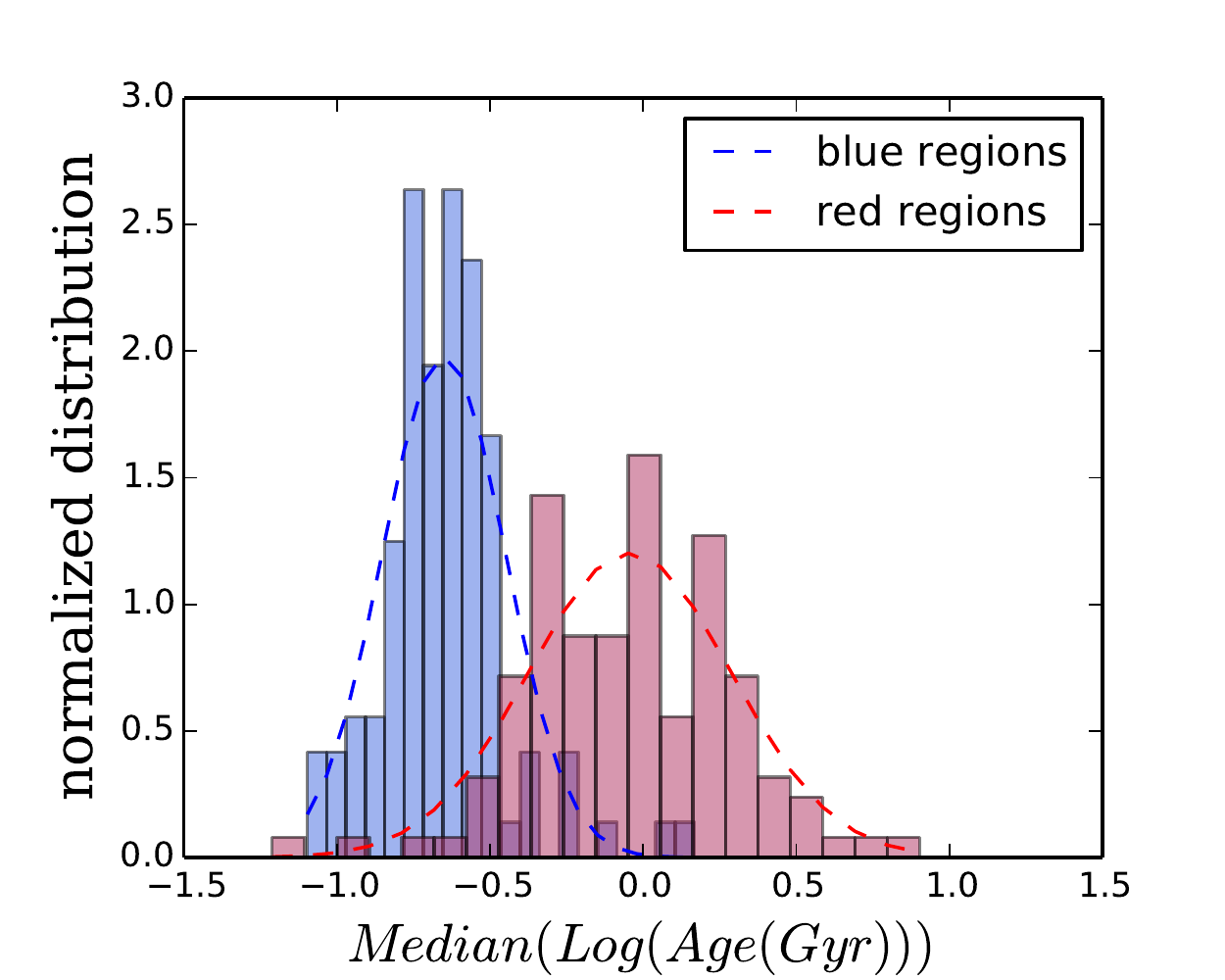}
\includegraphics[trim=0.3cm 0cm 0.3cm 0cm,width=2.8in]{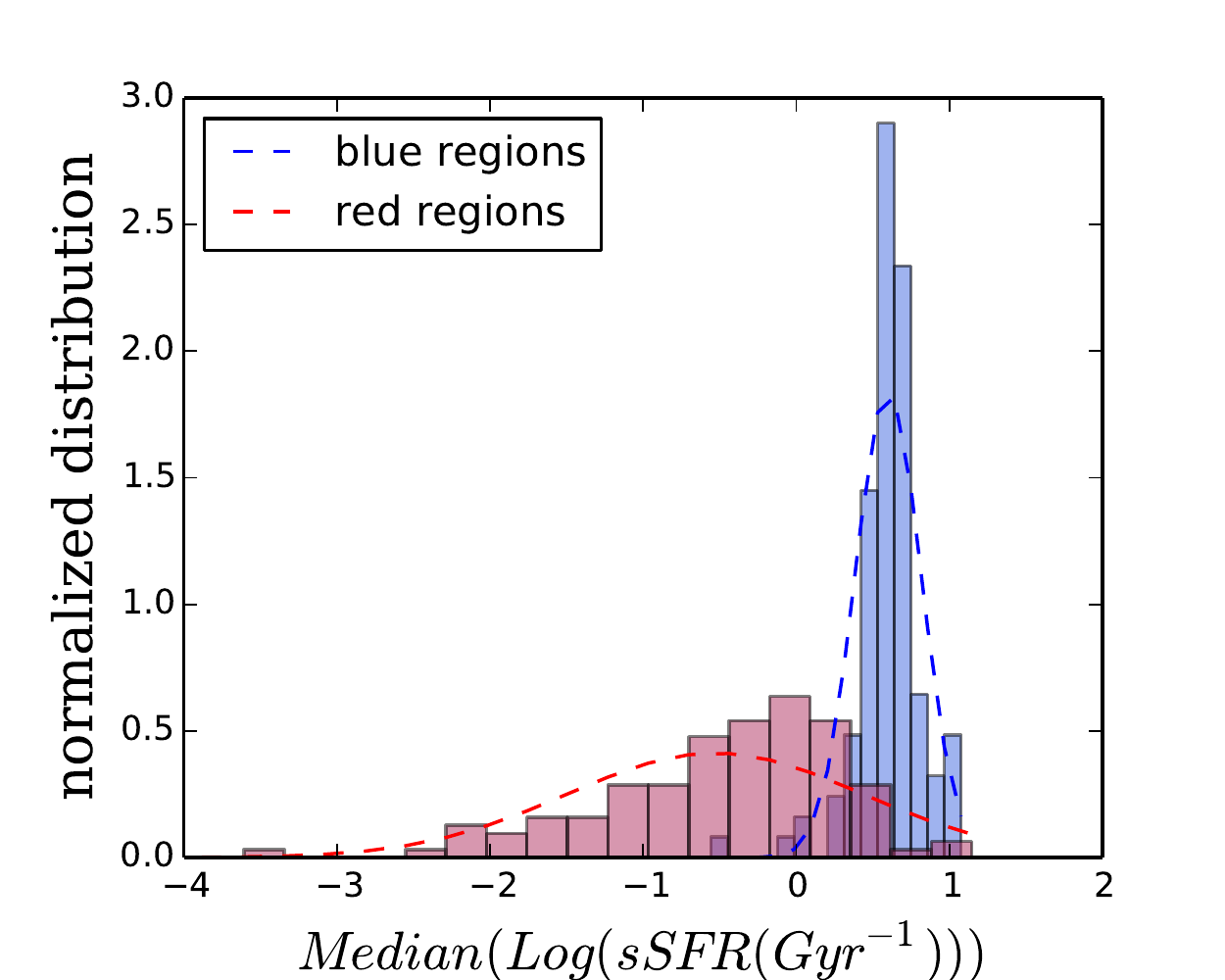}
\end{tabular}
\caption{Physical preperties of the blue and red regions in
  galaxies. Histogram of median values of stellar mass surface density, star formation
rate surface density, stellar age and sSFR  are plotted from top left
to bottom right respectively. Dashed curves are Gaussian functions fitted to each histogram.}
\end{figure*}

In order to study the physical nature of the red and blue regions, we
compare the distributions of median mass, SFR, age and sSFR for the regions
associated with each galaxy in our sample (Figure 11). To account for
the effect of redshift (a resolution element at low redshift corresponds
to a smaller area than at higher redshifts), we normalized each value
by the physical area  of a resolution element at the spectroscopic redshift of the
galaxy. While red and blue regions are identified solely based on rest-frame
(U-V) colors, there is a clear distinction between their stellar
mass surface density, age and sSFR distribution. However, they have similar SFR surface
densities. As in most of our galaxies, especially as mass increases, the
bulge, central pseudo-bulge, or central disk contributes to a
significant portion of the red regions. It is in the red regions
where the highest concentration of stellar mass exists. 
It is also clear that the distribution of stellar ages in blue regions
peaks at younger ages, even though the overall distribution 
of star formation rates are similar as that in red regions.

\subsection{Main sequence of star-forming galaxies}

There is a strong correlation between the SFR and stellar mass {\it
  ``main sequence''} in star-forming galaxies out to high redshifts
(e.g. \cite{Lilly2013}, \cite{Salmon2014}), with the
bulk of star formation occurring in massive galaxies than in less
massive systems (e.g. \cite{Noeske2007}, \cite{Bell2005},
\cite{Reddy2006}, \cite{Elbaz2007}). A population of passive galaxies
also exists, located below the main sequence, whereas starbursts lie
above. \cite{Wuyts2013} showed evidence that star formation and
assembled stellar mass also correlate at subgalactic scale. We present
the SFR-stellar mass relation for the blue and red regions in our galaxies
in Figure 12. Each red/blue point on the plot demonstrates the median
of SFR and mass over resolution elements associated to red/blue regions in any given galaxy.

\begin{figure}[htbp]
\centering
\includegraphics[trim=1.3cm 10.0cm 0.0cm 3.0cm,width=4.4in]{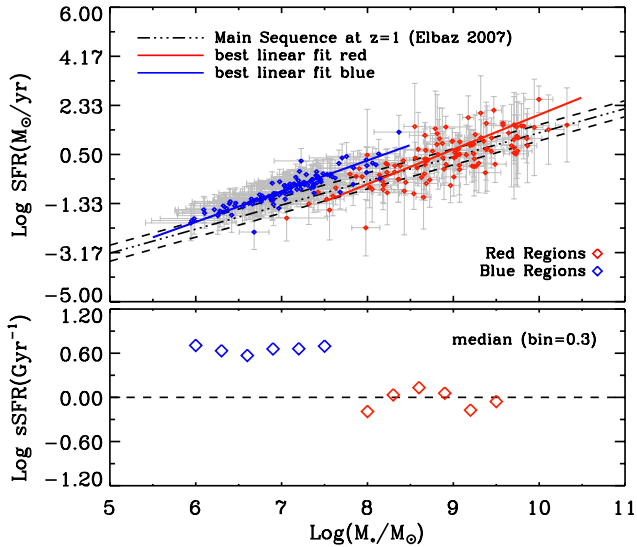}\\
\caption{The top panel is the median star formation rate with respect to median stellar
  mass for red (red diamonds) and blue (blue diamonds) regions in galaxies (1 red and 1 blue point
  for each galaxy in our sample). Linear fits to blue and
  red regions are plotted as solid lines. The dashed-dotted lines
  represent the extrapolation of the main sequence at $z\sim 1$
  \citep{Elbaz2007} (best fit and $1\sigma$). In the bottom panel the
  median sSFR in blue and red regions are plotted in stellar mass bins
  of 0.3 dex.}
\end{figure}

 There is a clear bimodality between the red and blue regions,
  with red regions having a higher stellar mass compared to blue
  regions at a fixed star formation rate. The blue regions form a tight
  relation with a scatter of $\sim0.2$ dex which is three times less
  than that seen in the red regions ($\sim0.6$ dex). This agrees with
  the spread in the SFR for red regions shown in Figure 11. The best linear
  fit to the blue and red regions, shown with solid lines,
  have slopes of $1.1\pm 0.1$ and $1.3\pm 0.1$ respectively. This is
  steeper compared to the relation seen for integrated main-sequence
  galaxies \citep{Elbaz2007} extrapolated to lower stellar
  masses. However, this is consistent with a recent work by \cite{Whitaker2014}, studying lower
  mass systems ($log(M_{*}/M_{sun})<10$) and finding a slope of $\sim1.0$
  roughly constant over the redshift range $0.5<z<2.5$. We also show
  the relation between the sSFR and median stellar mass, where blue regions have
  higher sSFR compared to red regions, a trend which does not appear
  to depend on median stellar mass of the region in this projection,
  however this changes as we account for stelar mass of the sample as
  a function of redshift.

\begin{figure*}[htbp]
\centering
\begin{tabular}{cc}
\includegraphics[trim=2cm 6.2cm 4cm 7cm,clip,width=8.3cm]{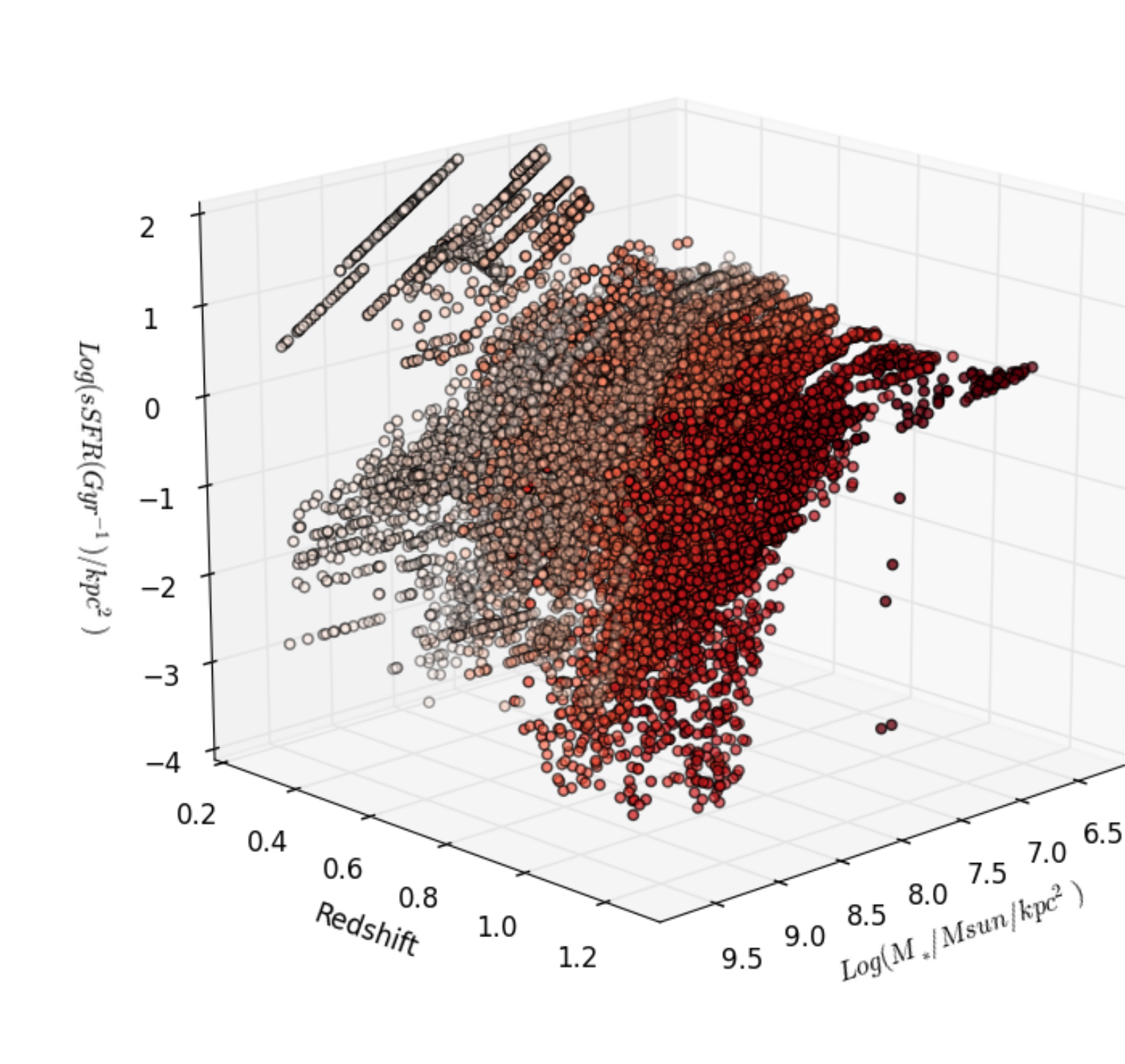}
\includegraphics[trim=3cm 6.2cm 3cm 7cm,clip,width=8.3cm]{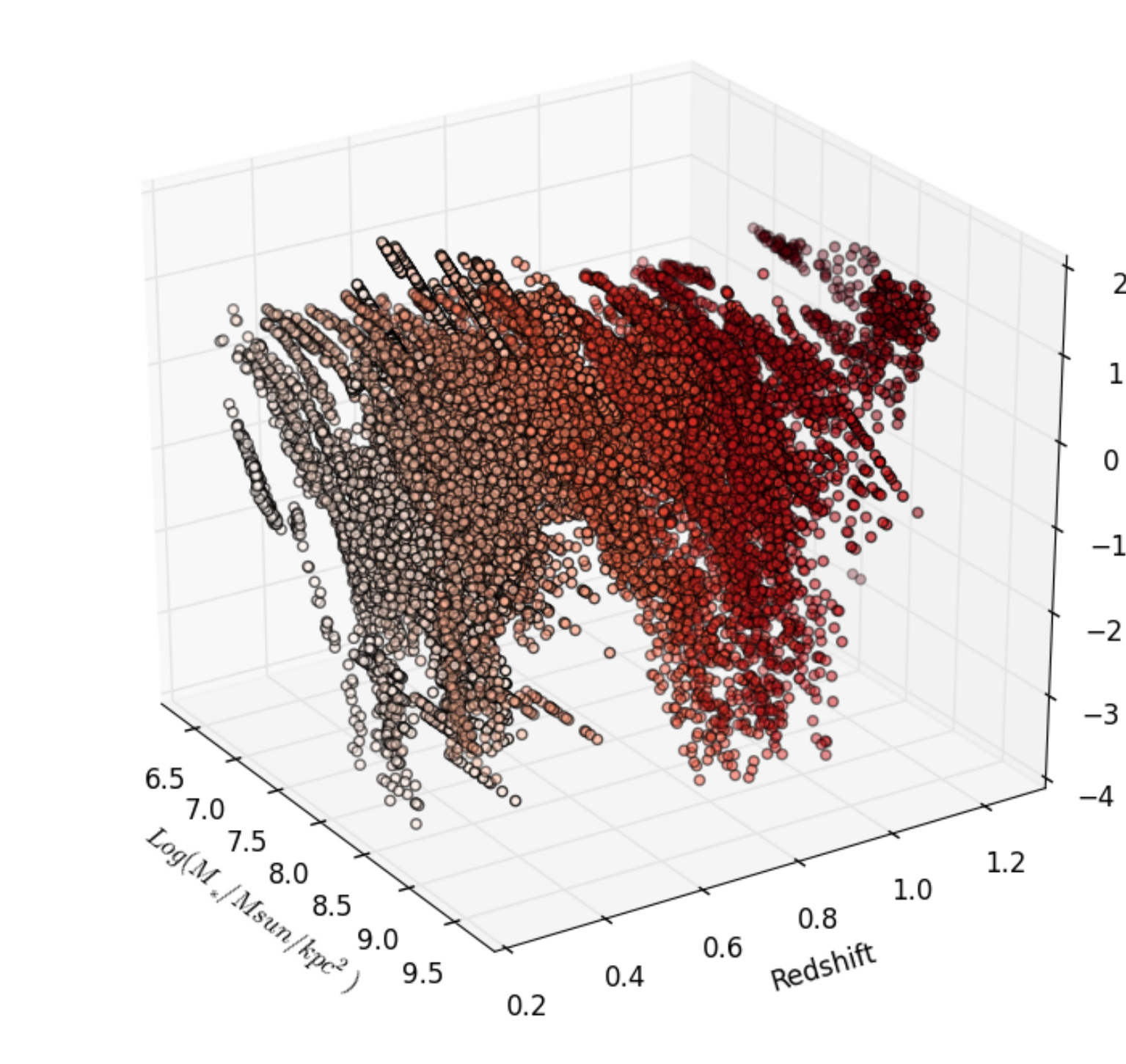}\\
\includegraphics[trim=2cm 6.2cm 4cm 7cm,clip,width=8.3cm]{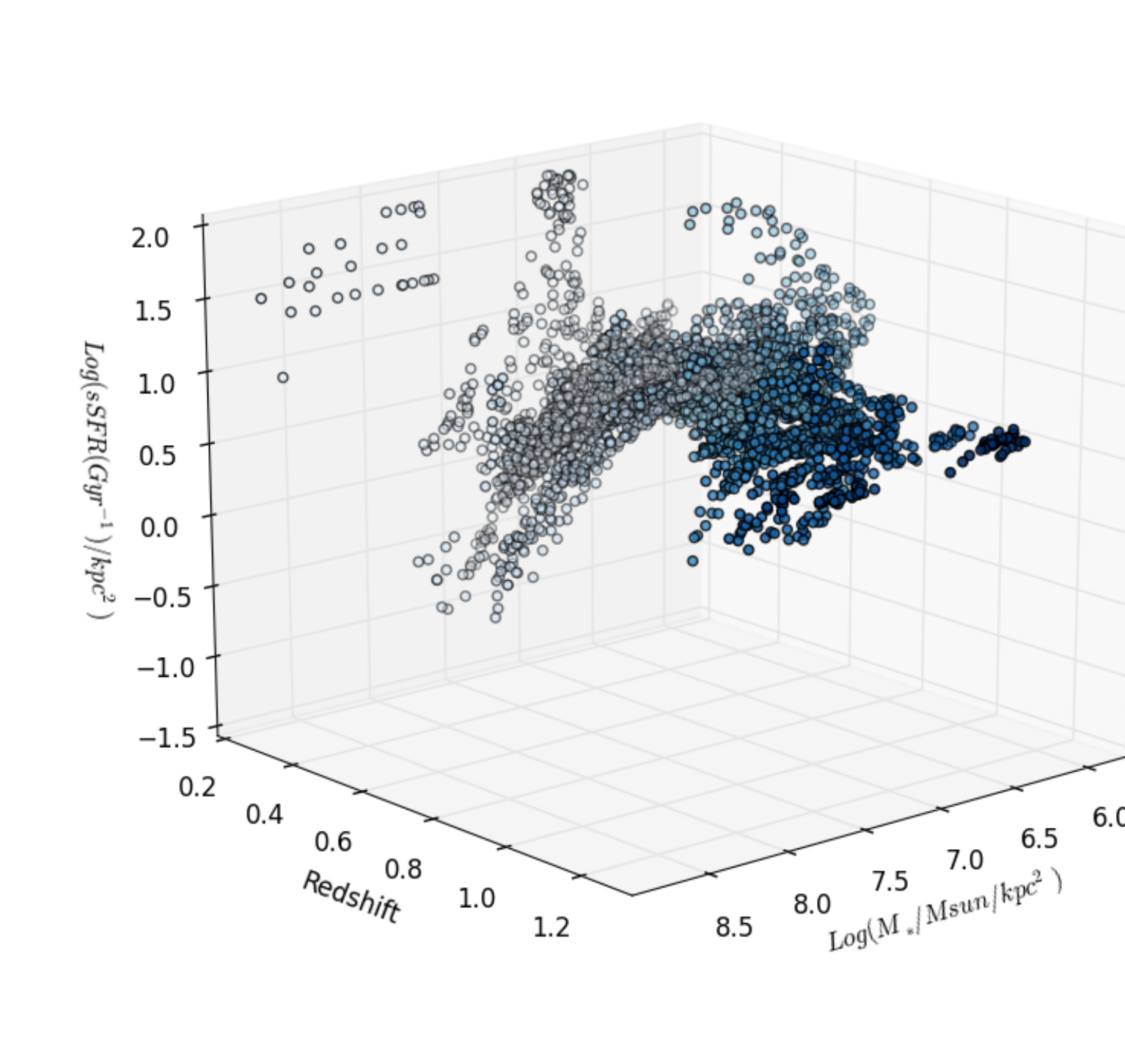}
\includegraphics[trim=3cm 6.2cm 3cm 7cm,clip,width=8.3cm]{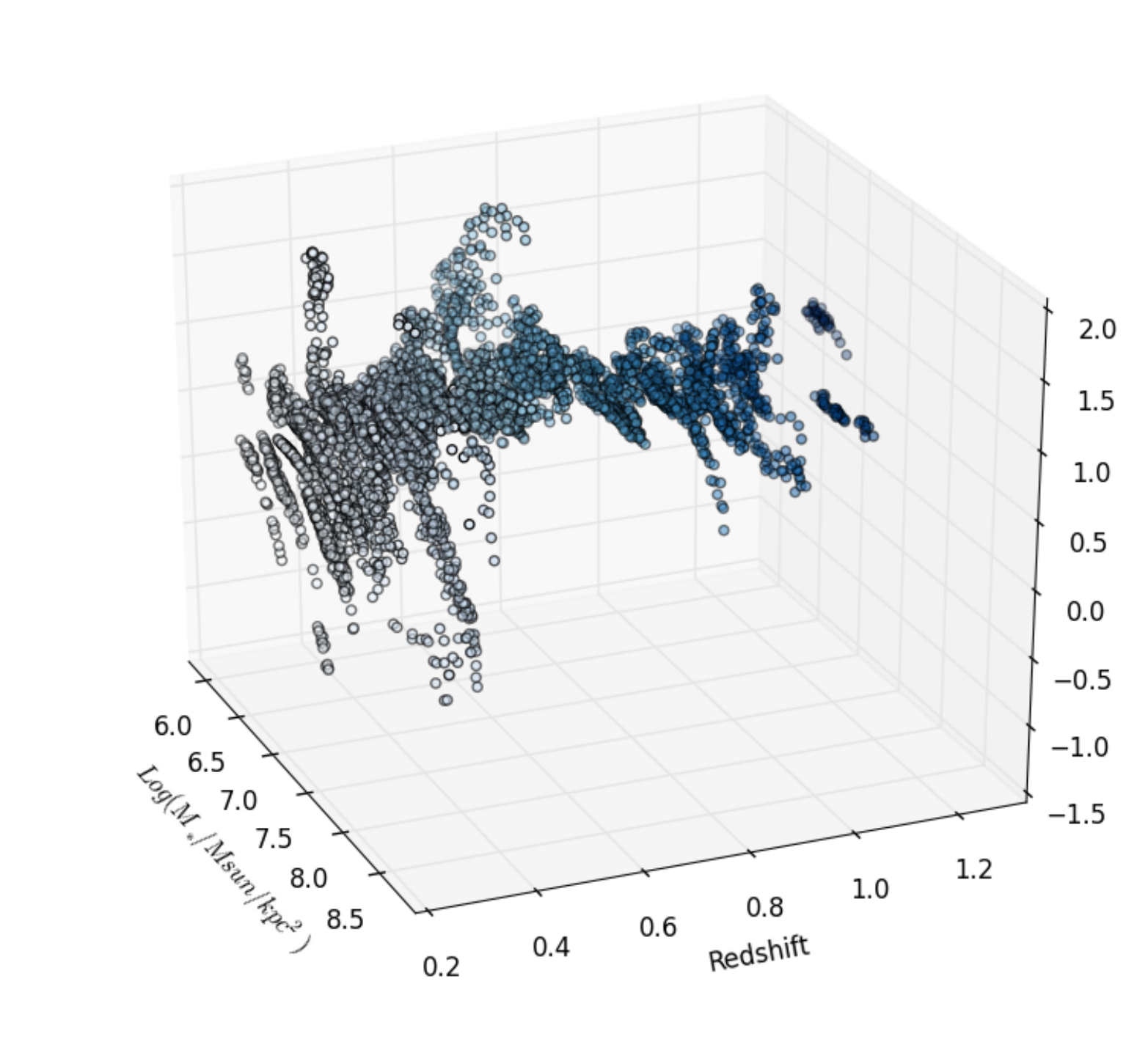}
\end{tabular}
\caption{sSFR plotted as a function of stellar mass and redshift for
  resolution elements associated with red(top) and blue(bottom)
  regions at two different projection (right/left). Each point is color coded based on redshift.}
\end{figure*}

 The specific star formation rate is known to evolve over intermediate
to low redshifts with both redshift and stellar mass of the
galaxies (e.g. \cite{Feulner2005}). In figure 13, we present a 3D plot
of the sSFR vs. redshift vs. stellar mass surface density for
individual resolution elements
belonging to red (left) or blue (right) regions. The sSFR of red
regions decreases as the stellar mass increases for all redshifts. The
same overall trend can be seen among resolution elements in blue regions. Binning
by stellar mass, we fit a function of the form $sSFR \propto
(1+z)^{b}$ to the sSFR per resolution element in red and blue regions. This is
plotted in Figure 14 where we see an increase in the $b$ value
with increasing stellar mass. It is known that at any given redshift, the mean sSFR is smaller for high mass
galaxies \citep{Damen2009}. In the context of ``downsizing'' more
massive galaxies formed their stars before less massive systems
\citep{Cowie1996}. The trend that we see here for substructures in
galaxies is very similar with bluer less massive regions having higher
sSFRs compared to red regions. This implies that the more massive
regions have assembled more of their mass at earlier
times.

\begin{figure}[htbp]
\centering
\includegraphics[trim=0.0cm 0.0cm 0cm 0cm, clip=true,width=3.7in]{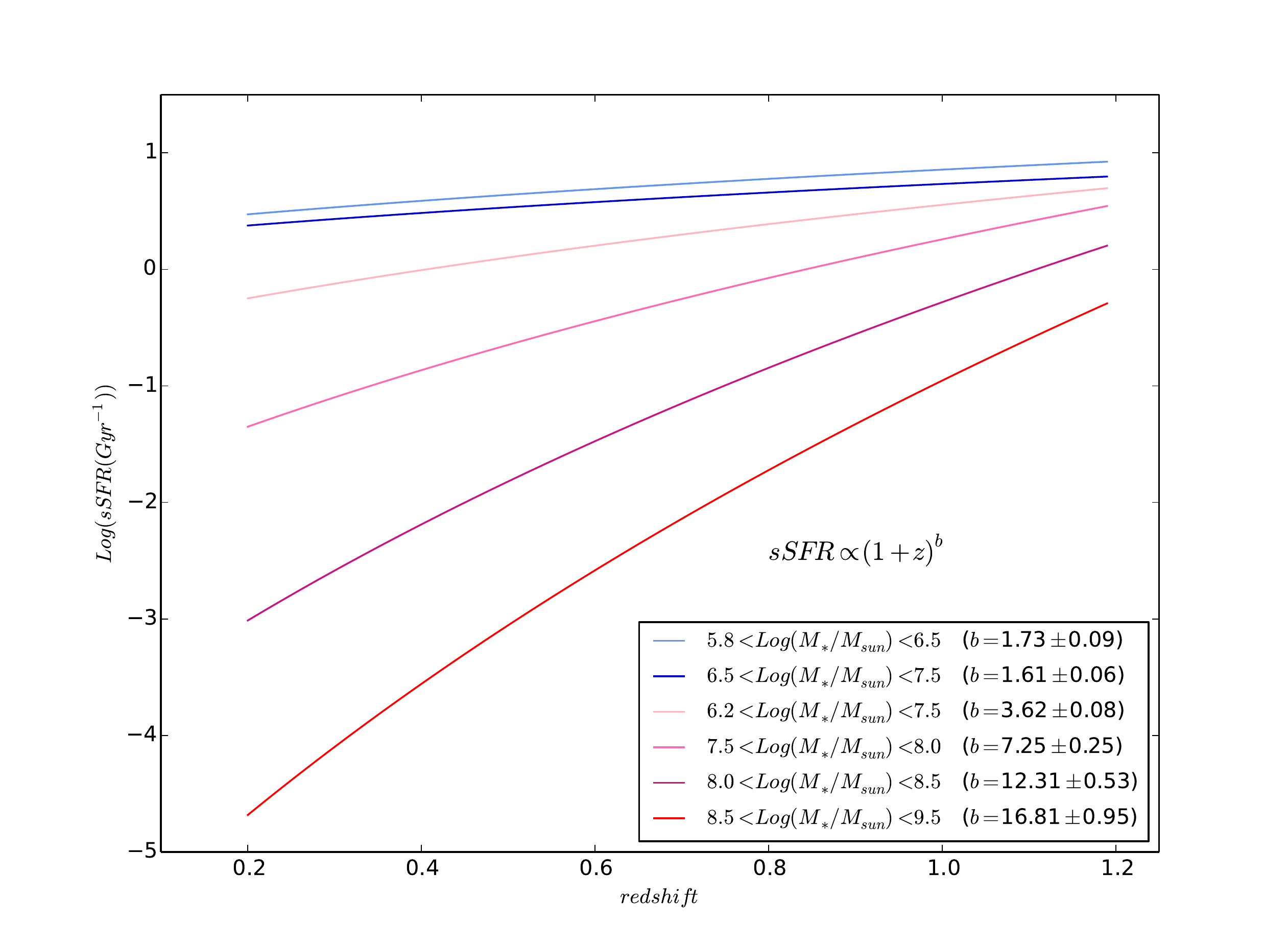}
\caption{Redshift evolution of specific SFR: red and blue
lines represent the best fit to the resolution elements in red and blue
regions binned by stellar mass surface density, respectively.}
\end{figure}

\section{Discussion}

 In studies of evolution of galaxies, samples are often divided into
 two broad populations: actively star-forming and passive. However,
 studying resolved maps of star-forming galaxies at kpc-scales,we
 found different modes of galaxy growth within a given galaxy, with
 specific star formation rates differing over an order of magnitude (sometimes
 even more) from region to region. 

Red regions on average have higher mass surface densities and are
older compared to bluer regions in galaxies out to $z\sim1.3$. Using our method, we find that the
 covering fraction of red regions is always greater than blue regions and that the covering fraction of these regions does not
evolve significantly with redshift or the total stellar mass of the
galaxy in the redshift ($0.2<z<1.3$) and stellar mass
($Log(M_{*}/M_{sun})\sim 8.5-11.0$) range examined in this study. We also
show that blue regions are further away from the center of the galaxy
(between 2-4 scale radii of the galaxy) compared
to red regions (typically within 1 scale radius) with blue regions
showing a larger dispersion. For both blue and red regions, the distance from the
center as normalized by the scale radius is not evolving with
redshift, however, more variation is seen with the stellar mass of the
galaxy. In more massive systems, the distribution of red regions tend to be tightly
confined to the center, with a much smaller dispersion compared to
less massive galaxies, while the dispersion remains larger
for blue regions. More generally, the clear distinction between the
radial distribution of red and blue regions becomes much less distinct
in the lowest mass galaxies as opposed to the high mass objects. In future works, we
will investigate how properties of these red and blue regions relate with the morphology of the host galaxy, specifically
substructures in galaxies with disturbed morphologies, including
instabilities and merger remnants.
 
 We find that red regions have already formed the bulk of their stars
 where relatively more star formation per stellar mass surface density persists in the blue regions.
 This appears to be driven by the difference in stellar mass surface
 density between red and blue regions, rather than systematic
 differences in the star-formation rate itself. The decline in
 specific star formation rate with redshift is thus seen for all
 regions but the declining rate depends on the stellar mass, with more
 massive regions declining with steeper slopes compared to less
 massive ones.

 The emerging picture is that centrally located red regions cover more
 galaxy surface area relative to the blue regions in the outer disk,
 as expected given that the highest stellar mass surface densities are found in
 galaxy centers (in bulges, pseudo-bulges, or simply the central peak
 of the disk profile) and regions of higher specific star formation
 rates are found further out in the relatively bluer galaxy/disks. We
 are indeed, seeing a difference in stellar mass surface density and
 dust combined with younger ages in blue regions, rather than elevated
 levels of star-formation, per se, as compared to red regions. 

Asked another way, if the star formation rates have similar
 distributions between red and blue regions, why do we still detect blue U-V
 color regions given the smoothness of stellar mass surface
 density maps? By process of elimination, it is the stellar ages, metallicities
 and/or dust distributions that are left as the probable causes of blue regions observed
 in the outer disks. While we know degeneracies exist in fitting these
 properties via SEDs, it is unclear if these compounded degeneracies are enough to falsely
 manufacture a $\sim1$ Gyr spread between median ages in red and blue
 regions (Fig. 11). Uneven distributions of metals in pockets left
 over from previous generations of stars could be at play, on their
 way to being smoothed out by the mixing orbits of typical disk
 dynamics. Ultimately though blue regions are either caused by a
 relatively higher fraction of bright, short-lived O and B stars being
 formed in the region, and/or windows in the dust just above the disk
 (dust that is elsewhere denser and absorbing the light escaping from,
 e.g., the denser, dustier galaxy core). The UV light of a small
 population of young stars could possibly be completely obscured from
 our view in red regions which are regions of thicker stellar
 disk/bulge and thus more affected by absorption. But if this is not
 the case, then why exactly are such different populations of stars
 forming in red regions vs. blue regions given a universal IMF? 
 
By overlooking the small-scale properties of galaxies when studying
their global properties, the whole picture is not revealed. In future work we plan to
 further address these questions by further application of the tools developed in this paper to leverage deep spectral profiles across galaxies in this sample. We aim to learn more about the
 evolution of metallicity profiles, rotation curve decomposition, and
 comparative dynamical and star-formation timescales derived from both
 the photometric and spectroscopic properties of our data.

\section{Summary}

 In this paper we have studied the resolved kpc-scale structures in a
unique sample of 119 disk-galaxies in the GOODS fields using the
highest resolution HST images from optical to near-infrared bands,
taken as a part of the CANDELS project. This sample has
extraordinarily deep spectroscopic observations with Keck DEIMOS,
including precise measures of their rotation curves, allowing for not
only a more detailed dynamical study using these maps, but also other
spectroscopically enabled studies using stellar absorption and
emission lines (e.g., metallicity gradients across the galaxies, etc.).
The main points of this first paper are summarized as follows:
\begin{itemize}
\item
We develop and test a method to generate spatially-resolved rest-frame color,
stellar mass, age, SFR and extinction maps of galaxies with HST resolution, 
allowing for differences in PSFs from different instruments and testing against
redshift-dependent biases. 
\item
We compared the stellar mass surface density, SFR surface density and stellar
ages measured by resolved SED fitting to the integrated stellar masses,
SFRs and ages. We found a good agreement with offsets of $0.04$
dex, $0.16$ dex and $0.02$ dex for stellar mass, SFR and age respectively. 
\item
We find that the stellar mass distribution in the disks are very
smooth and 30-90\% of the total stellar mass comes from the central
2.5 kpc of the galaxies of our sample.
\item
Developing a method to define statistically significant regions within
galaxies, we identify ``blue'' and ``red'' regions in the 2-D color
maps and investigated their physical properties.
In the majority of galaxies in our sample, the rest-frame (U-V) color maps covering the
Balmer/4000\AA\ break have red centers with most blue regions located in the outer parts of the
galaxies. We find few exceptions with blue nuclei (N=3 in this sample), which host
AGN given the X-ray detection within 1 arc second of their centers in 
each of these cases.
\item
We show that there is a bimodality between red and blue regions on the
SFR-mass plane, with the red regions having a higher stellar mass
surface densities compared to blue regions at each fixed SFR. The
relation is much tighter for blue regions with a slope of $1.1\pm 0.1$
compared to red regions with a slightly steeper slope of $1.3\pm0.1$.

\item
Fitting the sSFR as a function of redshift for different stellar mass
bins, we see more massive regions having smaller sSFRs at each
redshift. We quantify the rate of sSFR decline with redshift for
different mass bins in blue and red regions. We find that this decline
is driven primarily by the stellar mass surface densities rather than
the star formation rates at a given resolution element.

\end{itemize}
\acknowledgments
We thank the anonymous referee for the extensive review and constructive
comments on this manuscript. We thank Richard Ellis, Jeff Newman and
Behnam Darvish for helpful comments which improved the quality of this
work. This work is based on observations taken by the CANDELS
Multi-Cycle Treasury Program with the NASA/ESA HST, which is operated by the
Association of Universities for Research in Astronomy, Inc., under
NASA contract NAS5-26555. The authors would like to thank NASA and
STScI for HST Theory/Archival grant AR-13259. DK acknowledges partial
support from NSF grant AST-0808133.  We thank R. Ellis, K. Bundy,
T. Treu, and K. Chiu for their part in the initial assembly of the
spectroscopic dataset that this sample is based on, presented in
\citet{Miller2011}.

\begin{tiny}

\bibliography{shooby.bib}

\begin{thebibliography}{}
\expandafter\ifx\csname natexlab\endcsname\relax\def\natexlab#1{#1}\fi

\bibitem[{{Abraham} {et~al.}(1999){Abraham}, {Ellis}, {Fabian}, {Tanvir}, \&
  {Glazebrook}}]{Abraham1999}
{Abraham}, R.~G., {Ellis}, R.~S., {Fabian}, A.~C., {Tanvir}, N.~R., \&
  {Glazebrook}, K. 1999, \mnras, 303, 641

\bibitem[{{Arnouts} {et~al.}(1999){Arnouts}, {Cristiani}, {Moscardini},
  {Matarrese}, {Lucchin}, {Fontana}, \& {Giallongo}}]{Arnouts1999}
{Arnouts}, S., {Cristiani}, S., {Moscardini}, L., {et~al.} 1999, \mnras, 310,
  540

\bibitem[{{Barden} {et~al.}(2008){Barden}, {Jahnke}, \&
  {H{\"a}u{\ss}ler}}]{Ferengi2008}
{Barden}, M., {Jahnke}, K., \& {H{\"a}u{\ss}ler}, B. 2008, \apjs, 175, 105

\bibitem[{{Bell} {et~al.}(2006){Bell}, {Phleps}, {Somerville}, {Wolf}, {Borch},
  \& {Meisenheimer}}]{Bell2006}
{Bell}, E.~F., {Phleps}, S., {Somerville}, R.~S., {et~al.} 2006, \apj, 652, 270

\bibitem[{{Bell} {et~al.}(2004){Bell}, {Wolf}, {Meisenheimer}, {Rix}, {Borch},
  {Dye}, {Kleinheinrich}, {Wisotzki}, \& {McIntosh}}]{Bell2004}
{Bell}, E.~F., {Wolf}, C., {Meisenheimer}, K., {et~al.} 2004, \apj, 608, 752

\bibitem[{{Bell} {et~al.}(2005){Bell}, {Papovich}, {Wolf}, {Le Floc'h},
  {Caldwell}, {Barden}, {Egami}, {McIntosh}, {Meisenheimer},
  {P{\'e}rez-Gonz{\'a}lez}, {Rieke}, {Rieke}, {Rigby}, \& {Rix}}]{Bell2005}
{Bell}, E.~F., {Papovich}, C., {Wolf}, C., {et~al.} 2005, \apj, 625, 23

\bibitem[{{Bell} {et~al.}(2012){Bell}, {van der Wel}, {Papovich}, {Kocevski},
  {Lotz}, {McIntosh}, {Kartaltepe}, {Faber}, {Ferguson}, {Koekemoer}, {Grogin},
  {Wuyts}, {Cheung}, {Conselice}, {Dekel}, {Dunlop}, {Giavalisco},
  {Herrington}, {Koo}, {McGrath}, {de Mello}, {Rix}, {Robaina}, \&
  {Williams}}]{Bell2012}
{Bell}, E.~F., {van der Wel}, A., {Papovich}, C., {et~al.} 2012, \apj, 753, 167

\bibitem[{{Bertin} \& {Arnouts}(1996)}]{Bertin1996}
{Bertin}, E., \& {Arnouts}, S. 1996, \aaps, 117, 393

\bibitem[{{Bruzual} \& {Charlot}(2003)}]{BC03}
{Bruzual}, G., \& {Charlot}, S. 2003, \mnras, 344, 1000

\bibitem[{{Chabrier}(2003)}]{Chabrier2003}
{Chabrier}, G. 2003, \pasp, 115, 763

\bibitem[{{Conroy}(2013)}]{Conroy2013}
{Conroy}, C. 2013, \araa, 51, 393

\bibitem[{{Cowie} {et~al.}(1996){Cowie}, {Songaila}, {Hu}, \&
  {Cohen}}]{Cowie1996}
{Cowie}, L.~L., {Songaila}, A., {Hu}, E.~M., \& {Cohen}, J.~G. 1996, \aj, 112,
  839

\bibitem[{{Daddi} {et~al.}(2007){Daddi}, {Dickinson}, {Morrison}, {Chary},
  {Cimatti}, {Elbaz}, {Frayer}, {Renzini}, {Pope}, {Alexander}, {Bauer},
  {Giavalisco}, {Huynh}, {Kurk}, \& {Mignoli}}]{Daddi2007}
{Daddi}, E., {Dickinson}, M., {Morrison}, G., {et~al.} 2007, \apj, 670, 156

\bibitem[{{Daddi} {et~al.}(2010){Daddi}, {Bournaud}, {Walter}, {Dannerbauer},
  {Carilli}, {Dickinson}, {Elbaz}, {Morrison}, {Riechers}, {Onodera}, {Salmi},
  {Krips}, \& {Stern}}]{Daddi2010}
{Daddi}, E., {Bournaud}, F., {Walter}, F., {et~al.} 2010, \apj, 713, 686

\bibitem[{{Damen} {et~al.}(2009){Damen}, {F{\"o}rster Schreiber}, {Franx},
  {Labb{\'e}}, {Toft}, {van Dokkum}, \& {Wuyts}}]{Damen2009}
{Damen}, M., {F{\"o}rster Schreiber}, N.~M., {Franx}, M., {et~al.} 2009, \apj,
  705, 617

\bibitem[{{Dekel} {et~al.}(2009){Dekel}, {Sari}, \& {Ceverino}}]{Dekel2009}
{Dekel}, A., {Sari}, R., \& {Ceverino}, D. 2009, \apj, 703, 785

\bibitem[{{Dickinson} {et~al.}(2003){Dickinson}, {Papovich}, {Ferguson}, \&
  {Budav{\'a}ri}}]{Dickinson2003}
{Dickinson}, M., {Papovich}, C., {Ferguson}, H.~C., \& {Budav{\'a}ri}, T. 2003,
  \apj, 587, 25

\bibitem[{{Elbaz} {et~al.}(2007){Elbaz}, {Daddi}, {Le Borgne}, {Dickinson},
  {Alexander}, {Chary}, {Starck}, {Brandt}, {Kitzbichler}, {MacDonald},
  {Nonino}, {Popesso}, {Stern}, \& {Vanzella}}]{Elbaz2007}
{Elbaz}, D., {Daddi}, E., {Le Borgne}, D., {et~al.} 2007, \aap, 468, 33

\bibitem[{{Elmegreen} {et~al.}(2008){Elmegreen}, {Bournaud}, \&
  {Elmegreen}}]{Elmegreen2008}
{Elmegreen}, B.~G., {Bournaud}, F., \& {Elmegreen}, D.~M. 2008, \apj, 688, 67

\bibitem[{{Faber} {et~al.}(2003){Faber}, {Phillips}, {Kibrick}, {Alcott},
  {Allen}, {Burrous}, {Cantrall}, {Clarke}, {Coil}, {Cowley}, {Davis}, {Deich},
  {Dietsch}, {Gilmore}, {Harper}, {Hilyard}, {Lewis}, {McVeigh}, {Newman},
  {Osborne}, {Schiavon}, {Stover}, {Tucker}, {Wallace}, {Wei}, {Wirth}, \&
  {Wright}}]{Faber2003}
{Faber}, S.~M., {Phillips}, A.~C., {Kibrick}, R.~I., {et~al.} 2003, in Society
  of Photo-Optical Instrumentation Engineers (SPIE) Conference Series, Vol.
  4841, Instrument Design and Performance for Optical/Infrared Ground-based
  Telescopes, ed. M.~{Iye} \& A.~F.~M. {Moorwood}, 1657--1669

\bibitem[{{Faber} {et~al.}(2007){Faber}, {Willmer}, {Wolf}, {Koo}, {Weiner},
  {Newman}, {Im}, {Coil}, {Conroy}, {Cooper}, {Davis}, {Finkbeiner}, {Gerke},
  {Gebhardt}, {Groth}, {Guhathakurta}, {Harker}, {Kaiser}, {Kassin},
  {Kleinheinrich}, {Konidaris}, {Kron}, {Lin}, {Luppino}, {Madgwick},
  {Meisenheimer}, {Noeske}, {Phillips}, {Sarajedini}, {Schiavon}, {Simard},
  {Szalay}, {Vogt}, \& {Yan}}]{Faber2007}
{Faber}, S.~M., {Willmer}, C.~N.~A., {Wolf}, C., {et~al.} 2007, \apj, 665, 265

\bibitem[{{Feulner} {et~al.}(2005){Feulner}, {Goranova}, {Drory}, {Hopp}, \&
  {Bender}}]{Feulner2005}
{Feulner}, G., {Goranova}, Y., {Drory}, N., {Hopp}, U., \& {Bender}, R. 2005,
  \mnras, 358, L1

\bibitem[{{F{\"o}rster Schreiber} {et~al.}(2009){F{\"o}rster Schreiber},
  {Genzel}, {Bouch{\'e}}, {Cresci}, {Davies}, {Buschkamp}, {Shapiro},
  {Tacconi}, {Hicks}, {Genel}, {Shapley}, {Erb}, {Steidel}, {Lutz},
  {Eisenhauer}, {Gillessen}, {Sternberg}, {Renzini}, {Cimatti}, {Daddi},
  {Kurk}, {Lilly}, {Kong}, {Lehnert}, {Nesvadba}, {Verma}, {McCracken},
  {Arimoto}, {Mignoli}, \& {Onodera}}]{Forster2009}
{F{\"o}rster Schreiber}, N.~M., {Genzel}, R., {Bouch{\'e}}, N., {et~al.} 2009,
  \apj, 706, 1364

\bibitem[{{Genzel} {et~al.}(2011){Genzel}, {Newman}, {Jones}, {F{\"o}rster
  Schreiber}, {Shapiro}, {Genel}, {Lilly}, {Renzini}, {Tacconi}, {Bouch{\'e}},
  {Burkert}, {Cresci}, {Buschkamp}, {Carollo}, {Ceverino}, {Davies}, {Dekel},
  {Eisenhauer}, {Hicks}, {Kurk}, {Lutz}, {Mancini}, {Naab}, {Peng},
  {Sternberg}, {Vergani}, \& {Zamorani}}]{Genzel2011}
{Genzel}, R., {Newman}, S., {Jones}, T., {et~al.} 2011, \apj, 733, 101

\bibitem[{{Giavalisco} {et~al.}(2004){Giavalisco}, {Ferguson}, {Koekemoer},
  {Dickinson}, {Alexander}, {Bauer}, {Bergeron}, {Biagetti}, {Brandt},
  {Casertano}, {Cesarsky}, {Chatzichristou}, {Conselice}, {Cristiani}, {Da
  Costa}, {Dahlen}, {de Mello}, {Eisenhardt}, {Erben}, {Fall}, {Fassnacht},
  {Fosbury}, {Fruchter}, {Gardner}, {Grogin}, {Hook}, {Hornschemeier}, {Idzi},
  {Jogee}, {Kretchmer}, {Laidler}, {Lee}, {Livio}, {Lucas}, {Madau},
  {Mobasher}, {Moustakas}, {Nonino}, {Padovani}, {Papovich}, {Park},
  {Ravindranath}, {Renzini}, {Richardson}, {Riess}, {Rosati}, {Schirmer},
  {Schreier}, {Somerville}, {Spinrad}, {Stern}, {Stiavelli}, {Strolger},
  {Urry}, {Vandame}, {Williams}, \& {Wolf}}]{Giavalisco2004}
{Giavalisco}, M., {Ferguson}, H.~C., {Koekemoer}, A.~M., {et~al.} 2004, \apjl,
  600, L93

\bibitem[{{Grogin} {et~al.}(2011){Grogin}, {Kocevski}, {Faber}, {Ferguson},
  {Koekemoer}, {Riess}, {Acquaviva}, {Alexander}, {Almaini}, {Ashby}, {Barden},
  {Bell}, {Bournaud}, {Brown}, {Caputi}, {Casertano}, {Cassata}, {Castellano},
  {Challis}, {Chary}, {Cheung}, {Cirasuolo}, {Conselice}, {Roshan Cooray},
  {Croton}, {Daddi}, {Dahlen}, {Dav{\'e}}, {de Mello}, {Dekel}, {Dickinson},
  {Dolch}, {Donley}, {Dunlop}, {Dutton}, {Elbaz}, {Fazio}, {Filippenko},
  {Finkelstein}, {Fontana}, {Gardner}, {Garnavich}, {Gawiser}, {Giavalisco},
  {Grazian}, {Guo}, {Hathi}, {H{\"a}ussler}, {Hopkins}, {Huang}, {Huang},
  {Jha}, {Kartaltepe}, {Kirshner}, {Koo}, {Lai}, {Lee}, {Li}, {Lotz}, {Lucas},
  {Madau}, {McCarthy}, {McGrath}, {McIntosh}, {McLure}, {Mobasher},
  {Moustakas}, {Mozena}, {Nandra}, {Newman}, {Niemi}, {Noeske}, {Papovich},
  {Pentericci}, {Pope}, {Primack}, {Rajan}, {Ravindranath}, {Reddy}, {Renzini},
  {Rix}, {Robaina}, {Rodney}, {Rosario}, {Rosati}, {Salimbeni}, {Scarlata},
  {Siana}, {Simard}, {Smidt}, {Somerville}, {Spinrad}, {Straughn}, {Strolger},
  {Telford}, {Teplitz}, {Trump}, {van der Wel}, {Villforth}, {Wechsler},
  {Weiner}, {Wiklind}, {Wild}, {Wilson}, {Wuyts}, {Yan}, \& {Yun}}]{Grogin2011}
{Grogin}, N.~A., {Kocevski}, D.~D., {Faber}, S.~M., {et~al.} 2011, \apjs, 197,
  35

\bibitem[{{Guo} {et~al.}(2012){Guo}, {Giavalisco}, {Ferguson}, {Cassata}, \&
  {Koekemoer}}]{Guo2012}
{Guo}, Y., {Giavalisco}, M., {Ferguson}, H.~C., {Cassata}, P., \& {Koekemoer},
  A.~M. 2012, \apj, 757, 120

\bibitem[{{Hopkins} {et~al.}(2010){Hopkins}, {Bundy}, {Croton}, {Hernquist},
  {Keres}, {Khochfar}, {Stewart}, {Wetzel}, \& {Younger}}]{Hopkins2010}
{Hopkins}, P.~F., {Bundy}, K., {Croton}, D., {et~al.} 2010, \apj, 715, 202

\bibitem[{{Ilbert} {et~al.}(2006){Ilbert}, {Arnouts}, {McCracken},
  {Bolzonella}, {Bertin}, {Le F{\`e}vre}, {Mellier}, {Zamorani}, {Pell{\`o}},
  {Iovino}, {Tresse}, {Le Brun}, {Bottini}, {Garilli}, {Maccagni}, {Picat},
  {Scaramella}, {Scodeggio}, {Vettolani}, {Zanichelli}, {Adami}, {Bardelli},
  {Cappi}, {Charlot}, {Ciliegi}, {Contini}, {Cucciati}, {Foucaud}, {Franzetti},
  {Gavignaud}, {Guzzo}, {Marano}, {Marinoni}, {Mazure}, {Meneux}, {Merighi},
  {Paltani}, {Pollo}, {Pozzetti}, {Radovich}, {Zucca}, {Bondi}, {Bongiorno},
  {Busarello}, {de La Torre}, {Gregorini}, {Lamareille}, {Mathez}, {Merluzzi},
  {Ripepi}, {Rizzo}, \& {Vergani}}]{Ilbert2006}
{Ilbert}, O., {Arnouts}, S., {McCracken}, H.~J., {et~al.} 2006, \aap, 457, 841

\bibitem[{{Ilbert} {et~al.}(2010){Ilbert}, {Salvato}, {Le Floc'h}, {Aussel},
  {Capak}, {McCracken}, {Mobasher}, {Kartaltepe}, {Scoville}, {Sanders},
  {Arnouts}, {Bundy}, {Cassata}, {Kneib}, {Koekemoer}, {Le F{\`e}vre}, {Lilly},
  {Surace}, {Taniguchi}, {Tasca}, {Thompson}, {Tresse}, {Zamojski}, {Zamorani},
  \& {Zucca}}]{Ilbert2010}
{Ilbert}, O., {Salvato}, M., {Le Floc'h}, E., {et~al.} 2010, \apj, 709, 644

\bibitem[{{Kassin} {et~al.}(2012){Kassin}, {Weiner}, {Faber}, {Gardner},
  {Willmer}, {Coil}, {Cooper}, {Devriendt}, {Dutton}, {Guhathakurta}, {Koo},
  {Metevier}, {Noeske}, \& {Primack}}]{Kassin2012}
{Kassin}, S.~A., {Weiner}, B.~J., {Faber}, S.~M., {et~al.} 2012, \apj, 758, 106

\bibitem[{{Kauffmann} {et~al.}(2003){Kauffmann}, {Heckman}, {White}, {Charlot},
  {Tremonti}, {Brinchmann}, {Bruzual}, {Peng}, {Seibert}, {Bernardi},
  {Blanton}, {Brinkmann}, {Castander}, {Cs{\'a}bai}, {Fukugita}, {Ivezic},
  {Munn}, {Nichol}, {Padmanabhan}, {Thakar}, {Weinberg}, \&
  {York}}]{Kauffmann2003}
{Kauffmann}, G., {Heckman}, T.~M., {White}, S.~D.~M., {et~al.} 2003, \mnras,
  341, 33

\bibitem[{{Koekemoer} {et~al.}(2011){Koekemoer}, {Faber}, {Ferguson}, {Grogin},
  {Kocevski}, {Koo}, {Lai}, {Lotz}, {Lucas}, {McGrath}, {Ogaz}, {Rajan},
  {Riess}, {Rodney}, {Strolger}, {Casertano}, {Castellano}, {Dahlen},
  {Dickinson}, {Dolch}, {Fontana}, {Giavalisco}, {Grazian}, {Guo}, {Hathi},
  {Huang}, {van der Wel}, {Yan}, {Acquaviva}, {Alexander}, {Almaini}, {Ashby},
  {Barden}, {Bell}, {Bournaud}, {Brown}, {Caputi}, {Cassata}, {Challis},
  {Chary}, {Cheung}, {Cirasuolo}, {Conselice}, {Roshan Cooray}, {Croton},
  {Daddi}, {Dav{\'e}}, {de Mello}, {de Ravel}, {Dekel}, {Donley}, {Dunlop},
  {Dutton}, {Elbaz}, {Fazio}, {Filippenko}, {Finkelstein}, {Frazer}, {Gardner},
  {Garnavich}, {Gawiser}, {Gruetzbauch}, {Hartley}, {H{\"a}ussler},
  {Herrington}, {Hopkins}, {Huang}, {Jha}, {Johnson}, {Kartaltepe},
  {Khostovan}, {Kirshner}, {Lani}, {Lee}, {Li}, {Madau}, {McCarthy},
  {McIntosh}, {McLure}, {McPartland}, {Mobasher}, {Moreira}, {Mortlock},
  {Moustakas}, {Mozena}, {Nandra}, {Newman}, {Nielsen}, {Niemi}, {Noeske},
  {Papovich}, {Pentericci}, {Pope}, {Primack}, {Ravindranath}, {Reddy},
  {Renzini}, {Rix}, {Robaina}, {Rosario}, {Rosati}, {Salimbeni}, {Scarlata},
  {Siana}, {Simard}, {Smidt}, {Snyder}, {Somerville}, {Spinrad}, {Straughn},
  {Telford}, {Teplitz}, {Trump}, {Vargas}, {Villforth}, {Wagner}, {Wandro},
  {Wechsler}, {Weiner}, {Wiklind}, {Wild}, {Wilson}, {Wuyts}, \&
  {Yun}}]{Koekemoer2011}
{Koekemoer}, A.~M., {Faber}, S.~M., {Ferguson}, H.~C., {et~al.} 2011, \apjs,
  197, 36

\bibitem[{{Kriek} {et~al.}(2011){Kriek}, {van Dokkum}, {Whitaker}, {Labb{\'e}},
  {Franx}, \& {Brammer}}]{Kriek2011}
{Kriek}, M., {van Dokkum}, P.~G., {Whitaker}, K.~E., {et~al.} 2011, \apj, 743,
  168

\bibitem[{{Kriek} {et~al.}(2010){Kriek}, {Labb{\'e}}, {Conroy}, {Whitaker},
  {van Dokkum}, {Brammer}, {Franx}, {Illingworth}, {Marchesini}, {Muzzin},
  {Quadri}, \& {Rudnick}}]{Kriek2010}
{Kriek}, M., {Labb{\'e}}, I., {Conroy}, C., {et~al.} 2010, \apjl, 722, L64

\bibitem[{{Lang} {et~al.}(2014){Lang}, {Wuyts}, {Somerville}, {Forster
  Schreiber}, {Genzel}, {Bell}, {Brammer}, {Dekel}, {Faber}, {Ferguson},
  {Grogin}, {Kocevski}, {Koekemoer}, {Lutz}, {McGrath}, {Momcheva}, {Nelson},
  {Primack}, {Rosario}, {Skelton}, {Tacconi}, {van Dokkum}, \&
  {Whitaker}}]{Lang2014}
{Lang}, P., {Wuyts}, S., {Somerville}, R., {et~al.} 2014, ArXiv e-prints,
  arXiv:1402.0866

\bibitem[{{Lanyon-Foster} {et~al.}(2007){Lanyon-Foster}, {Conselice}, \&
  {Merrifield}}]{Lanyon2007}
{Lanyon-Foster}, M.~M., {Conselice}, C.~J., \& {Merrifield}, M.~R. 2007,
  \mnras, 380, 571

\bibitem[{{Lanyon-Foster} {et~al.}(2012){Lanyon-Foster}, {Conselice}, \&
  {Merrifield}}]{Lanyon2012}
---. 2012, \mnras, 424, 1852

\bibitem[{{Lilly} {et~al.}(2013){Lilly}, {Carollo}, {Pipino}, {Renzini}, \&
  {Peng}}]{Lilly2013}
{Lilly}, S.~J., {Carollo}, C.~M., {Pipino}, A., {Renzini}, A., \& {Peng}, Y.
  2013, \apj, 772, 119

\bibitem[{{Lilly} {et~al.}(1996){Lilly}, {Le Fevre}, {Hammer}, \&
  {Crampton}}]{Lilly1996}
{Lilly}, S.~J., {Le Fevre}, O., {Hammer}, F., \& {Crampton}, D. 1996, \apjl,
  460, L1

\bibitem[{{Lin} {et~al.}(2004){Lin}, {Koo}, {Willmer}, {Patton}, {Conselice},
  {Yan}, {Coil}, {Cooper}, {Davis}, {Faber}, {Gerke}, {Guhathakurta}, \&
  {Newman}}]{Lin2004}
{Lin}, L., {Koo}, D.~C., {Willmer}, C.~N.~A., {et~al.} 2004, \apjl, 617, L9

\bibitem[{{Lotz} {et~al.}(2011){Lotz}, {Jonsson}, {Cox}, {Croton}, {Primack},
  {Somerville}, \& {Stewart}}]{Lotz2011}
{Lotz}, J.~M., {Jonsson}, P., {Cox}, T.~J., {et~al.} 2011, \apj, 742, 103

\bibitem[{{Madau} {et~al.}(1998){Madau}, {Pozzetti}, \&
  {Dickinson}}]{Madau1998}
{Madau}, P., {Pozzetti}, L., \& {Dickinson}, M. 1998, \apj, 498, 106

\bibitem[{{Maraston} {et~al.}(2006){Maraston}, {Daddi}, {Renzini}, {Cimatti},
  {Dickinson}, {Papovich}, {Pasquali}, \& {Pirzkal}}]{Maraston2006}
{Maraston}, C., {Daddi}, E., {Renzini}, A., {et~al.} 2006, \apj, 652, 85

\bibitem[{{Maraston} {et~al.}(2010){Maraston}, {Pforr}, {Renzini}, {Daddi},
  {Dickinson}, {Cimatti}, \& {Tonini}}]{Maraston2010}
{Maraston}, C., {Pforr}, J., {Renzini}, A., {et~al.} 2010, \mnras, 407, 830

\bibitem[{{Miller} {et~al.}(2011){Miller}, {Bundy}, {Sullivan}, {Ellis}, \&
  {Treu}}]{Miller2011}
{Miller}, S.~H., {Bundy}, K., {Sullivan}, M., {Ellis}, R.~S., \& {Treu}, T.
  2011, \apj, 741, 115

\bibitem[{{Mobasher} {et~al.}(2009){Mobasher}, {Dahlen}, {Hopkins}, {Scoville},
  {Capak}, {Rich}, {Sanders}, {Schinnerer}, {Ilbert}, {Salvato}, \&
  {Sheth}}]{Mobasher2009}
{Mobasher}, B., {Dahlen}, T., {Hopkins}, A., {et~al.} 2009, \apj, 690, 1074

\bibitem[{{Nayyeri} {et~al.}(2014){Nayyeri}, {Mobasher}, {Hemmati}, {De
  Barros}, {Ferguson}, {Wiklind}, {Dahlen}, {Dickinson}, {Giavalisco},
  {Fontana}, {Ashby}, {Barro}, {Guo}, {Hathi}, {Kassin}, {Koekemoer}, \&
  {Willner}}]{Nayyeri2014}
{Nayyeri}, H., {Mobasher}, B., {Hemmati}, S., {et~al.} 2014, ArXiv e-prints,
  arXiv:1408.3684

\bibitem[{{Noeske} {et~al.}(2007){Noeske}, {Weiner}, {Faber}, {Papovich},
  {Koo}, {Somerville}, {Bundy}, {Conselice}, {Newman}, {Schiminovich}, {Le
  Floc'h}, {Coil}, {Rieke}, {Lotz}, {Primack}, {Barmby}, {Cooper}, {Davis},
  {Ellis}, {Fazio}, {Guhathakurta}, {Huang}, {Kassin}, {Martin}, {Phillips},
  {Rich}, {Small}, {Willmer}, \& {Wilson}}]{Noeske2007}
{Noeske}, K.~G., {Weiner}, B.~J., {Faber}, S.~M., {et~al.} 2007, \apjl, 660,
  L43

\bibitem[{{Oke} \& {Gunn}(1983)}]{Oke1983}
{Oke}, J.~B., \& {Gunn}, J.~E. 1983, \apj, 266, 713

\bibitem[{{Papovich} {et~al.}(2001){Papovich}, {Dickinson}, \&
  {Ferguson}}]{Papovich2001}
{Papovich}, C., {Dickinson}, M., \& {Ferguson}, H.~C. 2001, \apj, 559, 620

\bibitem[{{Patton} {et~al.}(2002){Patton}, {Pritchet}, {Carlberg}, {Marzke},
  {Yee}, {Hall}, {Lin}, {Morris}, {Sawicki}, {Shepherd}, \&
  {Wirth}}]{patton2002}
{Patton}, D.~R., {Pritchet}, C.~J., {Carlberg}, R.~G., {et~al.} 2002, \apj,
  565, 208

\bibitem[{{Pickles}(1998)}]{PICKLES1998}
{Pickles}, A.~J. 1998, \pasp, 110, 863

\bibitem[{{Reddy} {et~al.}(2012){Reddy}, {Pettini}, {Steidel}, {Shapley},
  {Erb}, \& {Law}}]{Reddy2012}
{Reddy}, N.~A., {Pettini}, M., {Steidel}, C.~C., {et~al.} 2012, \apj, 754, 25

\bibitem[{{Reddy} {et~al.}(2006){Reddy}, {Steidel}, {Fadda}, {Yan}, {Pettini},
  {Shapley}, {Erb}, \& {Adelberger}}]{Reddy2006}
{Reddy}, N.~A., {Steidel}, C.~C., {Fadda}, D., {et~al.} 2006, \apj, 644, 792

\bibitem[{{Rudnick} {et~al.}(2003){Rudnick}, {Rix}, {Franx}, {Labb{\'e}},
  {Blanton}, {Daddi}, {F{\"o}rster Schreiber}, {Moorwood}, {R{\"o}ttgering},
  {Trujillo}, {van der Wel}, {van der Werf}, {van Dokkum}, \& {van
  Starkenburg}}]{Rudnick2003}
{Rudnick}, G., {Rix}, H.-W., {Franx}, M., {et~al.} 2003, \apj, 599, 847

\bibitem[{{Salim} {et~al.}(2009){Salim}, {Dickinson}, {Michael Rich},
  {Charlot}, {Lee}, {Schiminovich}, {P{\'e}rez-Gonz{\'a}lez}, {Ashby},
  {Papovich}, {Faber}, {Ivison}, {Frayer}, {Walton}, {Weiner}, {Chary},
  {Bundy}, {Noeske}, \& {Koekemoer}}]{Salim2009}
{Salim}, S., {Dickinson}, M., {Michael Rich}, R., {et~al.} 2009, \apj, 700, 161

\bibitem[{{Salmon} {et~al.}(2014){Salmon}, {Papovich}, {Finkelstein}, {Tilvi},
  {Finlator}, {Behroozi}, {Dahlen}, {Dav{\'e}}, {Dekel}, {Dickinson},
  {Ferguson}, {Giavalisco}, {Long}, {Lu}, {Reddy}, {Somerville}, \&
  {Wechsler}}]{Salmon2014}
{Salmon}, B., {Papovich}, C., {Finkelstein}, S.~L., {et~al.} 2014, ArXiv
  e-prints, arXiv:1407.6012

\bibitem[{{Schade} {et~al.}(1995){Schade}, {Lilly}, {Crampton}, {Hammer}, {Le
  Fevre}, \& {Tresse}}]{Schade1995}
{Schade}, D., {Lilly}, S.~J., {Crampton}, D., {et~al.} 1995, \apjl, 451, L1

\bibitem[{Scott(1979)}]{Scott1979}
Scott, D.~W. 1979, Biometrika, 66, 605

\bibitem[{{Shapley} {et~al.}(2001){Shapley}, {Steidel}, {Adelberger},
  {Dickinson}, {Giavalisco}, \& {Pettini}}]{Shapley2001}
{Shapley}, A.~E., {Steidel}, C.~C., {Adelberger}, K.~L., {et~al.} 2001, \apj,
  562, 95

\bibitem[{{Tacconi} {et~al.}(2013){Tacconi}, {Neri}, {Genzel}, {Combes},
  {Bolatto}, {Cooper}, {Wuyts}, {Bournaud}, {Burkert}, {Comerford}, {Cox},
  {Davis}, {F{\"o}rster Schreiber}, {Garc{\'{\i}}a-Burillo}, {Gracia-Carpio},
  {Lutz}, {Naab}, {Newman}, {Omont}, {Saintonge}, {Shapiro Griffin}, {Shapley},
  {Sternberg}, \& {Weiner}}]{Tacconi2013}
{Tacconi}, L.~J., {Neri}, R., {Genzel}, R., {et~al.} 2013, \apj, 768, 74

\bibitem[{{van de Voort} {et~al.}(2011){van de Voort}, {Schaye}, {Booth}, \&
  {Dalla Vecchia}}]{Voort2011}
{van de Voort}, F., {Schaye}, J., {Booth}, C.~M., \& {Dalla Vecchia}, C. 2011,
  \mnras, 415, 2782

\bibitem[{{Walcher} {et~al.}(2011){Walcher}, {Groves}, {Budav{\'a}ri}, \&
  {Dale}}]{Walcher2011}
{Walcher}, J., {Groves}, B., {Budav{\'a}ri}, T., \& {Dale}, D. 2011, \apss,
  331, 1

\bibitem[{{Welikala} {et~al.}(2011){Welikala}, {Hopkins}, {Robertson},
  {Connolly}, {Tasca}, {Koekemoer}, {Ilbert}, {Bardelli}, {Kneib}, \&
  {Zentner}}]{Welikala2011}
{Welikala}, N., {Hopkins}, A.~M., {Robertson}, B.~E., {et~al.} 2011, ArXiv
  e-prints, arXiv:1112.2657

\bibitem[{{Whitaker} {et~al.}(2014){Whitaker}, {Franx}, {Leja}, {van Dokkum},
  {Henry}, {Skelton}, {Fumagalli}, {Momcheva}, {Brammer}, {Labbe}, {Nelson}, \&
  {Rigby}}]{Whitaker2014}
{Whitaker}, K.~E., {Franx}, M., {Leja}, J., {et~al.} 2014, ArXiv e-prints,
  arXiv:1407.1843

\bibitem[{{Wuyts} {et~al.}(2011){Wuyts}, {F{\"o}rster Schreiber}, {van der
  Wel}, {Magnelli}, {Guo}, {Genzel}, {Lutz}, {Aussel}, {Barro}, {Berta},
  {Cava}, {Graci{\'a}-Carpio}, {Hathi}, {Huang}, {Kocevski}, {Koekemoer},
  {Lee}, {Le Floc'h}, {McGrath}, {Nordon}, {Popesso}, {Pozzi}, {Riguccini},
  {Rodighiero}, {Saintonge}, \& {Tacconi}}]{wuyts2011}
{Wuyts}, S., {F{\"o}rster Schreiber}, N.~M., {van der Wel}, A., {et~al.} 2011,
  \apj, 742, 96

\bibitem[{{Wuyts} {et~al.}(2012){Wuyts}, {F{\"o}rster Schreiber}, {Genzel},
  {Guo}, {Barro}, {Bell}, {Dekel}, {Faber}, {Ferguson}, {Giavalisco}, {Grogin},
  {Hathi}, {Huang}, {Kocevski}, {Koekemoer}, {Koo}, {Lotz}, {Lutz}, {McGrath},
  {Newman}, {Rosario}, {Saintonge}, {Tacconi}, {Weiner}, \& {van der
  Wel}}]{Wuyts2012}
{Wuyts}, S., {F{\"o}rster Schreiber}, N.~M., {Genzel}, R., {et~al.} 2012, \apj,
  753, 114

\bibitem[{{Wuyts} {et~al.}(2013){Wuyts}, {F{\"o}rster Schreiber}, {Nelson},
  {van Dokkum}, {Brammer}, {Chang}, {Faber}, {Ferguson}, {Franx}, {Fumagalli},
  {Genzel}, {Grogin}, {Kocevski}, {Koekemoer}, {Lundgren}, {Lutz}, {McGrath},
  {Momcheva}, {Rosario}, {Skelton}, {Tacconi}, {van der Wel}, \&
  {Whitaker}}]{Wuyts2013}
{Wuyts}, S., {F{\"o}rster Schreiber}, N.~M., {Nelson}, E.~J., {et~al.} 2013,
  \apj, 779, 135

\bibitem[{{Zibetti} {et~al.}(2009){Zibetti}, {Charlot}, \& {Rix}}]{Zibetti2009}
{Zibetti}, S., {Charlot}, S., \& {Rix}, H.-W. 2009, \mnras, 400, 1181

\end{thebibliography}
\end{tiny}

\appendix
\section{Choice of Grid for model Libraries}

The physical parameters measured by fitting the SED of the integrated
light of galaxies or individual pixels are drawn from a library of
model SEDs. This means that choices in the discreate nature of model
libraries affects the estimated output parameters. Many studies look at the effect of these
choices, including their uncertainties, on the integrated (see e.g. \citealt{Conroy2013}, \citealt{Ilbert2010}) or resolved (see e.g. \citealt{Welikala2011})
SED fitting results. In this appendix, we analyze the effect of some of our assumptions regarding the
model library on the output parameters. 

To inspect how choosing a particular grid of input parameters for the library
affects our SED fitting output estimates and therefore the maps, we
select 9 representative galaxies from our sample spanning the whole
redshift range studied in this work ($0.2<z<1.2$).  We fit the SED per
resolution element following the methods described in \S 3, using
different libraries. We change one parameter in the library at a time
while keeping all the other library parameters the same (See table). We then compare the
measurements at resolution elements. The first run (Run1) is performed
using the library described in \S 3 and all the other measurements
will be compared to this one. The next two runs (Run2 and Run3) are
generated by changing the resolution in extinction and age. Ideally,
one would expect the highest resolution grids to produce smoother maps, but in reality increasing the
library size arbitrarily could lead to more degeneracies
(e.g. \citealt{Walcher2011}), in the $\chi^{2}$ fitting of the SEDs,
let alone the high computational cost. And finally in the last two
runs (Run4 and Run5) we examine the outcome by changing stellar
metallicity to solar (m62) and then to 20\% solar (m42).

\begin{center}
\begin{tabular}{|c|c|c|c|c|c|}
\hline
Run&SFH&$\tau$ (Gyr)&E(B-V)&Age(Gyr)&metallicity ($Z_{solar}$)\\\hline
1&declining&21:  0.01-10.0&15:  0-1.0&57:  0.01-13.5& 40\%\\\hline
2&declining&21:   0.01-10.0&{\bf 21:  0-1.0}&57:  0.01-13.5& 40\% \\\hline
3&declining&21:   0.01-10.0&15:  0-1.0&{\bf 40:  0.01-13.5} &40\%\\\hline
4&declining&21:  0.01-10.0&15:  0-1.0&57:  0.01-13.5& {\bf 100\%} \\\hline
5&declining&21:   0.01-10.0&15:  0-1.0&57:  0.01-13.5& {\bf 20\%} \\\hline
\end{tabular}
\end{center}

We compare the E(B-V) maps from the first two  runs, in which we
increase the number of steps in extinction by 25\%. Subtracting the
two maps we see a median offset of 0.0 and a scatter in the range of
$0.0-0.02$. The small median difference and scatter should not be a
surprise as we have only changed the resolution and not the range. In Figure 15, we plot the
extinction maps for one of the galaxies using the two libraries and
their difference. This Figure demonstrates the point that extinction
maps produced from the two runs are visually indistinguishable. The
same has been seen for all the test galaxies. Even though at some
individual resolution elements there are small differences, the overall structure and
patterns remain intact. 
\begin{figure*}[htbp]
\centering
\includegraphics[trim=0.0cm 15.0cm 0.0cm 6.0cm,clip,width=6in]{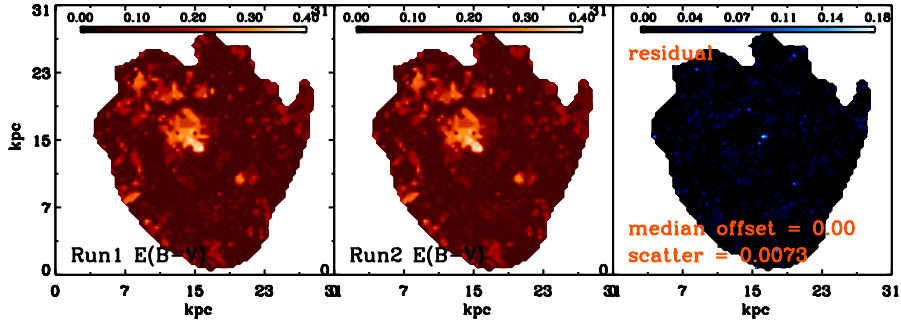}
\caption{Comparing E(B-V) maps from Run1 and Run2 for a galaxy in our
  sample. Run 2 has a higher resolution in extinction levels compared
  to Run1. }
\end{figure*}

\begin{figure*}[htbp]
\centering
\includegraphics[trim=0.0cm 15.0cm 0.0cm 6.0cm,clip,width=6in]{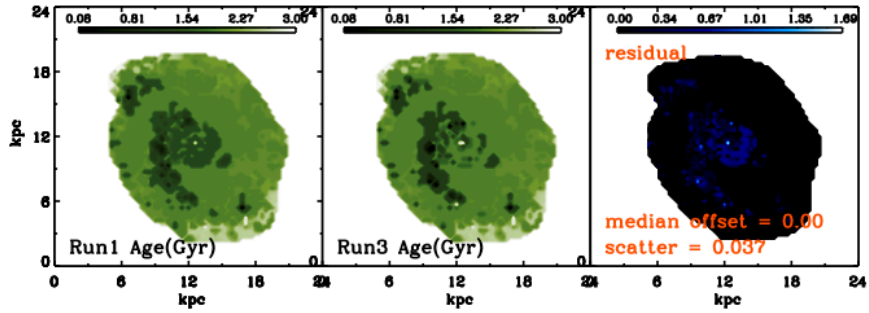}
\caption{Comparing Age maps from Run1 and Run3 for a galaxy in our
  sample. Run 3 has less resolution in age levels compared
  to Run1. }
\end{figure*}
The same situation holds, comparing the age maps from third
run to the first, with the overall trends not changing but slightly
larger offsets resolution element by resolution element (Figure 16). The median offset is 0.0 for all the
test galaxies, but there is a larger scatter in the range of
$0.01-0.1$. The difference in each resolution element is always well below its $1\sigma$
uncertainty of that resolution element in both runs. This does not necessarily mean
that the difference is small for all resolution elements, but rather it shows
that for resolution elements with less constrained PDFs the resolution can play an
important role.

\begin{figure*}[htbp]
\centering
\includegraphics[width=7in]{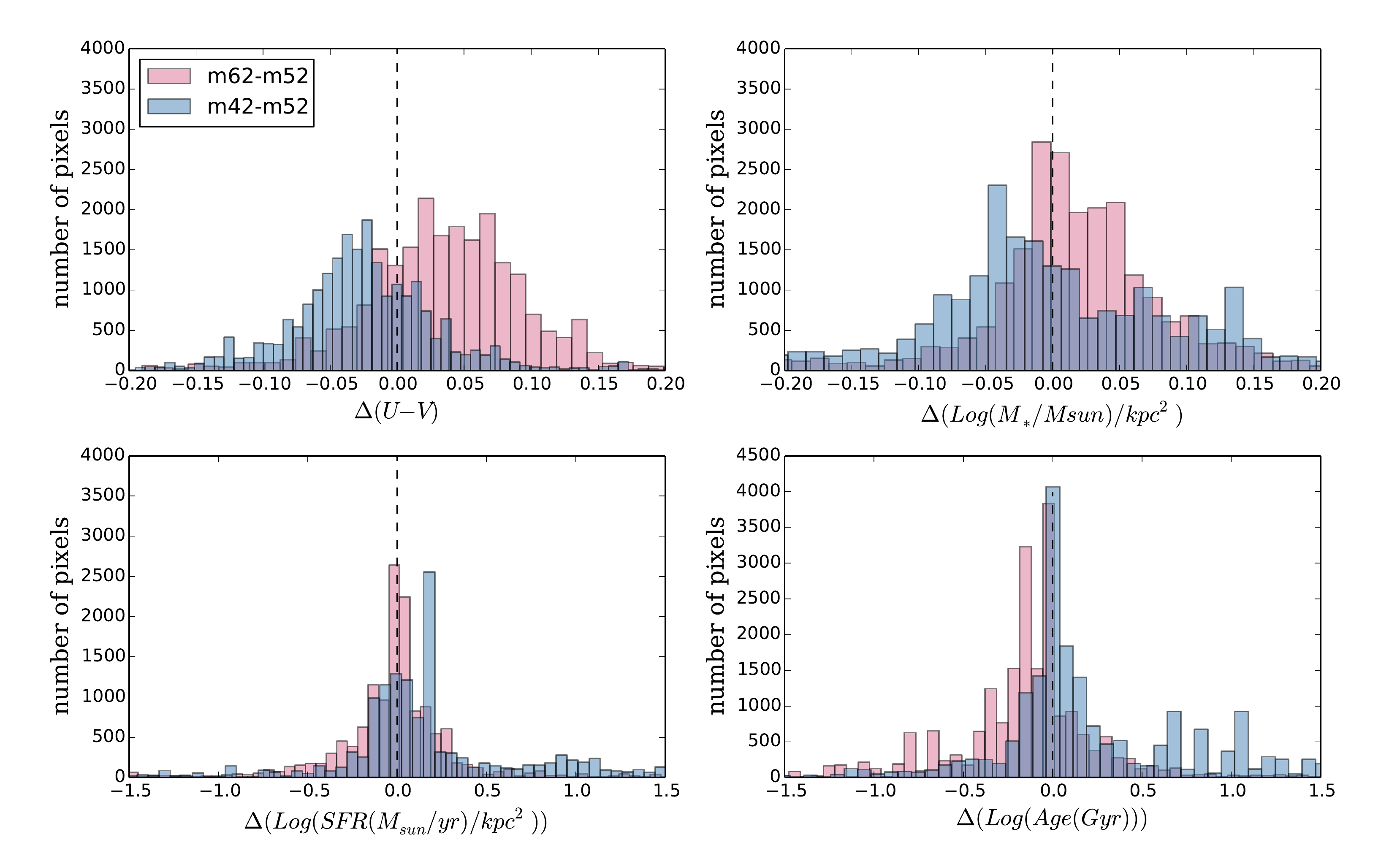}
\caption{Effect of metallicity on rest-frame (U-V), stellar mass
  surface density, SFR surface density and stellar age. Red and blue colors correspond to
$\Delta(m62-m52)$ (Run4-Run1) and $\Delta
(m42-m52)$ (Run5-Run1), respectively.}
\end{figure*}

In the last two runs we study the dependence of the parameters on variations in
metallicity. We change the metallicity once to solar (m62) and once to
20\% solar (m42) and compare the physical parameters with the first
run which has metallicity fixed to 40\% solar (m52). This is shown in
Figure 17, where red and blue histograms correspond to
$\Delta(m62-m52)$ (Run4-Run1) and $\Delta
(m42-m52)$ (Run5-Run1), respectively. The median offset and
dispersion in rest-frame (U-V) color, stellar mass, SFR and stellar
age over all resolution elements in the test galaxies are listed in the following table.

\begin{center}
\begin{tabular}{|l|c|c|c|c|}
\hline
&(U-V)&Log($\Sigma M_{*}$)&Log($\Sigma$SFR)&Log(Age)\\
&(rest-frame) & $(M_{sun} /\text{kpc}^{2})$ & $(M_{sun}yr^{-1}  /\text{kpc}^{2})$ & (Gyr) \\
\hline
Run4 - Run1 median offset&0.04&0.01&-0.01&-0.12\\
Run4 - Run1 dispersion&0.06&0.09&0.53&1.14\\
Run5 - Run1 median offset&-0.03&-0.01&0.14&0.11\\
Run5 - Run1 dispersion&0.06&0.14&0.77&1.6\\
\hline
\end{tabular}
\end{center}

In this work, we fixed the metallicity to 40\% solar and produced maps
of physical properties of galaxies. We selected regions based on (U-V)
rest-frame color. Having fixed the metallicity to a higher or lower
value, the same regions would have been identified but with slightly offseted
properties, as quantified in the above table. This is because the
region selection is based on the color distribution in each galaxy and
does not assume a predefined cut. This is of course different from having the metallicity as a
free parameter, which will expand the degeneracies, especially with
age. We find that assuming stellar metallicity leads to redder (U-V) rest-frame colors, slightly larger masses,
no significant change in median SFR (although significant scatter) and younger ages. Finally, further work is needed to fully uncover the effects of all
different grid parameter choices (from IMF and SSPs to star formation histories) on resolved SED fitting results.
\end{document}